\documentclass[12pt]{article}
\pdfoutput=1
\usepackage[colorlinks,linkcolor=Blue,citecolor=Blue,bookmarks,bookmarksnumbered]{hyperref}
\usepackage[scaled=0.85]{helvet}
\usepackage{amsmath,amssymb,accents,mathrsfs,XoohmE}
\usepackage{graphicx,color}
\usepackage{booktabs}
\usepackage{multirow}
\usepackage{placeins}
\usepackage{amsmath}

\definecolor{Green}  {rgb}{0.10,0.70,0.10} 
\definecolor{Orange} {rgb}{1.00,0.50,0.15} 
\definecolor{Red}    {rgb}{0.90,0.00,0.12} 
\definecolor{Purple} {rgb}{0.50,0.25,0.55} 
\definecolor{Turque} {rgb}{0.00,0.65,0.85} 
\definecolor{Blue}   {rgb}{0.00,0.00,1.00} 
\definecolor{Magenta}{rgb}{1.00,0.00,1.00} 
\definecolor{Gold}   {rgb}{1.00,0.75,0.25} 
\definecolor{Seaweed}{rgb}{0.01,0.24,0.09} 
\definecolor{Brown}  {rgb}{0.43,0.26,0.32} 
\definecolor{grey1}  {rgb}{0.20,0.20,0.20} 
\definecolor{grey2}  {rgb}{0.40,0.40,0.40} 
\definecolor{grey3}  {rgb}{0.60,0.60,0.60} 
\definecolor{grey4}  {rgb}{0.80,0.80,0.80} 
\definecolor{grey5}  {rgb}{0.90,0.90,0.90} 
\def\C#1#2{{\ifcase#1\or
             \color{Green}\or \color{Orange}\or \color{Red}\or
              \color{Purple}\or \color{Turque}\or \color{Blue}\or
               \color{Magenta}\or \color{Gold}\or \color{Seaweed}\or
                \color{Brown}\or\color{grey1}\or\color{grey2}\or
                 \color{grey3}\else\color{grey4}\fi#2}}

\definecolor{Slate} {rgb}{0.00,0.45,0.55}

\def\rI{{\rm I}}
\def\rJ{{\rm J}}
\def\rK{{\rm K}}
\def\rL{{\rm L}}

\def\hi{{\hat\imath}}
\def\hj{{\hat\jmath}}
\def\hk{{\hat{k}}}
\def\hl{{\hat\ell}}

\def\fracm#1#2{\hbox{\large{${\frac{{#1}}{{#2}}}$}}}

\def\vCent#1{\vcenter{\hbox{\hss#1\hss}}}
\def\be{\begin{equation}}
\def\ee{\end{equation}}
\newcommand{\bea}{\begin{eqnarray}}
\newcommand{\eea}{\end{eqnarray}}
\newcommand{\ena}{\end{eqnarray}}

\def\pp{{\mathchoice
          {
              \kern 1pt%
              \raise 1pt
              \vbox{\hrule width5pt height0.4pt depth0pt
                    \kern -2pt
                    \hbox{\kern 2.3pt
                          \vrule width0.4pt height6pt depth0pt
                          }
                    \kern -2pt
                    \hrule width5pt height0.4pt depth0pt}%
                    \kern 1pt
           }
            {
              \kern 1pt%
              \raise 1pt
              \vbox{\hrule width4.3pt height0.4pt depth0pt
                    \kern -1.8pt
                    \hbox{\kern 1.95pt
                          \vrule width0.4pt height5.4pt depth0pt
                          }
                    \kern -1.8pt
                    \hrule width4.3pt height0.4pt depth0pt}%
                    \kern 1pt
            }
            {
              \kern 0.5pt%
              \raise 1pt
              \vbox{\hrule width4.0pt height0.3pt depth0pt
                    \kern -1.9pt  
                    \hbox{\kern 1.85pt
                          \vrule width0.3pt height5.7pt depth0pt
                          }
                    \kern -1.9pt
                    \hrule width4.0pt height0.3pt depth0pt}%
                    \kern 0.5pt
            }
            {
              \kern 0.5pt%
              \raise 1pt
              \vbox{\hrule width3.6pt height0.3pt depth0pt
                    \kern -1.5pt
                    \hbox{\kern 1.65pt
                          \vrule width0.3pt height4.5pt depth0pt
                          }
                    \kern -1.5pt
                    \hrule width3.6pt height0.3pt depth0pt}%
                    \kern 0.5pt
            }
        }}

\def\mm{{\mathchoice
                       {
                             \kern 1pt
               \raise 1pt    \vbox{\hrule width5pt height0.4pt depth0pt
                                  \kern 2pt
                                  \hrule width5pt height0.4pt depth0pt}
                             \kern 1pt}
                       {
                            \kern 1pt
               \raise 1pt \vbox{\hrule width4.3pt height0.4pt depth0pt
                                  \kern 1.8pt
                                  \hrule width4.3pt height0.4pt depth0pt}
                             \kern 1pt}
                       {
                            \kern 0.5pt
               \raise 1pt
                            \vbox{\hrule width4.0pt height0.3pt depth0pt
                                  \kern 1.9pt
                                  \hrule width4.0pt height0.3pt depth0pt}
                            \kern 1pt}
                       {
                           \kern 0.5pt
             \raise 1pt  \vbox{\hrule width3.6pt height0.3pt depth0pt
                                  \kern 1.5pt
                                  \hrule width3.6pt height0.3pt depth0pt}
                           \kern 0.5pt}
                       }}

\def\ad{{\kern0.5pt
                   \alpha \kern-5.05pt \raise5.8pt\hbox{$\textstyle.$}\kern
0.5pt}}

\def\bd{{\kern0.5pt
                   \beta \kern-5.05pt \raise5.8pt\hbox{$\textstyle.$}\kern
0.5pt}}

\def\qd{{\kern0.5pt
                   q \kern-5.05pt \raise5.8pt\hbox{$\textstyle.$}\kern
0.5pt}}
\def\Dot#1{{\kern0.5pt
     {#1} \kern-5.05pt \raise5.8pt\hbox{$\textstyle.$}\kern
0.5pt}}

\catcode`@=11
\def\un#1{\relax\ifmmode\@@underline#1\else
        $\@@underline{\hbox{#1}}$\relax\fi}
\catcode`@=12

\def\a{\alpha}
\def\b{\beta}
\def\c{\chi}
\def\d{\delta}

\def\g{\gamma}

\def\i{\iota}
\def\j{\psi}
\def\k{\kappa}
\def\l{\lambda}
\def\m{\mu}

\def\o{\omega}

\def\r{\rho}
\def\s{\sigma}
\def\t{\tau}

\def\L{\Lambda}
\def\O{\Omega}
\def\P{\Pi}

\def\S{\Sigma}

\def\ca{{\cal A}}

\def\cf{{\cal F}}

\def\dslash{\not{\hbox{\kern-2pt $\partial$}}}
\def\Dslash{\not{\hbox{\kern-4pt $D$}}}
\def\pslash{\not{\hbox{\kern-2.3pt $p$}}}
 \newtoks\slashfraction
 \slashfraction={.13}
 \def\slash#1{\setbox0\hbox{$ #1 $}
 \setbox0\hbox to \the\slashfraction\wd0{\hss \box0}/\box0 }
 
\def\kcr{{\hbox{\ro \char'170}}}                
\def\ktl{{\hbox{\ro \char'170}}}        
\def\ktr{{\hbox{\ro \char'170}}}        
\def\kbl{{\hbox{\ro \char'170}}}        
\def\kbr{{\hbox{\ro \char'170}}}        

\def\plpl{\raise-2pt\hbox{$\raise3pt\hbox{$_+$}\hskip-6.67pt\raise0.0pt
\hbox{$^+$}\hskip 0.01pt$}}
\def\mimi{\raise-2pt\hbox{$\raise3pt\hbox{$_-$}\hskip-6.67pt\raise0.0pt
\hbox{$^-$}\hskip 0.01pt$}} 

\def\bo{{\raise.15ex\hbox{\large$\Box$}}}               
\def\TH{{\raise.2ex\hbox{$\displaystyle \bigodot$}\mskip-4.7mu \llap H \;}}
\def\face{{\raise.2ex\hbox{$\displaystyle \bigodot$}\mskip-2.2mu \llap {$\ddot
        \smile$}}}                                      

\def\dt#1{\on{\hbox{\bf .}}{#1}}                
\def\Dot#1{\dt{#1}}

\def\Tilde#1{\widetilde{#1}}                    
\def\Hat#1{\widehat{#1}}                        
\def\Bar#1{\overline{#1}}                       
\def\leftrightarrowfill{$\mathsurround=0pt \mathord\leftarrow \mkern-6mu
        \cleaders\hbox{$\mkern-2mu \mathord- \mkern-2mu$}\hfill
        \mkern-6mu \mathord\rightarrow$}
\def\dvec#1{\vbox{\ialign{##\crcr
        \leftrightarrowfill\crcr\noalign{\kern-1pt\nointerlineskip}
        $\hfil\displaystyle{#1}\hfil$\crcr}}}           
\def\dt#1{{\buildrel {\hbox{\LARGE .}} \over {#1}}}     

\def\fracm#1#2{\hbox{\large{${\frac{{#1}}{{#2}}}$}}}
\def\sfrac#1#2{{\vphantom1\smash{\lower.5ex\hbox{\small$#1$}}\over
        \vphantom1\smash{\raise.4ex\hbox{\small$#2$}}}} 
\def\bfrac#1#2{{\vphantom1\smash{\lower.5ex\hbox{$#1$}}\over
        \vphantom1\smash{\raise.3ex\hbox{$#2$}}}}       
\def\afrac#1#2{{\vphantom1\smash{\lower.5ex\hbox{$#1$}}\over#2}}    

\let\bm\relax
\newcommand{\bm}[1]{{\boldsymbol{#1}}}

\def\ad{{\dot{\alpha}}}
\def\bd{{\dot{\beta}}}

 \font\rOpe=cmsy10                        
 \def\ktl{{\hbox{\rOpe\char'170}}}        
 \def\kbl{{\hbox{\rOpe\char'170}}}        
 \def\kcr{{\reflectbox{\rOpe\char'170}}}        
 \def\ktr{{\reflectbox{\rOpe\char'170}}}        
 \def\kbr{{\reflectbox{\rOpe\char'170}}}        
 \def\Border{\vbox{\hsize0pt
        \setlength{\unitlength}{1mm}
        \newcount\xco
        \newcount\yco
        \xco=-21
        \yco=12
        \begin{picture}(0,0)(-7.5,0)
        \put(\xco,\yco){$\ktl$}
        \advance\yco by-1
        {\loop
        \put(\xco,\yco){$\kcr$}
        \advance\yco by-2
        \ifnum\yco>-240
        \repeat
        \put(\xco,\yco){$\kbl$}}
        \xco=170
        \yco=12
        \put(\xco,\yco){$\ktr$}
        \advance\yco by-1
        {\loop
        \put(\xco,\yco){$\kcr$}
        \advance\yco by-2
        \ifnum\yco>-240
        \repeat
        \put(\xco,\yco){$\kbr$}}
        \put(-19.5,13){\scalebox{.6065}{%
         University of Maryland Center for String and Particle  Theory \&\ Physics Department%
        |University of Maryland Center for String and Particle  Theory \&\ Physics Department}}
        \put(-19.5,-241.5){\scalebox{.5835}{%
         ****University of Maryland * Center for String and
         Particle  Theory* Physics Department****University of Maryland *Center
        for String and Particle  Theory* Physics Department}}
        \end{picture}
        \par\vskip-8mm}}
\definecolor{UMred}{rgb}{.9,.05,.2}
\definecolor{HUblue}{rgb}{.0,.3,.7}
 \def\UMbanner{\vbox{\hsize0pt
        \setlength{\unitlength}{.4mm}
        \thicklines  
        \begin{picture}(0,0)(-30,-10)
        \put(165,2){\line(1,0){4}}
        \put(170,2){\line(1,0){4}}
        \put(180,2){\line(1,0){4}}
        \put(175,-14){\line(1,0){4}}
        \put(180,-14){\line(1,0){4}}
        \put(185,-14){\line(1,0){4}}
        \put(169,-14){\line(0,1){16}}
        \put(170,-14){\line(0,1){16}}
        \put(179,-14){\line(0,1){16}}
        \put(180,-14){\line(0,1){16}}
        \put(184,-14){\line(0,1){16}}
        \put(185,-14){\line(0,1){16}}
        \put(169,2){\oval(8,32)[bl]}
        \put(170,2){\oval(8,32)[br]}
        \put(179,-14){\oval(8,32)[tl]}
        \put(185,-14){\oval(8,32)[tr]}
        \end{picture}
        \par\vskip-6.5mm
        \thicklines}}

\definecolor{Red}    {rgb}{0.90,0.00,0.12} 
\definecolor{Blue}   {rgb}{0.00,0.00,1.00} 
\definecolor{Green}  {rgb}{0.10,0.70,0.10} 
\definecolor{Turque} {rgb}{0.00,0.65,0.85} 
\definecolor{Orange} {rgb}{1.00,0.50,0.15} 
\definecolor{Magenta}{rgb}{1.00,0.00,1.00} 
\definecolor{Gold}   {rgb}{1.00,0.75,0.25} 
\definecolor{Seaweed}{rgb}{0.01,0.24,0.09} 
\definecolor{Purple} {rgb}{0.50,0.25,0.55} 
\definecolor{Brown}  {rgb}{0.43,0.26,0.32} 
\definecolor{grey1}  {rgb}{0.20,0.20,0.20} 
\definecolor{grey2}  {rgb}{0.40,0.40,0.40} 
\definecolor{grey3}  {rgb}{0.60,0.60,0.60} 
\definecolor{grey4}  {rgb}{0.80,0.80,0.80} 
\definecolor{grey5}  {rgb}{0.90,0.90,0.90} 
\def\C#1#2{{\ifcase#1\or
             \color{Red}\or \color{Green}\or \color{Blue}\or\
              \color{Turque}\or \color{Orange}\or \color{Magenta}\or 
               \color{Gold}\or \color{Seaweed}\or \color{Purple}\or
                \color{Brown}\or\color{grey1}\or\color{grey2}\or
                 \color{grey3}\else\color{grey4}\fi#2}}

\definecolor{Slate} {rgb}{0.00,0.45,0.55}

\newdimen\parshift\parshift=\parindent
\catcode`@=11
 \long\def\@footnotetext#1{\insert\footins{\reset@font\footnotesize
           \interlinepenalty\interfootnotelinepenalty\splittopskip%
            \footnotesep\splitmaxdepth\dp\strutbox\floatingpenalty\@MM%
             \hsize\columnwidth\addtolength{\hsize}{-2\parindent}
              \@parboxrestore\protected@edef\@currentlabel%
              {\csname p@footnote\endcsname\@thefnmark}%
                \color@begingroup%
                 \@makefntext{\rule\z@\footnotesep\ignorespaces#1%
                  \@finalstrut\strutbox}%
                \color@endgroup}}
 \long\def\@makefntext#1{\hglue\parshift%
           \vbox{\noindent\baselineskip=11pt plus.5pt minus.5pt\hb@xt@0em{\hss\@makefnmark\kern1pt}#1}}
\catcode`@=12

\newskip\humongous \humongous=0pt plus 1000pt minus 1000pt
\def\caja{\mathsurround=0pt}
\def\eqalign#1{\,\vcenter{\openup2\jot \caja
        \ialign{\strut \hfil$\displaystyle{##}$&$
        \displaystyle{{}##}$\hfil\crcr#1\crcr}}\,}
\newif\ifdtup
\def\panorama{\global\dtuptrue \openup2\jot \caja
        \everycr{\noalign{\ifdtup \global\dtupfalse
        \vskip-\lineskiplimit \vskip\normallineskiplimit
        \else \penalty\interdisplaylinepenalty \fi}}}

\makeatletter
\def\section{\@startsection{section}{1}{\z@}
        {3ex plus-1ex minus-.2ex}{1pt plus1pt}{\large\sf\bfseries\boldmath}}
\def\subsection{\@startsection{subsection}{2}{\z@}
         {1.5ex plus-1ex minus-.2ex}{0.01pt plus1pt}{\sf\slshape}}
\def\subsubsection{\@startsection{subsubsection}{3}{\z@}
          {1.5ex plus-1ex minus-.2ex}{0.01pt plus0.2pt}{\sf\boldmath}}
\def\paragraph{\@startsection{paragraph}{4}{\z@}
           {.75ex \@plus.5ex \@minus.2ex}{-2mm}{\sf\bfseries\boldmath}}
\makeatother

 \allowdisplaybreaks
 \seceq

\begin{document}

\thispagestyle{empty}
\vbox{\Border\UMbanner}
\noindent{\small
\hfill{PP--017-011 \\ 
$~~~~~~~~~~~~~~~~~~~~~~~~~~~~~~~~~~~~~~~~~~~~~~~~~~~~~~~~~~~~$
$~~~~~~~~~~~~~~~~~~~~\,~~~~~~~~~~~~~~~~~~~~~~~~~\,~~~~~~~~~~~~~~~~$
 {HET-1678}
}
\vspace*{8mm}
\begin{center}
{\large \bf
Adinkras From Ordered Quartets of   \\[2pt]
 $\bm {\rm BC{}_4}$
Coxeter Group Elements and Regarding \\[6pt] 1,358,954,496 Matrix 
Elements of the Gadget}   \\   [12mm]
{\large {
S.\ James Gates, Jr.,\footnote{gatess@wam.umd.edu}$^{a, \, b}$
Forrest Guyton, \footnote{guytof@rpi.edu}${}^c$}
Siddhartha Harmalkar,\footnote{sharmalk@umd.edu }$^{a}$
David S. Kessler,\footnote{3.14159.david@gmail.com}$^{a}$ 
Vadim Korotkikh,\footnote{va.korotki@gmail.com}$^{a}$
and Victor A. Meszaros\footnote{victorameszaros@gmail.com}$^{a}$}
\\*[12mm]
\emph{
\centering
$^a$Center for String and Particle Theory-Dept.\ of Physics,
University of Maryland, \\[-2pt]
4150 Campus Dr., College Park, MD 20472,  USA
\\[12pt]         
$^{b}$Department of Physics, Brown University,
\\[1pt]
Box 1843, 182 Hope Street, Barus \& Holley 545,
Providence, RI 02912, USA 
\\[12pt]
and
\\[12pt] 
$^c$ Physics, Applied Physics, and Astronomy,
Rensselaer Polytechnic Institute,
\\[1pt]
Jonsson Rowland Science Center, Room 1C25
110 Eigth Street
\\[1pt]
Troy, NY 12180, USA
}
 \\*[10mm]
{ ABSTRACT}\\[2mm]
\parbox{142mm}{\parindent=2pc\indent\baselineskip=14pt plus1pt
We examine values of the Adinkra Holoraumy-induced Gadget 
representation space metric over {\it {all}} possible four-color, 
four-open node, and four-closed node adinkras.  Of the 1,358,954,496 
gadget matrix elements, only 226,492,416 are non-vanishing 
and take on one of three values: $-1/3$, $1/3$, or $1$ and 
thus a subspace isomorphic to a description of a body-centered 
tetrahedral molecule emerges.
 }
 \end{center}
\vfill
\vfill
\clearpage

\section{Introduction}

Over the course of some number of years 
\cite{GRana1,GRana2,ENUF1,ENUF2}, one of the authors (SJG) of this current 
work noted a series of what appeared to be ``curious'' hints that the very 
representation space of spacetime supersymmetry, even without consideration 
of dynamics, might contain an exquisite yet hidden mathematical structure.  This 
suggested that whatever this hidden structure might be, it warranted careful study.  
This direction has at this point finally begun to yield a trove of unexpected 
connections to deep mathematical structures...Riemann surfaces and 
algebraic geometry, among others.  This can be seen in two very illuminating recent 
works.

The works ``Geometrization of N-extended 1-dimensional supersymmetry 
algebras (I \& II)"  \cite{adnkGEO1} and \cite{adnkGEO2} conclusively
describe the not generally appreciated nor previously recognized connections 
between spacetime supersymmetry representations, as described by 
adinkra graphs, and a raft of mathematical structures that include:
\newline \indent
$~~~~~~$ (a.) Grothendieck's ``dessin d'enfant,''
\newline \indent
$~~~~~~$ (b.) Belyi pairs, 
\newline \indent
$~~~~~~$ (c.) Cimasoni-Reshetikhin dimer models on Riemann surfaces,
\newline \indent
$~~~~~~$ (d.)  Donagi-Witten parabolic structure/ramified coverings of 
super Riemann surface,
\newline \indent
$~~~~~~$ (e.) Morse divisors,
\newline \indent
$~~~~~~$ (f.) Fuchsian uniformization, and
\newline \indent
$~~~~~~$ (g.) elliptic curves.

Built on the observation of the ubiquitous appearance of an algebraic
structure (eventually given the name of ``Garden Algebras'') 
\cite{GRana1,GRana2,ENUF1,ENUF2}, that seem to universally exist 
in all linear realizations of spacetime supersymmetry, a type of graph 
\cite{adnk1} (thus permitting use of graph theory techniques \cite{GrphThry}) 
was proposed in order to study the properties of ``Garden Algebras'' in a 
more general manner.  These graphs were christened as ``adinkras'' and 
come extraordinarily close, if not achieving, the goal of providing a 
coordinate-independent description of one dimensional spacetime 
supersymmetry representations.

The introduction of adinkra graphs was an important milestone in the effort
that was envisioned as in the inaugural works \cite{GRana1,GRana2} in this
direction.  The importance of achieving such  coordinate-independent
descriptions of spacetime supersymmetry algebras was especially emphasized 
in the response to a presentation given by SJG at the 2006 workshop on 
``Affine Hecke algebras, the Langlands Program, Conformal Field Theory 
and Matrix Models'' at the Centre International de Rencontres Math\' ematiques 
(CIRM) in Luminy/France. 

Independent researchers have also utilized concepts that arose from the 
study of adinkras.  For example, the concept of ``the root superfield'' 
\cite{G-1} has found applications such as the construction 
of new models and the classification of N-extended supersymmetric 
quantum mechanical systems in the research of \cite{i1,i2,i3,i4,i5,i6,i7}.  The 
program has also created links to purely mathematical discussions such as 
combinatorics \cite{WIAA}, Coxeter Groups \cite{CXgrp}, and Klein's Vierergruppe 
\cite{4Grp}, and spectral geometry \cite{CTa}.  The adinkra concept 
has generated at least one publication purely in the mathematical literature 
\cite{YZ} and uncovered other surprising structures \cite{adnkM1,adnkM2}.

The complete specification of adinkras at the level of a rigorous mathematical 
formulation has led to an enhanced level of understanding of the relationships
between decorated cubical cohomology and the {\em {very}} surprising role 
of coding theory \cite{codes1,codes2,codes3} as it apparently controls the 
definitions and structure of adinkras with more than four colors that define 
the irreducible representations of spacetime supersymmetry.

The works of Doran, Iga, Kostiuk, Landweber, and Mendez-Diez
\cite{adnkGEO1,adnkGEO2} have now erected a sturdy and broad
causeway to increase and deepen a mathematical representation theory
of spacetime supersymmetry in a way that has never before existed.  In
spite of this major advance, however, there remain numbers of puzzles.

One of these is, how does higher dimensional spacetime Lorentz symmetry
manifest itself in the context of adinkra graphs?

Several works have taken steps toward investigating this problem.  It was 
conjectured \cite{ENUF2} there must exist some sort of ``holography'' that 
connects one dimensional adinkras to higher dimensional superfield theories.  
We currently have renamed this concept ``SUSY holography'' to distinguish 
it from other concepts that use the word ``holography.''   This viewpoint was 
strengthened in later work \cite{G-1}.  An obvious consequent possibility from 
such a viewpoint is it might be possible to start solely with an adinkra and 
perform the process of ``dimensional extension'' to reconstruct a higher 
dimensional supermultiplet.  One example, where looking at an adinkra-based 
structures were related to higher dimensional ones occurred in relation to 4D 
$\cal N$ = 2 supermultiplets \cite{HyprPlt1,HyprPlt2}.

The first concrete examples \cite{enhanc1,enhanc2} of how to accomplish this 
outcome of connecting adinkras to 4D, $\cal N$ = 1 supermultiplets used
calculations which showed by successively raising adinkra nodes from a valise 
configuration, one could examine when a Lorentz covariant (in the context of 
4D Minkowski space) interpretation was consistent.  These papers provide
a demonstration of proof of concept most certainly.  However, the steps of 
systematically lifting nodes is computationally expensive.  Thus, we will not 
take this route.

Another approach \cite{extnd1}, somewhat related, was taken with respect to the 
simpler problem of providing a dimensional extension of adinkras into the construction 
of 2D Minkowski space supersymmetric representations.  In particular, this work 
uncovered the ``no two-color ambidextrous bow-tie'' rule which governs the lifting of the 
adinkra into a 2D Minkowski space supermultiplet.  A ``two-color ambidextrous 
bow-tie'' is a structure that can be directly calculated in terms of closed four-cycles 
within the context of a valise adinkra once 2D helicity labels are adding to the links
in an adinkra.  If this obstruction occurs, the lifting of a node will remove it 
and the resulting adinkra can then be extended to the higher dimension.  Next, an 
approach \cite{extnd2}  was also created as an alternate efficient calculation for 
solving the problem of providing a dimensional extension of adinkras into the construction 
of 2D Minkowski space supersymmetric representations.  This approach is based 
on the underlying coding theory structures \cite{codes1,codes2,codes3} that were 
discovered within the structure of all adinkras.

In facing the problem of reconstructing 4D simple supermultiplets from 
adinkras,  unlike the successful paths shown for the analogous problem 
in 2D, another path has been pursued also 
\cite{KIAS1,KIAS2,HoLoRmY1,HoLoRmY2,HoLoRmY4D,G&G1,G&G2}.   
This alternate path is based on the fact that closed four-cycles 
for four color valise adinkras possess naturally an $SO(4)$ 
symmetry.  Since $SO(4)$ can be decomposed into $SU(2) \, \otimes SU(2)$,
and since the covering algebra of the 4D Dirac matrices is also $SO(4)$,
these explorations have endeavored to explore a variant of the concept of ``spin
from isospin'' \cite{Jckw}.  As adinkras naturally carry isospin, it is
indicated to ask if this property can be used in a way to make 4D Lorentz
symmetry an emergent symmetry arising from adinkras with at least
four colors.

It is the purpose of this paper to continue to explore the possibility that
4D Lorentz symmetry {\em {is an emergent symmetry}} arising from the isospin
symmetry of adinkras with at least four colors.  The outline of this paper 
is given below.

In chapters two and three, we present the new results of this work.  These
results support the concept of ``SUSY holography'' by showing the ``angles'' 
\cite{HoLoRmY1,HoLoRmY2,HoLoRmY4D,G&G1,G&G2} between any two 
adinkras constructed from ordered quartets of ${\rm BC}{}_4$ elements take on 
exactly and {\em {only}} the same values as the ``angles'' \cite{HoLoRmY4D} 
between any of the 4D, $\cal N$ = 1 supermultiplets with minimal numbers 
of bosons and fermions.  The presentation in chapter three is made in
terms of a visual/graphical representation.

In chapter four in comparison to our previous work \cite{HoLoRmY4D}, 
we find there is one additional value of the 4D ``angles'' between supermultiplets 
when one expands the space of supermultiplet representations to include 
the axial vector and axial tensor supermultiplets.  We note that certain
``parity flipped versions'' may be added to the list of representation 
studied in our previous work and we expand the values of the 4D
Gadget matrix to include the calculations related to these additional 
supermultiplets.

The approach to deriving our main results was a bifurcated one.  The
first part of our analysis and code-based exploration only covered the
case of a small subset of the adinkras.  This subset consisted only of
the adinkras described in detail in \cite{permutadnk} as in this case 
there was a previously established ``library'' of SUSY pairings between 
``Boolean Factors'' and permutation elements from ${\rm BC}{}_4$.

Chapter five describes our previous construction of adinkras with 
four colors, four open-nodes, and four closed nodes based on the elements 
of Coxeter Group ${\rm BC}{}_4$ \cite{permutadnk}.  Here we {\em {do}} {\em {not}} 
consider the role of ordered quartets.  This results in the possibility of
writing 1,536 adinkras that are potentially ``usefully inequivalent'' \cite{Neqv}.
We also concentrate on the role of a subset of the elements of the
permutation that sets the stage of separating the twenty-four elements
of the permutation group into a group of ``corrals" containing four elements
each.  We discuss the relation of reduced versions of the 4D $\cal N$ = 2
chiral and twisted chiral \cite{G-HR,GHR} supermultiplets and show in this
example that it is the distinct corrals that appear to play the dominant
role in determining the ``usefully inequivalent.''  

Chapter six goes to the consideration of looking at the effects of considering 
the role of ``twisting'' quartets by introducing relative sign factors among the 
components within quartets of the ``small ${\rm BC}{}_4$ library'' that results 
strictly from taking a single representation of the elements of ${\rm BC}{}_4$ 
and using them to construct supersymmetric quartets that satisfy the ``Garden 
Algebra" conditions.  This material has {\em {exactly}} appeared previously 
in \cite{G&G2} and the reader familiar with this discussion can skip this.

Also as was {\em {explicitly}} discussed \cite{G&G2}, given a complete ${\rm BC
}{}_4$ description of an adinkra, one can construct ``complements'' of any adinkra. 
These also include ``anti-podal'' representations where one representation can 
be obtained from the other by simply re-defining either {\em {all}} of closed nodes 
(or open nodes) by a minus sign.  So we can reduce this number by a factor of
two since ``anti-podal'' representations where one representation can be 
obtained from the other by simply re-defining either {\em {all}} closed nodes (or 
open nodes) by a minus sign are included.

In chapter seven, we describe the expansion of our previous construction of
adinkras with four colors, four open-nodes, and four closed nodes based 
on the elements of Coxeter Group ${\rm BC}{}_4$ \cite{permutadnk}.  Whereas
 before we {\em {did}} {\em {not}} consider the role of ordered quartets, 
 in this work we {\em {do}} consider such quartets.  This is also another
 source for the expansion in the number of adinkras to consider.   Without 
 removing ``anti-podal'' representations this results in the possibility of 
 writing 36,864 adinkras that are potentially ``usefully inequivalent.'' 

In the short discussion of chapter eight, we review the counting arguments
that show how the 384 elements in the ${\rm BC}{}_4$ Coxeter group at
first naively leads to a total of ninety six quartets, but that by considering
the effects of sign flips and expanded to ordered quartets this number
increases almost a hundred-fold.

The work of chapter nine includes the description of the codes that were
created to attack the problem of calculating the ``angles'' between the
adinkra with four colors, four open-nodes, and four closed nodes.  There
were actually four different codes created independently that support the
work undertaken in chapter four.  These were created using C++,  
\textsl{Mathematica$^{\sss\text{TM}}$} , MATLAB, and Python.  We 
report the results of the calculations with regards to the results related 
to the discussion in chapter four.

This same chapter next turns to the main task of describing what was
undertaken in computing the ``angles" between all 36,864 adinkras 
based on ordered quartets of ${\rm BC}{}_4$ elements.  This was done 
by once more creating an entirely independent code using the 3.5 
version of Python together with the Numpy module for Python which 
allows for the matrix calculations.  After running for 1.86 hours calculating 
all the entries in a 36,864 $\times$ 36,864 matrix,  the only values found 
were those given in chapter two.
 
The final chapter includes a discussion of our conclusions and a 
number of appendices are included to provide details about all
the computations.  Appendix A consists of a set of multiplication
tables showing the action of the multiplication of element of the
permutation group on a quartet of permutations denoted by $\bm 
{\{ {\cal V}_{(4)} \}} $, which is a realization of the the Vierergruppe
over the 2 $\times$ 2 identity and first Pauli matrix.  Appendix B
has appeared in some of previous work, and contains the assignment
of ``Boolean Factors'' with quartets of permutations that are required
to construct supersymmetry representations.  Appendix C is drawn
from previous work and reports the values of the ``$\ell$ and ${\Tilde 
\ell}$'' parameters associated with permutation factor and Boolean
Factors.  The final Appendix D reports on the outcome of the evaluation
of the values of the Gadget function over a 96 $\times$ 96 matrix
over the domain of the ``small ${\rm BC}{}_4$ library'' (the latter is
defined in this work also).

Any party seeking to obtain the codes described in this work
can contact any of the authors in order to receive the code of
interest.

\newpage
\section{Calculating The Gadget With Ordered Quartets}
\label{s2a}

An example of an image given the name of an ``adinkra graph'' \cite{adnk1} and mathematically
defined in subsequent works \cite{adnkM1,adnkM2,codes1,codes2,codes3,adnkM3}
is shown in Fig.\ 1.  
\begin{figure}[ht]
\begin{center}
\begin{picture}(70,30)
\put(-5,-10){\includegraphics[height =1.6in]{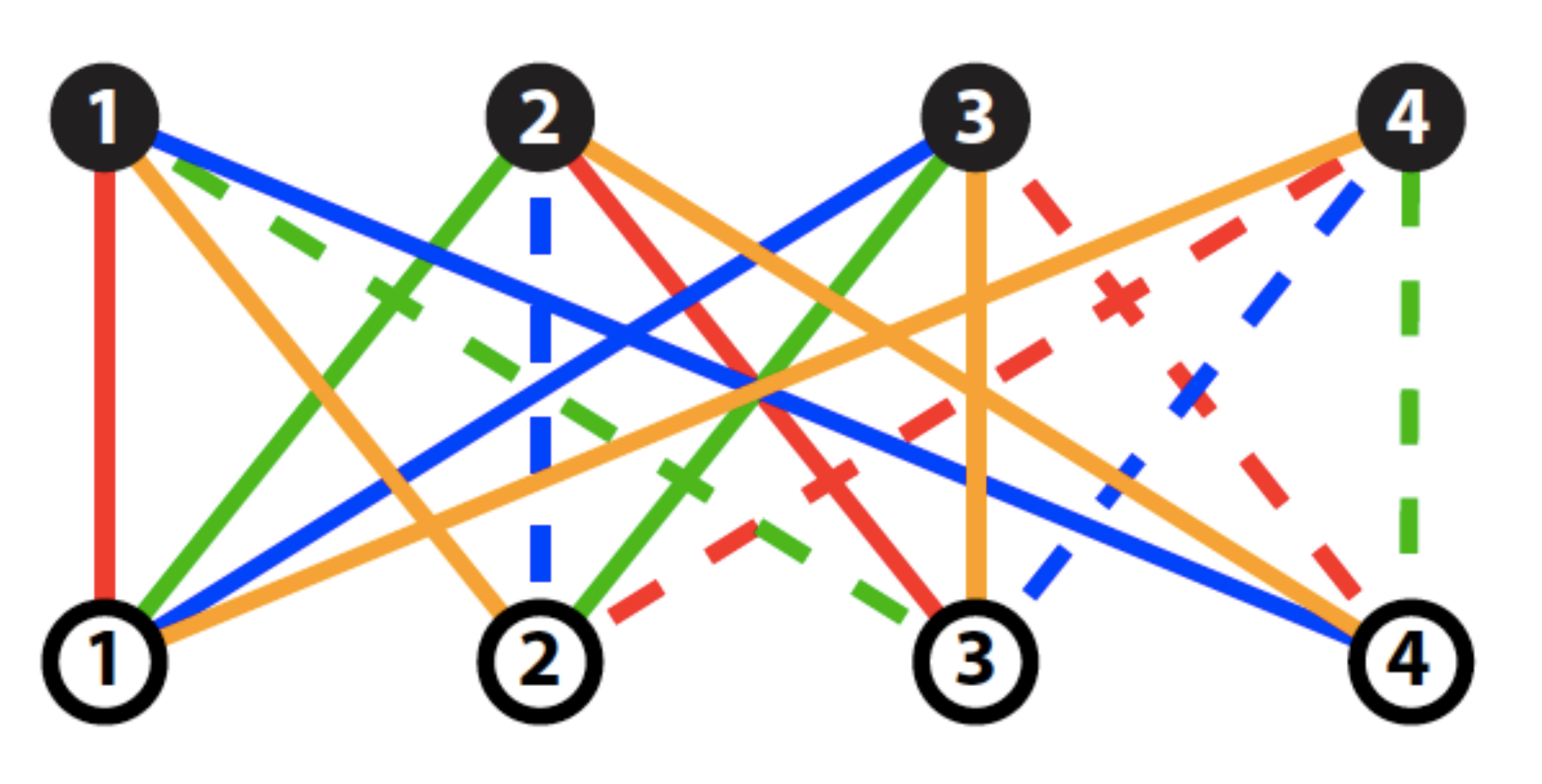}}
\put(-2,-18){{ $~~~$ {\bf {Figure}} {\bf {1:}} A ``Valise'' Adinkra}} 
\end{picture}
\end{center}
\label{f:CullR}
\end{figure}
 \vskip-17in
 \begin{center}
$~~~$
\end{center}
Every $N$-color adinkra representation $(\cal R)$ leads, via a set of Feynman-like 
rules, to sets of matrices denoted by ${\rm L}^{(\cal R)}_\rI\,$ and ${\rm R}^{
(\cal R)}_\rI$ which satisfy the ``Garden Algebra'' conditions,
\be { \eqalign{
 (\,{\rm L}^{(\cal R)}_\rI\,)_i{}^\hj\>(\,{\rm R}^{(\cal R)}_\rJ\,)_\hj{}^k ~+~ (\,{\rm L
 }^{(\cal R)}_\rJ\,)_i{}^\hj\>(\,{\rm R}^{(\cal R)}_\rI\,)_\hj{}^k &= 2\,\d_{\rI\rJ}\,\d_i{}^k
 ~~,\cr
(\,{\rm R}^{(\cal R)}_\rJ\,)_\hi{}^j\>(\, {\rm L}^{(\cal R)}_\rI\,)_j{}^\hk ~+~ (\,{\rm 
 R}^{(\cal R)}_\rI\,)_\hi{}^j\>(\,{\rm L}^{(\cal R)}_\rJ\,)_j{}^\hk  &= 2\,\d_{\rI\rJ}\,\d_\hi{
 }^\hk~~,  \cr
~~~(\,{\rm R}^{(\cal R)}_\rI\,)_\hj{}^k\,\d_{ik} = (\,{\rm L}^{(\cal R)}_\rI\,)_i{
}^\hk\,\d_{\hj\hk}&~~.
}}\label{GarDNAlg2}
\ee
The ``color-link rainbow to matrix'' assignments associated with Fig.\ 1 correspond to 
``red = ${\bm {\rm L}}{}_{1}$,'' 
``green = ${\bm {\rm L}}{}_{2}$,''
``blue = ${\bm {\rm L}}{}_{3}$,''  
and 
``orange = ${\bm {\rm L}}{}_{4}$'' 
and the explicit forms of the L-matrices associated with Fig.\ 1 are given by
$$
\left( {\rm L}{}_{1}\right) {}_{i \, {\hat k}}   ~=~
  \left[\begin{array}{cccc}
1&0&0&0\\ 0&0&0&-1\\ 0&1&0&0\\ 0&0&-1&0
\end{array}\right] 
~\,~~,~~~
\left( {\rm L}{}_{2}\right) {}_{i \, {\hat k}}   ~=~
  \left[\begin{array}{cccc}
0&1&0&0\\ 0&0&1&0\\ -1&0&0&0\\ 0&0&0&-1
\end{array}\right] 
 ~~~,
$$
\be
\left( {\rm L}{}_{3}\right) {}_{i \, {\hat k}}   ~=~
  \left[\begin{array}{cccc}
0&0&1&0\\ 0&-1&0&0\\ 0&0&0&-1\\ 1&0&0&0
\end{array}\right] 
~~~,~~~
\left( {\rm L}{}_{4}\right) {}_{i \, {\hat k}}   ~=~
\left[\begin{array}{cccc}
~0&~0&~0&~1\\ 1&0&0&0\\ 0&0&1&0\\ 0&1&0&0 
\end{array}\right] 
~~~.
 \label{chiD0F}
\ee

A set of L-matrices and R-matrices for a specified adinkra representation $(\cal 
R)$, can be used to define two additional sets of matrices.   We have given the 
name of ``bosonic holoraumy matrices'' and  ``fermionic holoraumy matrices,'' 
respectively, to the sets denoted by $\bm {V_{\rI\rJ}}^{(\cal R)}$ and $\bm{{\Tilde 
V}_{\rI\rJ}}{}^{(\cal R)}$ \cite{KIAS1,KIAS2,HoLoRmY1,HoLoRmY2,HoLoRmY4D} 
and defined via the equations
\be { \eqalign{
 (\,{\rm L}^{(\cal R)}_\rI\,)_i{}^\hj\>(\,{\rm R}^{(\cal R)}_\rJ\,)_\hj{}^k ~-~ (\,{\rm L
 }^{(\cal R)}_\rJ\,)_i{}^\hj\>(\,{\rm R}^{(\cal R)}_\rI\,)_\hj{}^k &= i\, 2\,  (V_{\rI\rJ}^{(\cal 
 R)})_i{}^k~~,\cr
 (\,{\rm R}^{(\cal R)}_\rI\,)_\hi{}^j\>(\, {\rm L}^{(\cal R)}_\rJ\,)_j{}^\hk ~-~ (\,{\rm 
 R}^{(\cal R)}_\rJ\,)_\hi{}^j\>(\,{\rm L}^{(\cal R)}_\rI\,)_j{}^\hk  &= i \, 2\,  
 ({\Tilde V}_{\rI\rJ}^{(\cal R)})_\hi{}^\hk ~~.
}}\label{GarDVs}
\ee

Given two adinkras denoted by $({ {\cal R}})$  and $( {\cal R}^{\prime})$ (which
possess $N$ colors, and $d$ open nodes) along with their associated fermionic 
holoraumy matrices $\bm{{\Tilde V}_{\rI\rJ}}{}^{(\cal R)}$ and $\bm{{\Tilde V}_{
\rI\rJ}}{}^{({\cal R}^{\prime})}$ we can form a scalar, ``the gadget value'' between 
two representations $({ {\cal R}})$  and $( {\cal R}^{\prime})$ defined by
\be
{{\cal G}} [  ({ {\cal R}}) , ( {\cal R}^{\prime}) ] ~=~ \left[ \, {1 \over {~ N \, (N-1) \, 
d_{\min}(N)}}  \,   \right] \, \sum_{\rI , \, \rJ} \,{\rm {Tr}} \,  \left[ \, \bm{{\Tilde V}_{
\rI\rJ}}{}^{(\cal R)}  \, \bm{{\Tilde V}_{\rI\rJ}}{}^{({\cal R}^{\prime})}  \right]  ~~~,
\label{Gdgt1}
\ee
where
\be
d_{\min}(N)=\begin{cases}
{~~}2^{\frac{N-1}{2}}\,~,&N\equiv 1,\,7 $~~~~~~~\,~~~$ \bmod{(8)}\\
{~~}2^{\frac{N}{2}}{~~~~},&N\equiv 2,\,4,\,6 $~~~~~~~$ \bmod{(8)}\\
{~~}2^{\frac{N+1}{2}}~\, ,&N\equiv 3,\,5 $~~~~\,~~~~~~$ \bmod{(8)}\\
{~~}2^{\frac{N - 2}{2}}~\, ,&N\equiv  8 $~~~~~\,~\,~~\,~~\,~~$ \bmod{(8)}\\
\end{cases}
{~~~~~~~~~~~~~~~}
\label{eqn:dmin}
\ee
and we exclude the case of $N$ = 0 (i.e.\ no supersymmetry) in these expressions.  
For every adinkra \cite{adnk1,GrphThry} based on the Coxeter Group ${\rm {BC}}_4$, 
the L-matrices and R-matrices \cite{GRana1,GRana2} must have 
four colors (${\rm I} = 1,\dots, 4$), four open nodes ($i = 1,\dots, 4$), and four closed 
nodes ($\hk = 1,\dots, 4$).


\subsection{The $  {\ell}$ and $  {\Tilde \ell}$ Coefficients Basis}
\label{ELLs}

Since we will only be looking at adinkras and associated quantities related to 
 ${\rm {BC}}_4$, there is one special feature related to the fact that for such
 adinkras, the holoraumy matrices are necessarily related to SO(4).  Due to this,
 the holoraumy matrices may be expanded in the ``$\a$-$\b$ basis where six
 matrices define by 
 \be {
\begin{array}{cccc}
&{\bm {\a}}{}^{\,\Hat 1} ~=~ {\bm \s}^2 \otimes {\bm \s}^1 ~~, & ~~ {\bm {\a}}{}^{\,\Hat 2} ~=~  {\bm 
{\rm I}}{}_{2 \times 2}  \otimes {\bm \s}^2  ~~, &  
~~{\bm {\a}}{}^{\,\Hat 3}  ~=~ {\bm \s}^2 \otimes {\bm \s}^3   ~~, \\
&{\bm {\b}}{}^{\,\Hat 1}  ~=~  {\bm \s}^1 \otimes {\bm \s}^2~~, & ~~ {\bm {\b}}{}^{\,\Hat 2} ~=~ {\bm 
\s}^2 \otimes  {\bm {\rm I}}{}_{2 \times 2} ~~, &  ~~{\bm {\b}}{}^{\,\Hat 3} ~=~  {\bm \s}^3 
\otimes {\bm \s}^2  ~~, \\ \end{array}
} \label{aLbE}
 \ee
 can be chosen as a basis over which to expand $ ({\Tilde V}_{\rI\rJ}^{(\cal 
 R)})_\hi{}^\hk$ as
\be   \eqalign{ {~~~~~~~}
 {\bm {\Tilde V}{}_{\rI\rJ}^{(\cal R)}} ~=~ &\Big[ \,\ell^{({\cal R}){\Hat 1}}_{\rI\rJ}\, 
 {\bm {\a}}{}^{\,\Hat 1}  \, + \,  \ell^{({\cal R}){\Hat 2}}_{\rI\rJ}\, {\bm {\a}}{}^{\,\Hat 2} 
 \,+\,  \ell^{({\cal R}){\Hat 3} }_{\rI\rJ}\, {\bm {\a}}{}^{\,\Hat 3}    \, \Big]     
 ~+~  \Big[ \,    {{\Tilde \ell}^{(\cal R)}}_{\rI\rJ}{
 }^{\Hat 1}\,  {\bm {\b}}{}^{\,\Hat 1}  \,+\, \, {{\Tilde \ell}^{(\cal R)}}_{\rI\rJ}{}^{\Hat 2}\, 
  {\bm {\b}}{}^{\,\Hat 2} 
 \,+\, {{\Tilde \ell}^{(\cal R)}}_{\rI\rJ}{}^{\Hat 3}\,  {\bm {\b}}{}^{\,\Hat 3}    \, \Big]   ~~~,
}  \label{Veq}
\ee
written in terms of the thirty-six coefficients 
$\ell^{({\cal R}){\Hat 1}}_{\rI\rJ}$, $\ell^{({\cal R}){\Hat 2}}_{\rI\rJ}$,
$\ell^{({\cal R}){\Hat 3}}_{\rI\rJ}$, ${\Tilde \ell}^{({\cal R}){\Hat 1}}_{\rI\rJ}$, ${\Tilde \ell}^{({\cal 
R}){\Hat 2}}_{\rI\rJ}$, and ${\Tilde \ell}^{({\cal R}){\Hat 3}}_{\rI\rJ}$.

Using this approach, the values of the ``gadget,'' expressed in terms of the ${ 
\ell}$ and ${ {\Tilde \ell}}$ coefficients, are defined by
$$
\eqalign{
{ {\cal G}} \left[  ({\cal R}) \, , \,  ({\cal R}^{\prime}) \right]{}_{\ell} ~&\equiv ~~~\frc{1}{6}
\sum_{{\Hat a} } \, {\Big [} ~ 
{\ell}_{1 \, 2}^{\,({\cal R}) \Hat{a}} \,  {\ell}_{1 \, 2}^{\, ({\cal R}^{\prime}) \Hat{a}}  ~+~ 
{\ell}_{1 \, 3}^{\,({\cal R}) \Hat{a}} \,  {\ell}_{ 1 \, 3}^{\, ({\cal R}^{\prime}) \Hat{a}} ~+~
{\ell}_{1\, 4}^{\,({\cal R}) \Hat{a}} \,  {\ell}_{1\, 4}^{\, ({\cal R}^{\prime}) \Hat{a}}    ~+~ \cr
&{~~~~~~~~~~~~~~~~~}      
{\ell}_{2 \, 3}^{\, ({\cal R}) \Hat{a}}  {\ell}_{2 \, 3}^{\, ({\cal R}^{\prime}) \Hat{a}} \, ~+~ 
{\ell}_{2 \, 4}^{\,({\cal R}) \Hat{a}} \,  {\ell}_{ 2 \, 4}^{\, ({\cal R}^{\prime}) \Hat{a}} ~+~
{\ell}_{3 \, 4}^{\,({\cal R}) \Hat{a}} \,  {\ell}_{3\, 4}^{\, ({\cal R}^{\prime}) \Hat{a}}  ~ {\Big ]}  ~+~
~~~~~~~~~
}$$
\be  
\eqalign{~~~~~~~~\,~~~~~~~~~&
\frc{1}{6} \sum_{{\Hat a} } \, {\Big [} ~ 
{\Tilde {\ell}}_{1 \, 2}^{\,({\cal R}) \Hat{a}} \,  {\Tilde {\ell}}_{1 \, 2}^{\, ({\cal R}^{\prime}) \Hat{a}}  ~+~ 
{\Tilde {\ell}}_{1 \, 3}^{\,({\cal R}) \Hat{a}} \,  {\Tilde {\ell}}_{ 1 \, 3}^{\, ({\cal R}^{\prime}) \Hat{a}} ~+~
{\Tilde {\ell}}_{1\, 4}^{\,({\cal R}) \Hat{a}} \,  {\Tilde {\ell}}_{1\, 4}^{\, ({\cal R}^{\prime}) \Hat{a}}    ~+~ \cr
{~~~~~~~~~~~~~~~}& {~~~~~~~~~~}
{\Tilde {\ell}}_{2 \, 3}^{\, ({\cal R}) \Hat{a}}  {\Tilde {\ell}}_{2 \, 3}^{\, ({\cal R}^{\prime}) \Hat{a}}~+~ 
{\Tilde {\ell}}_{2 \, 4}^{\,({\cal R}) \Hat{a}} \,  {\Tilde {\ell}}_{ 2 \, 4}^{\, ({\cal R}^{\prime}) \Hat{a}} ~+~
{\Tilde {\ell}}_{3 \, 4}^{\,({\cal R}) \Hat{a}} \,  {\Tilde {\ell}}_{3\, 4}^{\, ({\cal R}^{\prime}) \Hat{a}}  
~ {\Big ]}
~~~,  } \label{Gdgt2}\ee
and necessarily
\be {
cos \left\{ \theta [({{\cal R}})\,  , \, ({ {\cal R}}^{\prime} )] {}_{\ell} \right\} ~=~
{{{ {\cal G}} [ \, ({ {\cal R}}) , ({ {\cal R}}^{\prime}) \,]{}_{\ell} } \over {~ {\sqrt{{ 
{\cal G}} [ \, ({ {\cal R}}) , ({ {\cal R}}) \, ]{}_{}}} \, {\sqrt{ { {\cal G}} [ \, ({ {\cal 
R}}^{\prime}) , ({ {\cal R}}^{\prime}) ]{}_{\ell}}}~~ } } ~~~.  }   \label{M4gL}
\ee

The existence of the ${\ell}$ and ${\Tilde \ell}$ coefficients also means that 
for each BC${}_4$ adinkra representation $({\cal R})$ there is a vector in an 
associated thirty-six dimensional vector space defined by all the values of 
the coefficients.   If we denote the elements in this space by the symbol 
${\vec \g} {}_{\ell \, {\Tilde \ell} }({\cal R})$ we can lexicographically order 
the $I$, $J$, and $\hat \a$ indices according to 
\be
{\vec \g} {}_{\ell \, {\Tilde \ell} }({\cal R}) ~=~ \left( \, {\ell}_{1 \, 2}^{\,(
{\cal R}) \hat{1}}, \, {\ell}_{1 \, 2}^{\,({\cal R}) \hat{2}},  \, {\ell}_{1 \, 2}^{
\,({\cal R}) \hat{3}}, \, {\ell}_{2 \, 3}^{\,({\cal R}) \hat{1}}, \, {\ell}_{2 \, 3
}^{\,({\cal R}) \hat{2}},  \, {\ell}_{2 \, 3}^{\,({\cal R}) \hat{3}},  \, \dots \, , \, 
{\Tilde {\ell}}_{3 \, 4}^{\,({\cal R}) \hat{1}} , \, {\Tilde {\ell}}_{3 \, 4}^{\,(
{\cal R}) \hat{2}}, \, {\Tilde {\ell}}_{3 \, 4}^{\,({\cal R}) \hat{3}} \right)  ~~~,
\label{gees}
\ee
to create a convention for the definition of the components of the vector.  The 
Gadget acts as the ``dot product'' on this space.  Alternately, we can regard 
the components of ${\vec \g} {}_{\ell \, {\Tilde \ell} }({\cal R})$ as forming the 
components of a rank one tensor $ {{\g {}_{\ell \, {\Tilde \ell}}^{A^*}}}({\cal R})$
where the index ${A^*}$ enumerates the components in (\ref{gees}) in the 
order shown.  Also given ${{\g {}_{\ell}^{{A}_1^*}}}({\cal R}_{1})$, $\dots$, $
{{\g {}_{\ell}^{A_{p}^*}}}({\cal R}_{p})$ corresponding to representations $({\cal 
R}_{1})$, $\dots$, $({\cal R}_{p})$, we can form a $p$-th order simplex via 
the definition
\be
{\rm Y}{}^{A_{1}^* \, \cdots \,  A_{p}^*} ({\cal R}_{1},\,  \dots, \, {\cal R}_{p})
~=~ {{\g {}_{\ell \, {\Tilde \ell}}^{{A}_1^*}}}({\cal R}_{1}) \, \wedge \,   {{\g {}_{
\ell \, {\Tilde \ell}}^{A_{2}^*}}}({\cal R}_{2}) \,  \cdots \, \wedge \, {{\g {}_{\ell 
\, {\Tilde \ell}}^{A_{p}^*}}}({\cal R}_{p}) ~~~~. 
\label{gees2}
\ee


\subsection{The Results}
\label{RSlts}

In the work of \cite{G&G2}, there was presented and released a list of the values 
of the ${\ell}$ and ${ {\Tilde \ell}}$ parameters, though the calculations of these 
occurred contemporaneously with work of \cite{permutadnk}.  On the basis of 
this ``library'' (that we will subsequently call the ``small ${\rm BC}{}_4$ library,'' 
algorithms and codes, to be discussed later,  were written in order to calculate 
the values of the quadratic forms (\ref{Gdgt2}) on the $\Tilde \ell$ and $\ell$ 
adinkra parameter spaces. However, for the current work, codes were also 
developed to carry out the calculation directly that follow from the expression 
in (\ref{Gdgt1}).  The results of these calculation provide the foundation for the 
statements made subsequently.

One may regard the Gadget, ${{\cal G}}$, as a mapping that assigns the value 
of a real number to a pairs of adinkras.  In this sense the Gadget acts as a 
metric on the space of adinkras.  Each adinkra may be associated with a representation 
label $({\cal R}_1)$,  $({\cal R}_2)$, $\dots$,  $({\cal R}_{\rm T})$ where T is a integer.  In
the case of adinkras based on ordered quartets of elements of ${\rm BC}{}_4$, the value of
T is 36,864.  We can now report the main result of this current study.

The values of the Gadget over all the adinkras based on ordered quartets of elements 
of ${\rm BC}{}_4$ may be regarded as a square matrix with the rank of the matrix being 36,864.  
Thus, the task of calculating all (36,864) $\times$ (36,864) = 1,358,954,496 values of the 
matrix elements seems impossible.  However, with the aid of modern computational 
capacities, this task has been completed.

The Adinkra ``Gadget'' Representation Matrix (AGRM) over the 36,864 ordered-pair 
${\rm BC}{}_4$-based adinkras is very sparse.  Just over 83\% of the entries, or 1,132,462,080, 
are zero.  Among the remaining non-vanishing entries only three numbers appear: 
$\pm 1/3$, and $1$.   The frequencies of appearance with which these three numbers 
together with 0 appear are shown in Table 1.
\vskip3pt
\begin{table}[h]
\begin{center}
\footnotesize
\begin{tabular}{|c|c|}
\hline 
~~~ Gadget Value ~~ & ~~~~Count \\  \hline   \hline  
- 1/3 & ~~\,~\,127,401,984 ~  \\   \hline   
~ 0 & ~\, 1,132,462,080 ~~  \\ \hline   
 ~ 1/3 &  ~~~~\, 84,934,656 ~  \\ \hline   
 ~1 &  ~\,~~\,\, 14,155,776 ~  \\  \hline 
\end{tabular}
\end{center}
\end{table}
$~~~~~$ $~~~~~~~~~~~~~~~~~~~~$ {\bf {Table}} {\bf {1:}} Frequency 
of Appearance of AGRM Elements
\newline $~$ \newline
The sum of the counts add up to 1,358,954,496 which means one of our 
programs calculated every single entry in the AGRM.  
It is clear that the independent values that appear and the multiplicities of
their appearance in Table 1 can be written in the form of a polynomial
\be
{\cal P}{}_{\cal G} (y) ~=~ 1,132,462,080 \, y \, [ \, 127,401,984 \, (\, y \,+\, \fracm 13 \,)]
\, [ \, 84,934,656 \, (\, y \,-\, \fracm 13 \,)] \, [ \, 14,155,776 \, (\, y \,-\, 1  \,)] 
\ee
where the roots of the polynomial correspond to the non-vanishing
entries.

The normalization
given in (\ref{Gdgt1}) is such that for all these  representations $({\cal R})$ 
satisfy the condition
\be
 {\cal G} [ \, ({ {\cal R}}) , ({ {\cal R}}) ]  ~=~ 1 ~~~.
 \label{UntV}
 \ee
Thus, it is possible for any two adinkra $({{\cal R}})$ and $({ {\cal R}}^{\prime})$
to define an angle $\theta [({{\cal R}})\,  , \, ({ {\cal R}}^{\prime})]$ between
them via the equation 
\be {
cos \left\{ \theta [({{\cal R}})\,  , \, ({ {\cal R}}^{\prime})]{}_{} \right\} ~=~
{{{ {\cal G}} [ \, ({ {\cal R}}) , ({ {\cal R}}^{\prime}) \,]{}_{} } \over {~ {\sqrt{{ 
{\cal G}} [ \, ({ {\cal R}}) , ({ {\cal R}}) \, ]{}_{}}} \, {\sqrt{ { {\cal G}} [ \, ({ {\cal 
R}}^{\prime}) , ({ {\cal R}}^{\prime}) ]{}_{}}}~~ } } ~~~.  }   \label{M4g}
\ee
and this may be applied to adinkras based on ordered quartets of elements 
of ${\rm BC}{}_4$. 

Since the Gadget values only take on the four values shown in the table, 
the corresponding angles are: $arccos\left( - 1/3 \right)$, $\pi /2$, $arccos
\left(1/3 \right)$, and 0.  The result in (\ref{UntV}) implies we can regard 
the 36,864 adinkras as defining an equivalent number of unit vectors.  Next 
the angles defined via (\ref{M4g}) informs us that these unit vectors only meet
at the angles of  $arccos\left( - 1/3 \right)$, $\pi /2$, $arccos\left(1/3 \right)$, 
and 0.  The fact that 14,155,776 - 36,864 = 14,118,912 implies  that among 
the 36,864 associated unit vectors many are colinear to one another.  The
angle $arccos\left( - 1/3 \right)$ has been noted for some time in our past
research papers \cite{KIAS1,KIAS2,HoLoRmY1,HoLoRmY2,HoLoRmY4D}.  
In fact, with the exception of the angle of $arccos\left(1/3 \right)$ all the other 
angles have been found in our previous calculations.  We will discuss in a 
later section this exception.

The presence of the angle $arccos\left( - 1/3 \right)$
is amusing from the view of tetrahedral geometry.  In Fig.\ 2, there appears a 
regular tetrahedron inscribed within a sphere.  Referring to the labelling of 
the vertices shown in the figure, $\angle$ ACD has the value of $arccos\left(1/2 
\right)$ while the angles  $\angle$ AOD, $\angle$ AOC, and $\angle$ AOB 
all have the same value of $arccos\left(- 1/3 \right)$. 
\begin{figure}[ht]
\begin{center}
\begin{picture}(70,36)
\put(-7,-52){\includegraphics[height =3.4in]{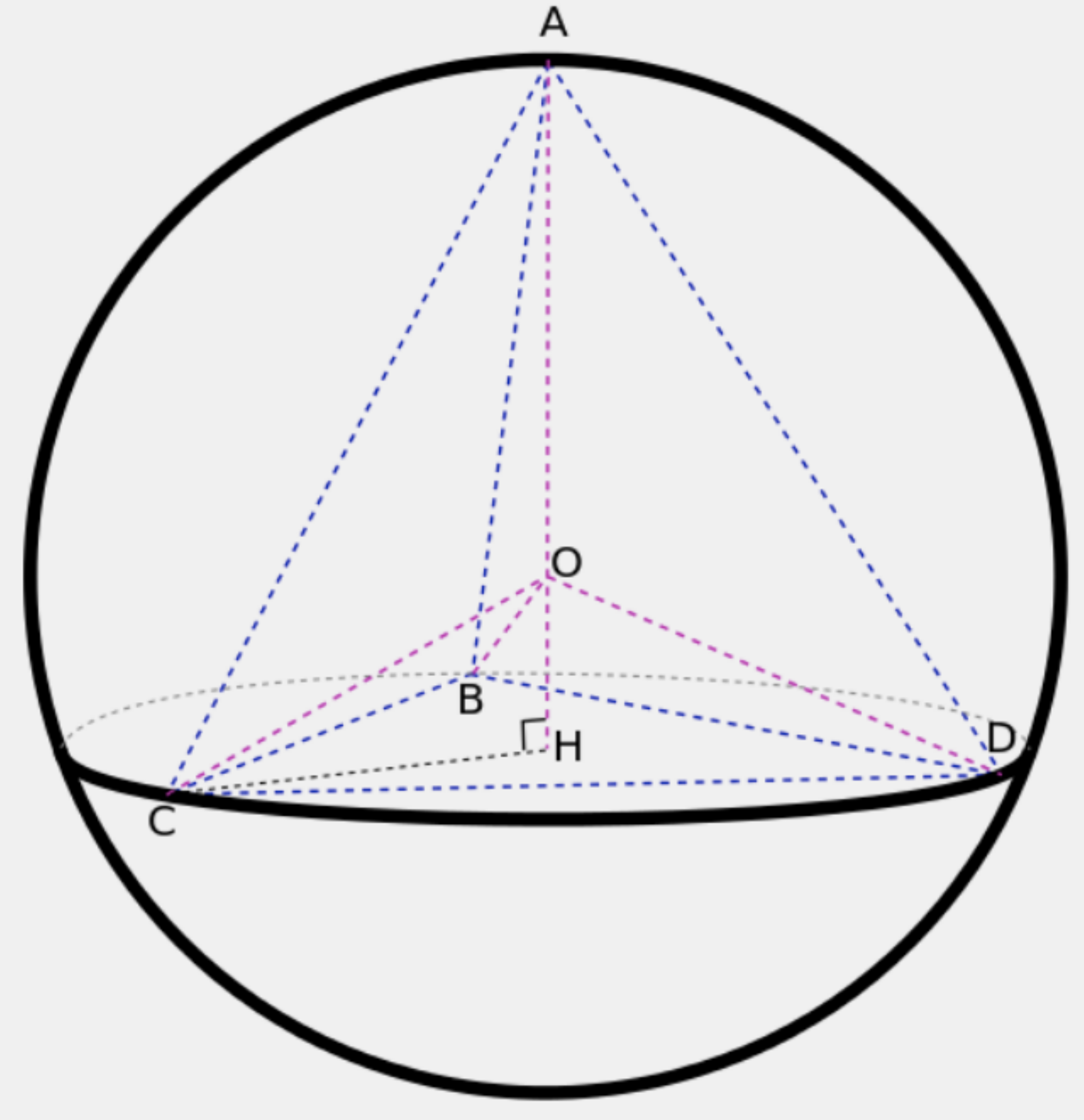}}
\put(-37,-60){{{\bf {Figure}} {\bf {2:}}
The adinkra representations with angle $arccos\left( - 1/3 \right)$ are equidistant}}
\put(-16,-65){{from each other within the considered 3D space in the metric defined}}
\put(-16,-70){{by the gadget.}}
\end{picture}
\end{center}
\label{f:CullR2}
\end{figure}
 \vskip2.2in
 \begin{center}
$~~~$
\end{center}
As it will prove useful in a later chapter, let us observe the points on the
sphere that lie respectively along the line segments $\Bar {OA}$,  $\Bar {OB}$,
$\Bar {OC}$, and  $\Bar {OD}$, have the coordinates given by
$\left( \, 0, \, 0, \, 1 \,\right)$, 
$\left(  \, - \, \fracm {\sqrt{2}}3 , \,  - \, \fracm {\sqrt{6}}3  , \, - \, \fracm 1 3 \,\right)$, 
$\left(  \,  \fracm {2 \sqrt{2}}3 , \,  0 , \, - \, \fracm 1 3 \,\right)$, and 
$\left(  \, - \, \fracm {\sqrt{2}}3 , \,  \fracm {\sqrt{6}}3  , \, - \, \fracm 1 3 \,\right)$
in the same order.  Each of these (considered as vectors) has length one, and
a dot product between any pair is equal to - 1/3.  It should be noted the 
condition in (\ref{UntV}) implies these all must be unit vectors.

\newpage
\section{Visual Graph of Adinkra Gadget Values Over The ``Small  ${\rm BC}_4$
Library"}
\label{s2ab}

\indent
$~~~~$ Any attempt to present the results that describe the entries in a 36,864 
$\times$ 36,864 symmetrical matrix obviously presents some challenges.  In
fact, we shall not even attempt this.  We will provide copies of all our codes
to any interested party upon request.

However as a ``peek" into one tiny (96 $\times$ 96) sector of the total 1.3+ 
billion results, we will here describe the results for the ``small ${\rm BC}_4$ library" 
described in detail in a later chapter.  One very accessible way to present this 
data is in the form of an array, but where the entries in the array are colored 
squares that play the role of pixels.  As there are only four values found in the 
entirety of the range of our calculations, we only need pixels of four colors.  
We make the numerical assignments between the calculated gadget values 
and the colors according to:

$$
\vCent
{\setlength{\unitlength}{1mm}
\begin{picture}(-20,0)
\put(-84,-154){\includegraphics[width=6.6in]{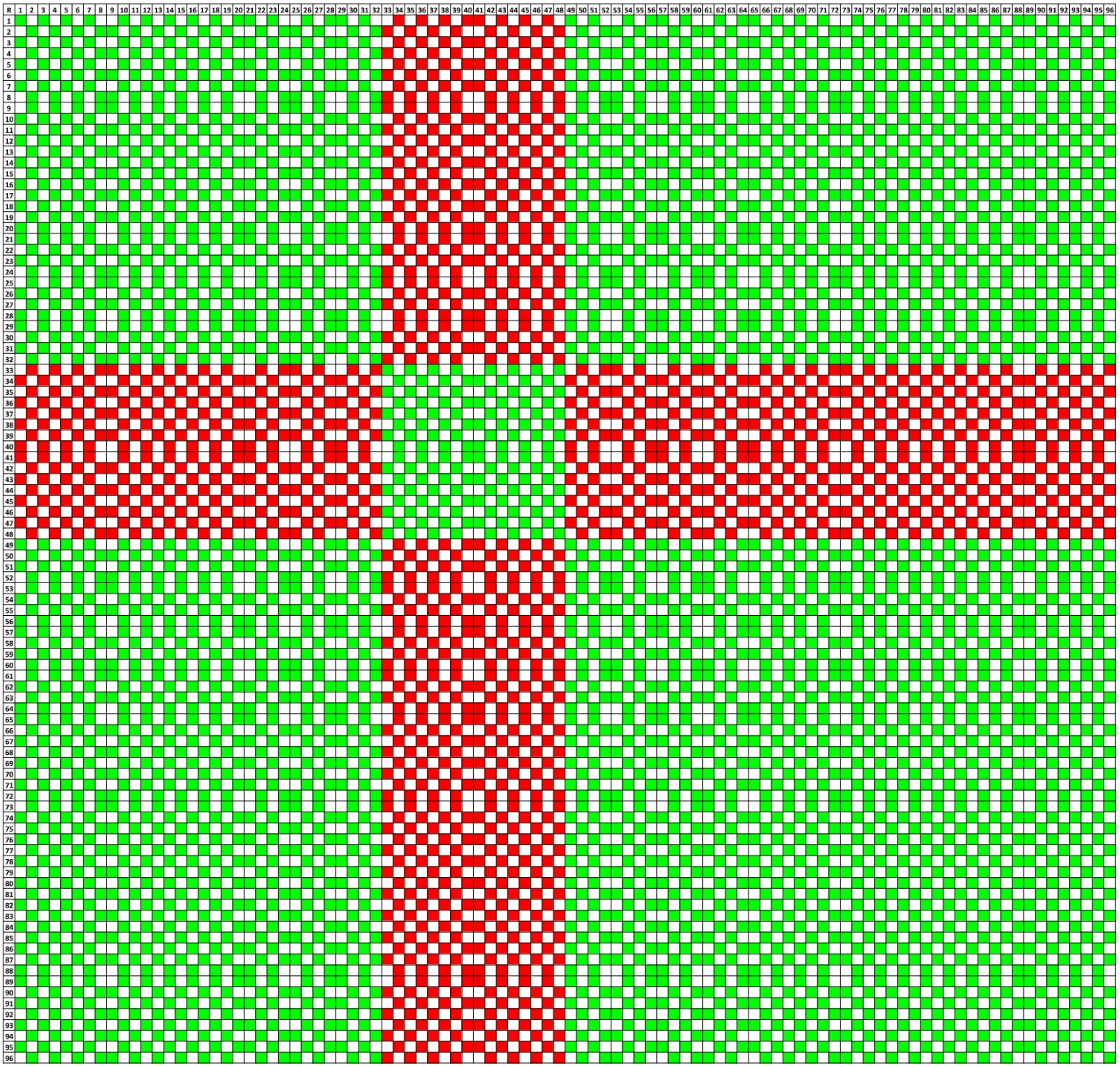}}
\put(-82,-159){{{\bf {Figure}} {\bf {3:}}
Graphical Representation of the Values in 
the Adinkra Gadget Representation Matrix }}
\end{picture}}
\nonumber
$$
\vskip5.9in

\newpage
\noindent
red = -1/3, white = 0, black = + 1/3, and green = 1.  When this assignment is
done, we find the image\footnote{The image in Fig.\ 3 is a scalable pdf file in the native
format of this document which implies it can be magnified as desired
to obtain greater detail.} in Fig.\ 3 over the 96 $\times$ 96 entries of the small
${\rm BC}_4$ library.  The image in Fig.\ 3 possesses no black ``pixels'' as
the value of + 1/3 does not occur in the context of the small library.  However,
it might be an interesting problem in computer visualization to extend this
graphical presentation beyond the small library.

Let us also note that the lack of appearance of the black pixels seems correlated
with the method by which the small library was constructed.  Its construction
began from the dimensional reduction of supermultiplets in four dimensions
which were then extended using signed permutation elements to conjugate
elements of this original set.

\newpage
\section{Computing 4D Gadget Values}
\label{s2ac}
We have proposed \cite{HoLoRmY4D} that the analog of the adinkra Gadget 
exists in the form of a supermultiplet Gadget denoted by ${\widehat  {\cal G}}$.
Unlike the adinkra Gadget, the supermultiplet Gadget is defined by calculations
{\em {solely}} involving supermultiplets in higher dimensions.  As such its arguments,
denoted by $({\widehat {\cal R}})$, and $({\widehat {\cal R}}^{\prime})$, refer to
higher dimensional supermultiplets\footnote{We use the symbol ``$\widehat {~~~}$'' 
in order to distinguish when a representation and associated Gadget are related 
to adinkras or to supermultiplets. Those related to adinkra representations and 
associated Gadgets appear without the ``hat,'' while those associated with supermultiplets
representations and associated Gadgets appear with the ``hat.'' }.  Like the adinkra
Gadget ${{\cal G}}$, the supermultiplet Gadget ${\widehat  {\cal G}}$ assigns a real 
number to the pair of supermultiplets denoted by $({\widehat {\cal R}})$, and 
$({\widehat {\cal R}}^{\prime})$.

The most general expression for a four dimensional ``Gadget"  defined in previous 
work \cite{HoLoRmY4D}, for minimal $\cal N$ = 1 supermultiplets that is Lorentz
covariant is given by
\be
 \eqalign{ {~~}
{\widehat  {\cal G}} [  ({\widehat {\cal R}}) , ({\widehat {\cal R}}^{\prime}) ] 
~&=~ m_1 \,  [{ {\bm H}}{}^{\mu}{}^{({\widehat {\cal R}})}
]{}_{a \, b \, c}{}^d  \, [ { {\bm H}}{}_{\mu}{}^{({\widehat {\cal 
R}}^{\prime})} ]{}^{a \, b}{}_{d}{}^c   \cr
&~~~~+~ m_2 \,   (\gamma^{\a})_c^{~e}  \,  [{ {\bm H}}{}^{\mu}{
}^{({\widehat {\cal R}})} ]{}_{a \, b \, e}{}^f  \, (\gamma_{\a})_f^{
~d} \, [ { {\bm H}}{}_{\mu}{}^{({\widehat {\cal R}}^{\prime})} ]{}^{a \, 
b}{}_{d}{}^c                \cr
&~~~~+~ m_3 \,   ([\, \gamma^{\a}  ~,~ \gamma^{\b} \,])_c^{~e} \,  
[{ {\bm H}}{}^{\mu}{}^{({\widehat {\cal R}})} ]{}_{a \, b \, 
e}{}^f  \, ([\, \gamma_{\a}  ~,~ \gamma_{\b} \,])_f^{~d} \, [ { {\bm 
H}}{}_{\mu}{}^{({\widehat {\cal R}}^{\prime})} ]{}^{a \, b}{}_{d}{}^c  
{~~~~~~}  \cr
&~~~~+~ m_4 \,    (\gamma^5 \gamma^{\a})_c^{~e}  \,  [{ {\bm 
H}}{}^{\mu}{}^{({\widehat {\cal R}})} ]{}_{a \, b \, e}{}^f  \, (\gamma^5
\gamma_{\a})_f^{~d} \, [ { {\bm H}}{}_{\mu}{}^{({\widehat {\cal R}}^{\prime
})} ]{}^{a \, b}{}_{d}{}^c    \cr
&~~~~+~ m_5 \,   (\gamma^5)_c^{~e}  \,  [{ {\bm H}}{
}^{\mu}{}^{({\widehat {\cal R}})} ]{}_{a \, b \, e}{}^f  \, (\gamma^5
)_f^{~d} \, [ { {\bm H}}{}_{\mu}{}^{({\widehat {\cal R}}^{\prime})} ]{}^{a 
\, b}{}_{d}{}^c  ~~~,
}  \label{GdGET1}
\ee
in terms of constants $m_1$, $m_2$, $m_3$, $m_4$, and
$m_5$.  The expressions in (\ref{GdGET1}) define a mapping that
assigns the value of a real number given by ${\widehat  {\cal G}}$ 
to the two supermultiplet representations denoted by $({\widehat 
{\cal R}})$ and, $({\widehat {\cal R}}^{\prime})$.  

The constant rank four tensors $[{ {\bm H}}{}^{\mu}{}^{({\widehat 
{\cal R}})} ]{}_{a \, b \, c}{}^d$, and $[ { {\bm H}}{}^{\mu}{}^{({\widehat {\cal R}}^{\prime
})} ]{}^{a \, b}{}_{d}{}^c$ are called the ``holoraumy'' tensors for the supermultiplet 
representations $({\widehat {\cal R}})$, and $({\widehat {\cal R}}^{\prime})$, respectively.   
The means for calculating the ``holoraumy'' tensors for the minimal supermultiplet 
representations was given previously \cite{HoLoRmY4D} with results that were
reported.

All 4D  holoraumy tensors can be subjected to a transformation of the 
form
\be   \eqalign{
\left[{ {\bm H}}{}^{\mu}{}^{({\widehat {\cal R}})}\right]{}_{a \, b \, c}{}^d \, ~&\to~ 
(\gamma^{5} )_{c}{}^{e} \, \left[{ {\bm H}}{}^{\mu}{}^{({\widehat {\cal R}})}\right]
{}_{a \, b \, e}{}^f ( \gamma^{5} )_{f}{}^{d} \, 
}  \label{HTTswap}
\ee
which is equivalent to a ``parity-swap'' for all the bosonic fields in each
supermultiplets.  Accordingly, all scalars are interchanged with 
pseudo-scalars and vice-versa, all vectors are interchanged with axial-vectors
and vice-versa, etc.

Since the chiral supermultiplet has equal numbers of bosons of opposite
parity, its holoraumy tensor is unchanged by the transformation in (\ref{HTTswap}).  
This is not the case for the latter two supermultiplets.  The transformation 
in (\ref{HTTswap}) replaces the vector in the vector supermultiplet by an axial 
vector and similarly replaces the tensor in the tensor supermultiplet by an axial 
tensor, thus leading to new holoraumy tensors that follow.
\be   \eqalign{
\left[{ {\bm H}}{}^{\mu}{}^{(AVS )}\right]{}_{a \, b \, c}{}^d \, ~&=~ 
(\gamma^{5} )_{c}{}^{e} \, \left[{ {\bm H}}{}^{\mu}{}^{(VS)}\right]
{}_{a \, b \, e}{}^f ( \gamma^{5} )_{f}{}^{d} \,  ~~~,  \cr
\left[{ {\bm H}}{}^{\mu}{}^{(ATS)}\right]{}_{a \, b \, c}{}^d \, ~&=~ 
(\gamma^{5} )_{c}{}^{e} \, \left[{ {\bm H}}{}^{\mu}{}^{(TS)}\right]
{}_{a \, b \, e}{}^f ( \gamma^{5} )_{f}{}^{d} \,  ~~~~.
}  \label{HTTswap1}
\ee
Including the two new representation implies all can be expressed as
in a single formula by writing\footnote{Here we have changed slightly the conventions
used in \cite{HoLoRmY4D}.}
\be  \eqalign{ {~~~~~~~~~~~}
\left[{ {\bm H}}{}^{\mu}( {\rm p}_{({\cal R})},  {\rm q}_{({\cal R})}, {\rm r}_{({\cal R})},   
{\rm s}_{({\cal R})}  \,) \right]{}_{a \, b \, c}{}^d  ~&=~ -i 2 \, {\Big [} \, \,  {\rm p}_{({\cal 
R})} \, C_{ab} \, (\gamma^{\m})_c{}^{d}  ~+~   {\rm q}_{({\cal R})} \, (\gamma^{5}
)_{ab} (\gamma^{5}\gamma^{\m})_{c}{}^{d}     \cr 
&~~~~~~~~~~~~
 ~+~ {\rm r}_{({\cal R})} \,  (\gamma^5 \gamma^{\mu} )_{ab} \, 
 (\gamma^5)_{c}{}^{d}     \cr
 &~~~~~~~~~~~~~+~  \fracm 12 \,  {\rm s}_{({\cal R})} \, (\gamma^5 \gamma^{\nu} )_{ab} \, (\gamma^5 \, [
 \gamma {}_{\nu}\,, \, \gamma^{\mu} ])_{c}{}^{d}    \, \, {\Big ]}    ~~~,
}   \label{4DHs}
\ee
where the representation labels in this formula takes on the range of values 
given by $(CS)$, $(VS)$, $(TS)$, $(AVS)$, and $(ATS)$.  The integers ${\rm 
p}_{({\cal R})}, \,{\rm q}_{({\cal R})}, \, {\rm r}_{({\cal R})}$, {\rm {and}} ${\rm s}_{({\cal R})}$ 
are correlated with the representation labels according to the 
results shown in equations in (\ref{pees}).
\be \eqalign{
0 ~&=~ {\rm p}_{(CS)} ~=~  {\rm q}_{(CS)} ~=~  {\rm r}_{(CS)}   
~~~~~~~~~~~~~~~~,~~~  1 ~=~ {\rm s}_{(CS)}  ~~~, \cr
1 ~&=~  {\rm p}_{(VS)} ~=~  {\rm q}_{(VS)} ~=~ {\rm r}_{(VS)}   
~~~~~~~~\,~~~~~\,~~,~~~  0 ~=~ {\rm s}_{(VS)}  ~~~, \cr
1 ~&=~  - \, {\rm p}_{(TS)} ~=~ {\rm q}_{(TS)} ~=~   - \, {\rm r}_{(TS)}  
 ~~~~~~~\,~,~~~  0 ~=~ {\rm s}_{(TS)} ~~~,  \cr
1 ~&=~  -\, {\rm p}_{(AVS)} ~=~  -\, {\rm q}_{(AVS)} ~=~ {\rm r}_{(AVS)} 
\,~~,~~~  0~=~ {\rm s}_{(AVS)}    ~~, \cr
1 ~&=~ {\rm p}_{(ATS)} ~=~  -\, {\rm q}_{(ATS)} ~=~  -\, {\rm r}_{(ATS)}   
~~~,~~~0 ~=~ {\rm s}_{(ATS)} ~\,~.~~~~   
} \label{pees}  \ee
When the result in (\ref{4DHs}) is substituted into (\ref{GdGET1}), it yields
\be
 \eqalign{ {~~}
{\widehat  {\cal G}} [  ({\widehat {\cal R}}) , ({\widehat {\cal R}}^{\prime}) ]
~&=~ 256 \,  {\Big \{}   ~  {\rm p}_{({\cal R})} \,   {\rm p}_{({\cal R}^{\prime})}  
{\big (} ~ - \, m_1~+~ 2 m_2 ~+~ 2 m_4 ~+~ m_5 ~ {\big )} ~+~  \cr
&{~~~~~~~~~~~~~}     {\rm q}_{({\cal R})} \,   {\rm q}_{({\cal R}^{\prime})}  
{\big (} ~  m_1~+~ 2 m_2 ~+~ 2 m_4 ~-~ m_5 ~ {\big )} ~+~  \cr
&{~~~~~~~~~~~~~}     {\rm r}_{({\cal R})} \,   {\rm r}_{({\cal R}^{\prime})}  
{\big (} ~  m_1 ~-~ 4 m_2 ~-~ 48 m_3 ~+~ 4 m_4 ~+~ m_5 ~ {\big )} ~+~  \cr
&{~~~~~~~~~~~~~}   3\,  {\rm s}_{({\cal R})} \,   {\rm s}_{({\cal R}^{\prime})}  
{\big (} ~  - m_1 ~-~ 16 m_3 ~-~ m_5 ~ {\big )} ~ {\Big \}} ~~~.
}  \label{GdGET3}
\ee
Let us now observe that there are five undetermined constants $m_1$,
$\dots$, $m_5$ on the four lines of the equation in (\ref{GdGET3}).  We
are therefore free to impose the following {\em {four}} conditions on these
constants,
\be  \eqalign{
\frc 1{768} ~&=~ - \, m_1~+~ 2 m_2 ~+~ 2 m_4 ~+~ m_5    ~~~~~~~~~~~,   \cr
\frc 1{768} ~&=~ m_1~+~ 2 m_2 ~+~ 2 m_4 ~-~ m_5   ~~~~\,~~~~~~~~~~~,   \cr
\frc 1{768} ~&=~  m_1 ~-~ 4 m_2 ~-~ 48 m_3 ~+~ 4 m_4 ~+~ m_5 ~~~,   \cr
\frc 1{768} ~&=~  - m_1 ~-~ 16 m_3 ~-~ m_5  ~~~~\,~~~~~~~~~~~~~~~~~,
}   \label{Cnstnst}    \ee
which possesses the solution
\be  \eqalign{
m_1  ~&=~   - \, m_4 ~=~m_5  ~=~ - \, \left[ \, \frc 1{1,536} ~+~ 8 m_3 \, \right] ~~~, ~~~
m_2 ~=~  -\,  8 m_3   ~~~,   \cr
}   
\ee
and further implies the result for the 4D, Lorentz covariant Gadget can be written as
\footnote{These results correct the previously reported ones in \cite{HoLoRmY4D}.}
\be
 \eqalign{ {~~}
{\widehat  {\cal G}} [  ({\widehat {\cal R}}) , ({\widehat {\cal R}}^{\prime}) ] 
~&=~ -\, \frc 1{1,536} \, {\Big \{ } \,\, [{ {\bm H}}{}^{\mu}{}^{({\widehat {\cal R}})}
]{}_{a \, b \, c}{}^d  \, [ { {\bm H}}{}_{\mu}{}^{({\widehat {\cal 
R}}^{\prime})} ]{}^{a \, b}{}_{d}{}^c   \cr
&~~~~~~~~~~~~~~~~~~~
-~ \,    (\gamma^5 \gamma^{\a})_c^{~e}  \,  [{ {\bm H}}{}^{\mu}{}^{({\widehat 
{\cal R}})} ]{}_{a \, b \, e}{}^f  \, (\gamma^5 \gamma_{\a})_f^{~d} \, [ { {\bm 
H}}{}_{\mu}{}^{({\widehat {\cal R}}^{\prime})} ]{}^{a \, b}{}_{d}{}^c    \cr
&~~~~~~~~~~~~~~~~~~
~+~ \,    (\gamma^5)_c^{~e}  \,  [{ {\bm H}}{}^{\mu}{}^{({\widehat {\cal R}})} 
]{}_{a \, b \, e}{}^f  \, (\gamma^5)_f^{~d} \, [ { {\bm H}}{}_{\mu}{}^{({\widehat 
{\cal R}}^{\prime})} ]{}^{a \, b}{}_{d}{}^c  ~ {\Big \} }  ~~~,  {~~~~~~~~~~~~~}
}  \label{GdGET2}
\ee
for $m_3$ = 0 or alternately as
\be
 \eqalign{ {~~}
{\widehat  {\cal G}} [  ({\widehat {\cal R}}) , ({\widehat {\cal R}}^{\prime}) ] 
~&=~  \frc 1{1,536} \, {\Big \{ } \,\, 
 (\gamma^{\a})_c^{~e}  \,  [{ {\bm H}}{}^{\mu}{
}^{({\widehat {\cal R}})} ]{}_{a \, b \, e}{}^f  \, (\gamma_{\a})_f^{
~d} \, [ { {\bm H}}{}_{\mu}{}^{({\widehat {\cal R}}^{\prime})} ]{}^{a \, 
b}{}_{d}{}^c                \cr
&~~~~~~~~~~~~~~~~~~~
-~ \, \frc 18   
  ([\, \gamma^{\a}  ~,~ \gamma^{\b} \,])_c^{~e} \,  
[{ {\bm H}}{}^{\mu}{}^{({\widehat {\cal R}})} ]{}_{a \, b \, 
e}{}^f  \, ([\, \gamma_{\a}  ~,~ \gamma_{\b} \,])_f^{~d} \, [ { {\bm 
H}}{}_{\mu}{}^{({\widehat {\cal R}}^{\prime})} ]{}^{a \, b}{}_{d}{}^c
  ~ {\Big \} }  ~~~,  {~~~~~~~~~~~~~}
}  \label{GdGET2a}
\ee
for 8$m_3$ = - 1/1,536.  Now independent of $m_1$, $\dots$,  $m_5$ due
to the four conditions (\ref{Cnstnst}), the expression in (\ref{GdGET3}) becomes
\be
 \eqalign{ {~~}
{\widehat  {\cal G}} [  ({\widehat {\cal R}}) , ({\widehat {\cal R}}^{\prime}) ]
~&=~ \frc 13 \,  {\Big [}   ~  {\rm p}_{({\cal R})} \,   {\rm p}_{({\cal R}^{\prime})} 
~+~    {\rm q}_{({\cal R})} \,   {\rm q}_{({\cal R}^{\prime})}  ~+~ {\rm r}_{({\cal R})} 
\, {\rm r}_{({\cal R}^{\prime})} ~+~  3\,  {\rm s}_{({\cal R})} \,   {\rm s}_{({\cal 
R}^{\prime})}  ~ {\Big ]}    ~~~.
}  \label{GdGET4}
\ee
It follows, from the allowed values of  ${\rm p}_{({\cal R})}$, ${\rm q}_{({\cal R})}$, 
$ {\rm r}_{({\cal R})}$, and $ {\rm s}_{({\cal R})}$ in (\ref{pees}), the supermultiplet
matrix analogous to the AGRM over these representations takes the form
\be
{\widehat  {\cal G}} [  ({\widehat {\cal R}}) , ({\widehat {\cal R}}^{\prime}) ] 
~=~ \left[\begin{array}{ccccc}
~1 & ~0 &  ~0  &~0 &  ~0 \\
~0 & ~~1 &  -\, \fracm 13 & -\, \fracm 13 &  -\, \fracm 13\\
~0 &  -\, \fracm 13 &  ~~ 1 & ~ -\, \fracm 13 &  -\, \fracm 13 \\
~0 &  -\, \fracm 13 &  -\, \fracm 13 & ~1 &  ~ -\, \fracm 13 \\
~0 &  -\, \fracm 13 &  -\, \fracm 13 & ~ -\, \fracm 13 &  ~1 \\
\end{array}\right]      ~~~.
\label{Gdgt3b}
\ee
This matrix has been obtained previously, but purely in the context of 
solely 1D arguments \cite{HoLoRmY2}.  Here we have proven this result
is obtained strictly on the basis of 4D  supermultiplet calculations also. 

Given the values of the supermultiplet Gadget representation matrix above, 
we can define a set of angles between the supermultiplets via the equation,
\be {
cos \left\{ \theta [({{\widehat {\cal R}}})\,  , \, ({ {\widehat {\cal R}}}^{\prime} )]   
{}_{} \right\} ~=~ {{{ {\widehat {\cal G}}} [ \, ({ {\widehat {\cal R}}}) , ({ {\widehat 
{\cal R}}}^{\prime}) \,]{}_{} } \over {~ {\sqrt{{ {\widehat {\cal G}}} [ \, ({ {\widehat 
{\cal R}}}) , ({ {\widehat {\cal R}}}) \, ]{}_{}}} \, {\sqrt{ { {\widehat {\cal G}}} [ \, ({ 
\widehat {\cal R}}^{\prime}) , ({ {\widehat {\cal R}}}^{\prime}) ]{}_{}}}~~ } } ~~~,  }   
\label{M4gSF}
\ee
and the angles thus found correspond to $arccos\left( - 1/3 \right)$, $\pi /2$, and 0 
via calculation directly in 4D, $\cal N$ = 1 supersymmetry.  The supermultiplet Gadget 
in (\ref{GdGET4}) allows us to map:

(a.) the vector supermultiplet into the point on the sphere along the line segment $\Bar {OA}$,

(b.) the axial-vector supermultiplet into the point on the sphere along the line segment $\Bar {OB}$,

(c.) the axial-tensor supermultiplet into the point on the sphere the along line segment $\Bar {OC}$,
and

(d.) the tensor supermultiplet into the point on the sphere along the line segment $\Bar {OD}$,
\vskip.1pt
\noindent
in the image shown in Fig.\ 2.  This is a ``weight space like'' diagram showing
these minimal 4D, $\cal N$= 1 supermultiplets while the chiral supermultiplet lies in
a direction orthogonal to this three dimensional subspace.  With this interpretation,
the formulae in (\ref{4DHs}) and (\ref{GdGET4}) define a space of minimal 
4D, $\cal N$ = 1 representations together with its metric.  The coordinates of points
for each superfield representation in the space are provided by the values 
of ${\rm p}_{({\cal R})}$, ${\rm q}_{({\cal R})}$, $ {\rm r}_{({\cal R})}$, and ${\rm 
s}_{({\cal R})}$.  So the the vector supermultiplet, the axial-vector supermultiplet,
the axial-tensor supermultiplet, and the tensor supermultiplet all reside in the
${\rm s}_{({\cal R})}$ = 0 three dimensional subspace.

Let us delve more deeply into this point. The quantity $\left( \, {\rm p}_{({\cal R})},
\, {\rm q}_{({\cal R})}, \, {\rm r}_{({\cal R})}, \, {\rm s}_{({\cal R})}  \, \right)$ defines
a vector in four-dimensional space.  For this space the expression in (\ref{GdGET4})
defines a metric or inner product.  The results already presented in the chapter
imply the following components of such a vector associated with each representation
label. \newpage

\vskip4pt
\begin{table}[h]
\begin{center}
\footnotesize
\begin{tabular}{|c|c|c|c|c|}\hline 
$(\widehat {\cal R})$  & ${\rm p}_{({\cal R})}$ & ${\rm q}_{({\cal R})}$ &
${\rm r}_{({\cal R})}$ & ${\rm s}_{({\cal R})}$ \\ \hline \hline
(CS) & 0 & 0 & 0 & 1  \\ \hline
(VS) & 1 & 1 & 1 & 0  \\ \hline
(AVS) & -1 & -1 & 1 & 0  \\ \hline
(ATS) &  1 & - 1 & - 1 & 0   \\ \hline
(TS) & - 1 & 1 & - 1 & 0   \\ \hline
\end{tabular}
\vskip10pt {{{\bf {Table}} {\bf {2:}}
Components For The (CS), (VS), (AVS) (ATS), and (TS) Supermultiplets}}
\end{center}
\end{table}  
\noindent
The components shown in this table indicate that the vectors associated 
with each of the representation labels are unit vectors with regard to the
metric defined in (\ref{GdGET4}).  This same metric implies that the
vector associated with (CS) is orthogonal to the vectors associated with
the four remaining representations.  These remaining representations all
`live' in a three dimensional subspace which is shown in Fig.\  4.
\begin{figure}[ht]
\begin{center}
\begin{picture}(70,26)
\put(-30,-33){\includegraphics[height =2.2in]{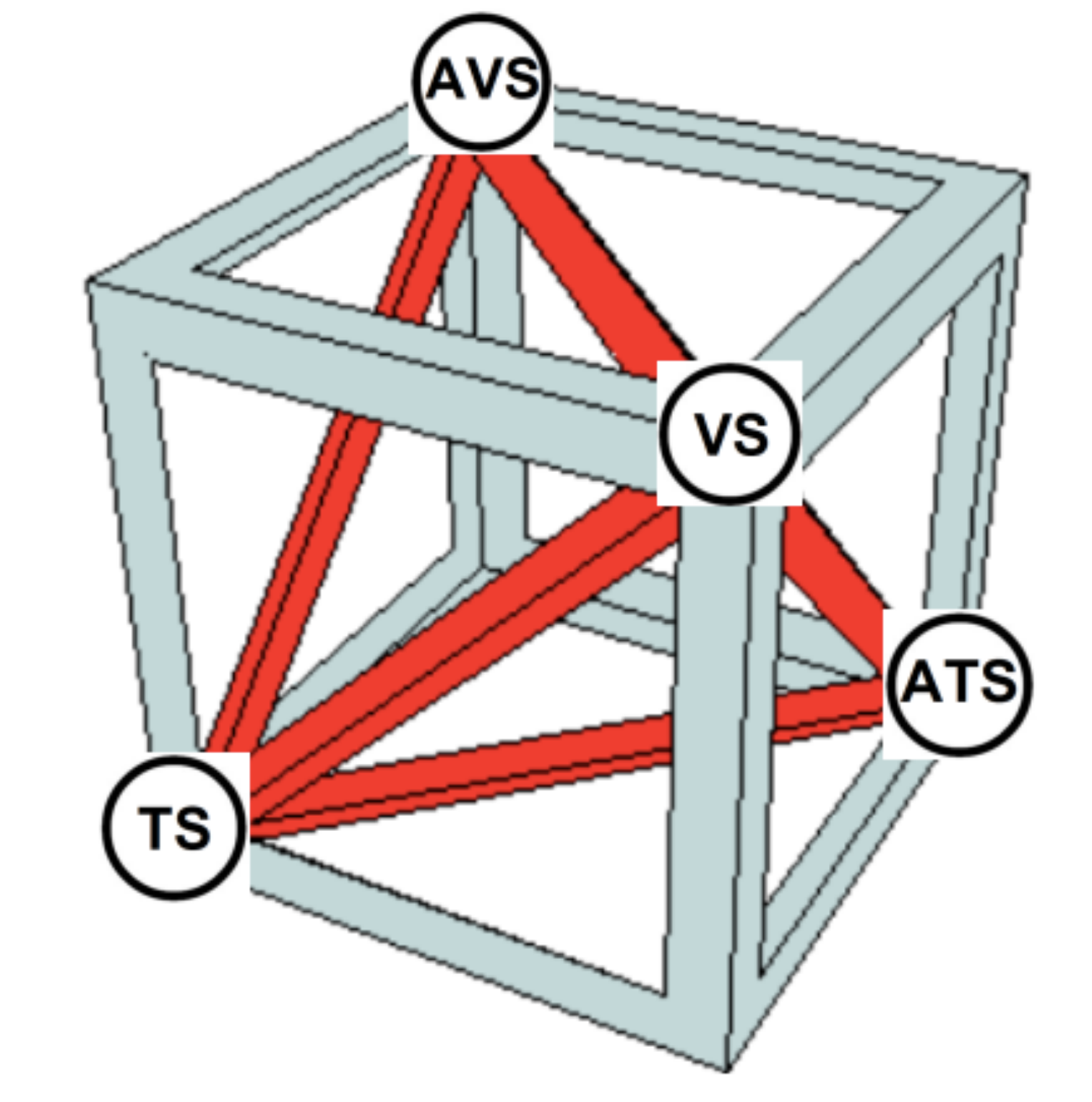}}
\put(-25,-22.8){\includegraphics[height =0.29in]{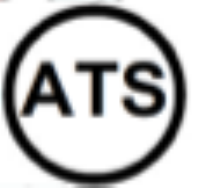}}
\put(15,-15.6){\includegraphics[height =0.29in]{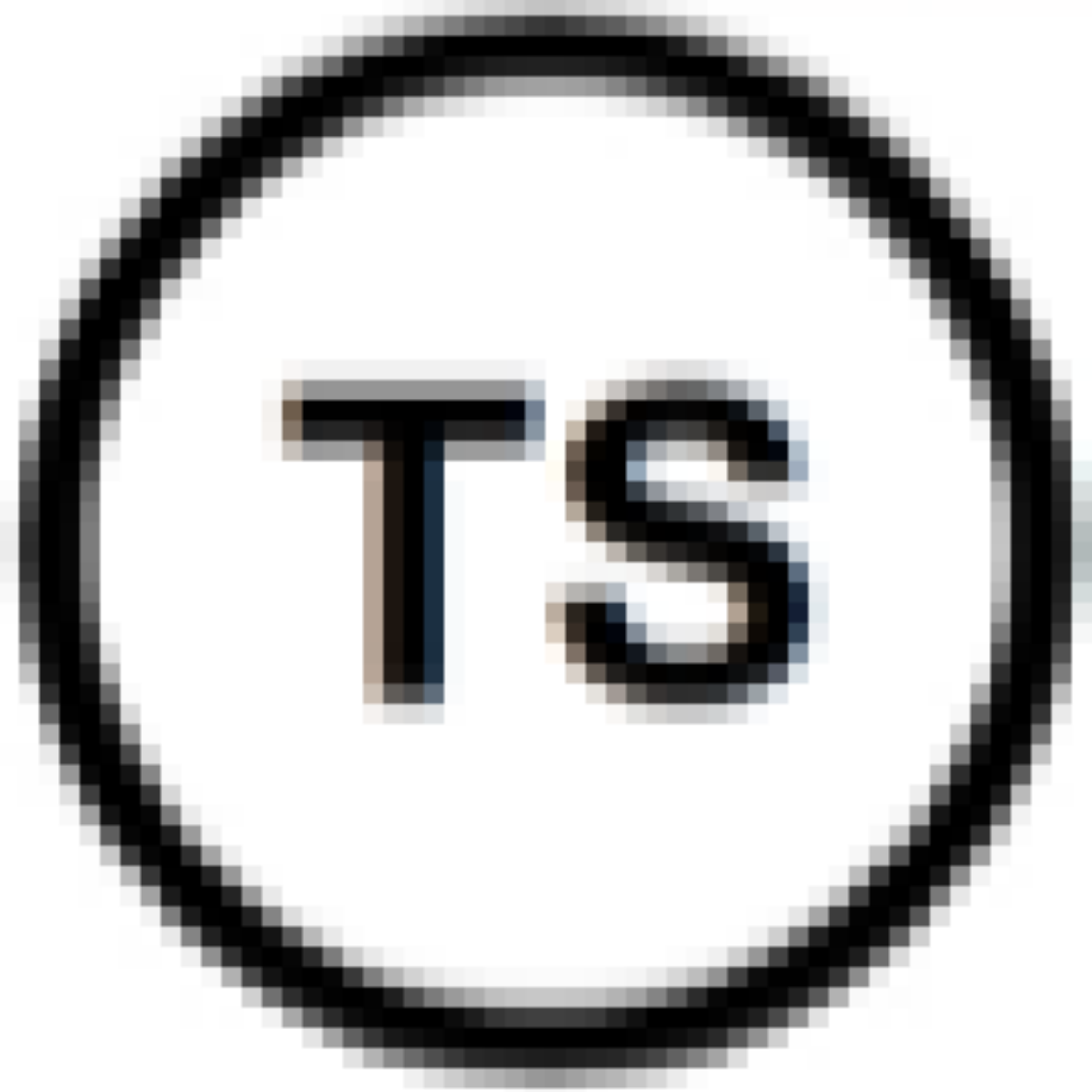}}
\put(48,-33){\includegraphics[height =2.2in]{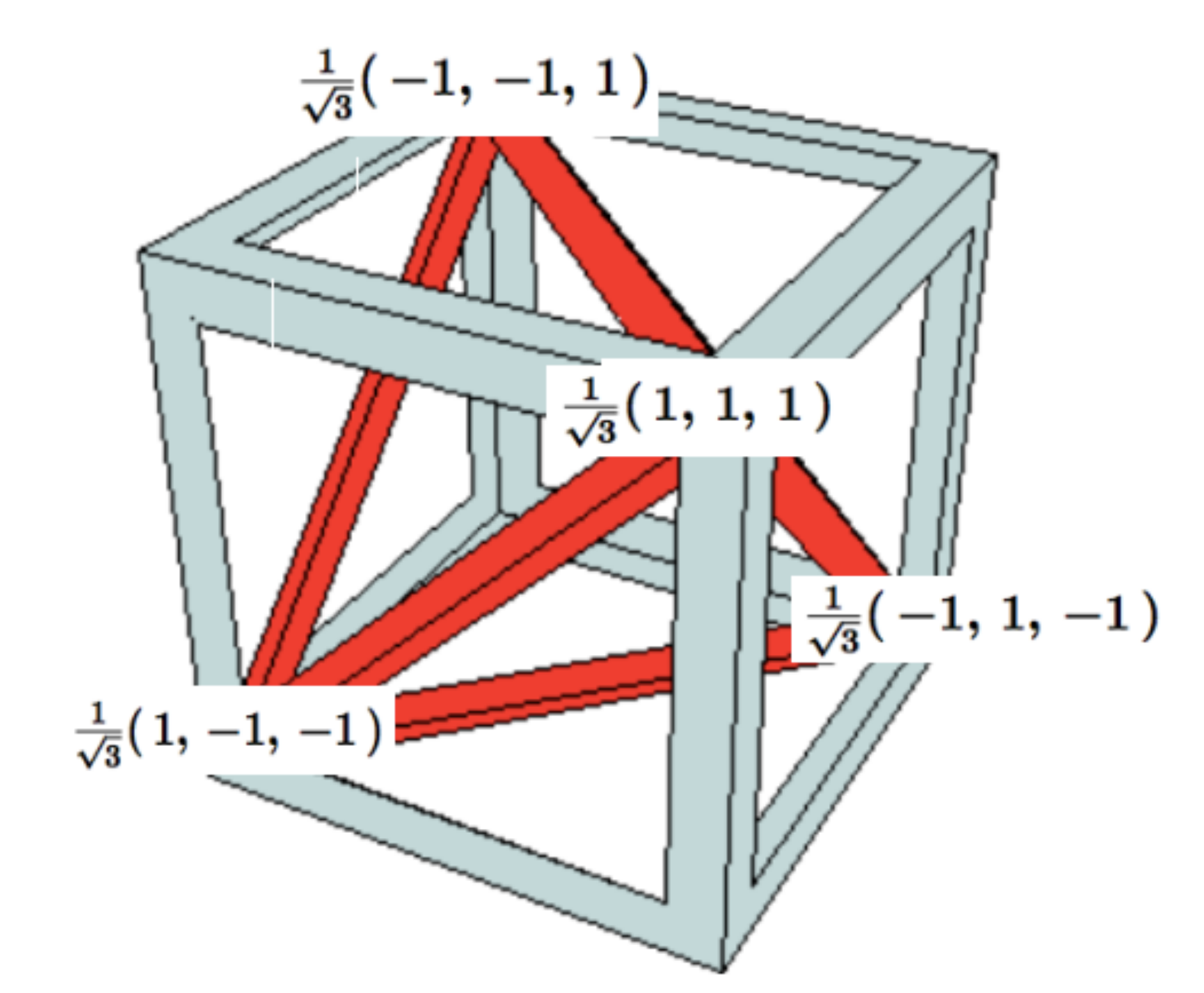}}
\put(-32,-40){{{\bf {Figure}} {\bf {4:}}
Cubically Inscribed Tetrahedron With Supermultiplets And Their Re- }}
\put(-10,-45){{Scaled
${\rm p}_{({\cal R})}$,
${\rm q}_{({\cal R})}$, and ${\rm r}_{({\cal R})}$ Coordinates At
Intersecting Vertices}}
\end{picture}
\end{center}
\label{TetMin}
\end{figure}
 \vskip1.2in
 \begin{center}
$~~~$
\end{center}
We note the diagonal entries in the matrix shown in (\ref{Gdgt3b})
imply that the distance defined by the metric in (\ref{GdGET4}) from
the center of the cube to any of its vertices must be equal to one.

The vector associated with the (CS) representation is not shown in this
diagram as it is not contained within this three dimensional-subspace.
We have re-scaled the ${\rm p}_{({\cal R})}$, ${\rm q}_{({\cal R})}$, and 
${\rm r}_{({\cal R})}$ components so that the standard Euclidean
metric implies that the vectors drawn from the center of the cube to
any vertex containing a supermultiplet vectors has length one.

It is now a simple matter to show the three dimensional rotation matrix 
${\bm {\rm R}}$ (satisfying $ {\bm {\rm R}} \,  {\bm {\rm R}}{}^T ~=~ {\bm 
{\rm R}}{}^T \,  {\bm {\rm R}} ~=~$ ${\bm {\rm I}}$, where the superscript T 
stands for transposed) described by\footnote{This matrix should not be 
confused with one of the R-matrices that appear in (\ref{GarDNAlg2}).}
\be
{\bm {\rm R}} ~=~ 
\left[\begin{array}{ccc}
- \, \fracm {\sqrt 6}6 & ~~ \fracm {\sqrt 6}3 &  -\, \fracm {\sqrt 6}6 \\
\fracm {\sqrt 2}2 & ~~0 &  -\, \fracm {\sqrt 2}2\\
\fracm {\sqrt 3}3 & \fracm {\sqrt 3}3 &  \fracm {\sqrt 3}3\\
\end{array}\right]
\ee
transforms the red tetrahedron (defined {\em {solely}} from superfield 
considerations above) to align it with the tetrahedron shown at the end 
of chapter two (defined {\em {solely}} from adinkra considerations).  We 
write the four vectors from the point O to the points of intersection 
indicated below Fig.\ 2 in the forms
\be
{~~~~~~~}
{\vec A}{}_1  ~=~ \left[\begin{array}{c}
~0  \\
~ 0\\
~1 \\
\end{array}\right]  ~~,~~
{\vec A}{}_2  ~=~  \left[\begin{array}{c}
- \, \fracm {\sqrt 2}3  \\
- \, \fracm {\sqrt 6}3   \\
- \, \fracm {1}3  \\
\end{array}\right] 
 ~~,~~
{\vec A}{}_3  ~=~  \left[\begin{array}{c}
 \fracm { 2{\sqrt 2}}3  \\
~0   \\
- \, \fracm {1}3  \\
\end{array}\right]  ~~,~~
{\vec A}{}_4  ~=~  \left[\begin{array}{c}
- \, \fracm {\sqrt 2}3  \\
 \fracm {\sqrt 6}3   \\
- \, \fracm {1}3  \\
\end{array}\right] 
 ~~~~,
\ee
and in a similar manner we denote four more vectors by the equations
\be
{~~~~~~~}
{\vec B}{}_1  ~=~  \fracm 1{\sqrt 3}  \left[\begin{array}{c}
~1  \\
~ 1\\
~1 \\
\end{array}\right]  ~~,~~
{\vec B}{}_2  ~=~ \fracm 1{\sqrt 3} \left[\begin{array}{c}
- \, 1  \\
- \, 1   \\
 ~1  \\
\end{array}\right] 
 ~~,~~
{\vec B}{}_3  ~=~  \fracm 1{\sqrt 3} \left[\begin{array}{c}
- \, 1  \\
~ 1   \\
 - \, 1  \\
\end{array}\right]  ~~,~~
{\vec B}{}_4  ~=~  \fracm 1{\sqrt 3} \left[\begin{array}{c}
~ 1  \\
- \, 1   \\
 - \, 1  \\
\end{array}\right] 
 ~~,
 \ee
having been obtained from the coordinates of the intersecting points that appear 
on the right hand side of Fig.\ 4.  Next one can note
\be
{\bm {\rm R}} \, {\vec B}{}_1 ~=~ {\vec A}{}_1  ~~~~,~~~~
{\bm {\rm R}} \, {\vec B}{}_2 ~=~ {\vec A}{}_2  ~~~~,~~~~
{\bm {\rm R}} \, {\vec B}{}_3 ~=~ {\vec A}{}_3 ~~~~,~~~~
{\bm {\rm R}} \, {\vec B}{}_4 ~=~ {\vec A}{}_4  ~~~~,
\ee
which implies the alignment of the tetrahedron in Fig.\ 2 and the tetrahedron
in Fig.\ 4. 

Let us close this section with two amusing idylls, with the second one possibly
hinting at tantalizing additional developments.

A first one is the observation that the presence of the sphere, the cube, and the 
tetrahedron (with the latter two being among the five platonic solids \cite{PlatS}) 
implies 4D, $\cal N$ = 1 space-time supersymmetry representation theory contains 
in a hidden manner a structure with some similarity to Kepler's ``Mysterium 
Cosmographicum'' \cite{MC}.  

The second comment relates to results by Nekrasov \cite{Nikit1,Nikit2,Nikit2a,Nikit3,Nikit4,Nikit5,Nikit6}.  
In the first
of these, it is noted that the tetrahedral angle $arccos\left( - 1/3 \right)$ can be uncovered 
by looking at Yang-Mills gauge theories in various dimensions in the presence of a supergravity 
background.  In relation to summing up instantons for 4D, $\cal N$ = 2 theories he observed that, 
``the theory is subject to a special supergravity background, which softly breaks super-Poincare
symmetry yet deforms some of the supercharges in such a way that they anti-commute onto 
spacetime rotations instead of translations. The supersymmetric field configurations then become 
(for gauge groups the products of unitary groups) enumerated by sequences of Young diagrams, 
i.e. two dimensional arrangements of squares. 

One can then study higher dimensional theories, e.g. maximal super-Yang-Mills in 6 or 7 dimensions (which should be defined quantum mechanically using D6 branes in IIA string theory) and then the instanton counting becomes the study of three dimensional Young diagrams aka the plane partitions. These can be visualized by projecting them onto a two-plane along the (1,1,1) axis, where the plane partitions look like the tessellations of the plane by three types of rombi.''
 
It is this final step that leads to the appearance of the angle $arccos\left( - 1/3 \right)$
as seen in adinkra gadget values.

In the work of \cite{Nikit6}, there is also one other tantalizing similarity between some
of Nekrasov's discussions and the structure uncovered in the work of \cite{permutadnk}.
In the former, there is defined a function $\varepsilon$ that maps the set of 2-element
subsets of the partitions of four objects to ${\bm Z}_2$.  The 2-element subsets,
which Nekrasov denoted by 
\be
{\underline 6} ~=~ \left( {4 \over 2} \right) ~=~ \{ 12, \, 13, \, 14, \, 23, \,  24, \, 34 \}
\label{Nk1}
\ee
correspond to the six distinct sets shown in the Venn diagram in Fig.\ 5.  
Furthermore, the work in \cite{permutadnk} explicitly seems to note a realization of Nekrasov's  
$\varepsilon$-map. When the 2-element subsets are represented by permutation 
matrices, the $\varepsilon$-map corresponds to a construction based on matrix transposition
seen in  \cite{permutadnk}.

\newpage
\section{The Coxeter Group $\bm {BC_4}$ \& The ``Small  ${\bm {\rm {BC}
{}_4}}$ Library"}
\label{s4}

We define elements of the Coxeter Group ${\rm BC}{}_4$ \cite{CXgrp} by 
consider the set of all real 4 $\times$ 4 matrices that arise as a bilinear 
product of the form
\cite{permutadnk}
\begin{equation}
 {\bm \rL} ~=~ 
     {\bm {\cal S}} \,{\bm  \cdot} \, {\bm {\cal P}}
\label{aas0}
\end{equation}
The  real $4\times4$ diagonal matrix ${\bm {\cal S}}$ is the ``Boolean 
Factor''  \cite{permutadnk} and squares to the identity.  The matrix ${\bm {
\cal P}}$ is a representation of a permutation of $4$ objects. There are 
$ 2^d \, d!$ =  $2^4 \times 4! =384$ ways to choose the factors which is 
the dimension of the Coxeter group ${\rm BC}{}_4$.  More explicitly this 
expression can be written as
\begin{equation}
(\rL_{\sss\rI} {}^{({\cal R})})_i{}^\hk ~=~  [{\cal S}^{\sss(\rI)} {}^{({\cal 
R})}]_i{}^\hl\, [{\cal P}_{\!\sss(\rI)} {}^{({\cal R})}]_\hl{}^\hk, \qquad 
\text{for each fixed }\rI=1,2, 3, 4 \text{ on the LHS.}
\label{aas1}
\end{equation}
This notation anticipates distinct adinkra representations exist and are
denoted by ``a representation label'' $({\cal R})$ that takes on values
from one to some integer, T. 

Our experience in the work of \cite{permutadnk} gave a very valuable
lesson...there is an smaller algebraic structure, the Vierergruppe, 
\be  \eqalign{
{\bm \{} {\cal V}  {\bm \} }  ~=~&{\bm \{} (), ~ (12)(34), ~ (13)(24), ~ (14)(23) 
{\bm \} }     ~~~,  \cr
} \label{Cgrp4}
\ee
whose role is critical.  The Vierergruppe elements above are written using 
cycle notation to indicate the distinct permutations and can be used in 
partitioning the permutation elements.  These partitions allow all 24
permutation elements to be gathered into six ``corrals'' which then provide 
a basis for constructing adinkras.  Since the elements of ${\bm \{} {\cal V}  
{\bm \}} $ can also be represented as ${4 \times 4}$ matrices as well, we 
can alternately express them in the form of outer products of the 2 $\times$ 
2 identity matrix ${\bm {\rm I}}{}_{2 \times 2}$ and the first Pauli matrix ${\bm 
\s}^1$,
\be  \eqalign{
\bm {\{ {\cal V}_{(4)} \}}  ~=&~ {\bm \{ }  ~  {\bm {\rm I}}{}_{2 \times 2} \otimes   
{\bm {\rm I}}{}_{2 \times 2} , ~  {\bm {\rm I}}{}_{2 \times 2}  \otimes  {\bm \s}^1, ~ 
{\bm \s}^1 \otimes   {\bm {\rm I}}{}_{2 \times 2} , ~  {\bm \s}^1 \otimes
{\bm \s}^1 ~  {\bm \} }  ~~~.
} \label{Cgrp4z}
\ee
Written in this form, we are able to connect this expression back to the first 
works \cite{GRana1,GRana2} that launched our efforts. 

Using either notation, one can show that for unordered quartets, the equations 
\be  \eqalign{
(12) \, \bm {\{ {\cal V}_{(4)} \}} ~&=~    (34) \, \bm {\{ {\cal V}_{(4)} \}} ~=~   (1324) 
\, \bm {\{ {\cal V}_{(4)} \}} ~=~  (1423) \, \bm {\{ {\cal V}_{(4)} \}}  ~~~~,  \cr
(13) \, \bm {\{ {\cal V}_{(4)} \}} ~&=~   (24) \, \bm {\{ {\cal V}_{(4)} \}}  ~=~   (1234)
\, \bm {\{ {\cal V}_{(4)} \}} ~=~ (1432) \, \bm {\{ {\cal V}_{(4)} \}} ~~~~,  \cr
(14) \, \bm {\{ {\cal V}_{(4)} \}} ~&=~   (23) \, \bm {\{ {\cal V}_{(4)} \}} ~=~   (1243) 
\, \bm {\{ {\cal V}_{(4)} \}} ~=~ (1342) \, \bm {\{ {\cal V}_{(4)} \}}  ~~~~,  \cr
(123) \, \bm {\{ {\cal V}_{(4)} \}} ~&=~   (134) \, \bm {\{ {\cal V}_{(4)} \}} ~=~   (142) 
\, \bm {\{ {\cal V}_{(4)} \}} ~=~   (243) \, \bm {\{ {\cal V}_{(4)} \}} ~~~\,~~,  \cr
(124) \, \bm {\{ {\cal V}_{(4)} \}} ~&=~   (132) \, \bm {\{ {\cal V}_{(4)} \}} ~=~   
(143) \, \bm {\{ {\cal V}_{(4)} \}} ~=~   (234) \, \bm {\{ {\cal V}_{(4)} \}} ~~~\,~~.
} \label{VLsets}
\ee
\be  \eqalign{
 \bm {\{ {\cal V}_{(4)} \}} \, (12)  ~&=~    \bm {\{ {\cal V}_{(4)} \}} \, (34)  ~=~  \bm 
 {\{ {\cal V}_{(4)} \}} \, (1324) \,  ~=~    \bm {\{ {\cal V}_{(4)} \}} \, (1423)   ~~~~, \cr
\bm {\{ {\cal V}_{(4)} \}} \, (13) ~&=~   \bm {\{ {\cal V}_{(4)} \}} \,(24)  ~=~    \bm {\{ 
{\cal V}_{(4)} \}} \, (1234) ~=~   \bm {\{ {\cal V}_{(4)} \}} \, (1432)    ~~~~,  \cr
\bm {\{ {\cal V}_{(4)} \}} \, (14) ~&=~    \bm {\{ {\cal V}_{(4)} \}} \, (23) ~=~    \bm 
{\{ {\cal V}_{(4)} \}} \,  (1243)  ~=~    \bm {\{ {\cal V}_{(4)} \}} \, (1342)  ~~~~,  \cr
\bm {\{ {\cal V}_{(4)} \}} \, (123)  ~&=~    \bm {\{ {\cal V}_{(4)} \}} \, (134) ~=~    
\bm {\{ {\cal V}_{(4)} \}} \, (142) ~=~    \bm {\{ {\cal V}_{(4)} \}} \, (243)  ~~~\,~~,  \cr
\bm {\{ {\cal V}_{(4)} \}} \, (124) ~&=~    \bm {\{ {\cal V}_{(4)} \}} \, (132)  ~=~    
\bm {\{ {\cal V}_{(4)} \}} \, (143) ~=~    \bm {\{ {\cal V}_{(4)} \}} \, (234) ~~~\,~~.
} \label{VRsets}
\ee
of (\ref{VLsets}) and (\ref{VRsets}) are satisfied.  These define five ``corrals''
of the permutation operators. The set defined by $ \bm {\{ {\cal V}_{(4)} \}}$
provides a sixth such corral.

In order to precede with explicit calculations, it is necessary to choose ``fiducial set''
quartets\footnote{The ``fiducial set'' is simply a historical artifact of our path of
discovery on this subject has taken.} where explicit choices are made for which 
quartets of permutation matrices are given a specific designation and  \newpage 
\noindent what is the order of the permutations in each designation.  Using the 
conventions of \cite{permutadnk} we assign the following definitions of these
`fiducial set'' quartets.
\be  \eqalign{
{~~}&  ~~~~ {\bm {\rm L}}{}_{1} ~~~\,\,~~ {\bm {\rm L}}{}_{2} ~~~\,\,~~ {\bm {\rm 
 L}}{}_{3}  ~~~\,\,~~ {\bm {\rm L}}{}_{4}  \cr
{\bm \{ } {\bm {\cal P}}{}_{[1]} {\bm \}}  ~=~&{\bm \{}   (243), ~ (123 ), ~ (134), ~ (142) 
{\bm \} }  ~~~, \cr
{\bm \{} {\bm {\cal P}}{}_{[2]}  {\bm \} } ~=~&{\bm \{}   (234) ,~(124), ~ (132), ~(143) 
{\bm \} }   ~~~, \cr
{\bm \{} {\bm {\cal P}}{}_{[3]}  {\bm \} } ~=~&{\bm \{}  (1243), ~ (23 ),~ (14),  ~  (1342) 
{\bm \} }   ~~~, \cr
{\bm \{} {\bm {\cal P}}{}_{[4]}  {\bm \} } ~=~&{\bm \{}   (1432), ~ (24) ,  ~  (1234), ~ 
(13) {\bm \} }   ~~~, \cr
{\bm \{} {\bm {\cal P}}{}_{[5]}  {\bm \} } ~=~&{\bm \{}  (1324), ~  (1423),~ (12), ~ 
(34) {\bm \} }   ~~~, \cr
{\bm \{} {\bm {\cal P}}{}_{[6]}  {\bm \} } ~=~&{\bm \{} (13)(24),  ~  (14)(23), ~ (), ~ (12)(34)
 {\bm \} }     ~~~. \cr
} \label{PermSets}
\ee
In fact, if one computes the cycles that are associated with the adinkra shown in
Fig.\ 1, the cycles that arise from such deductions are precisely the cycles shown
for ${\bm \{ } {\bm {\cal P}}{}_{[1]} {\bm \}}$ and in the same order.   The meaning of
the results shown in (\ref{PermSets}) is that one can obtain the L-matrices shown 
at the top once appropriate Boolean Factors are attached to the permutations in
each corral.

We collectively express the permutations subsets as $\bm { \{  {\cal P}_{[\L]} \} }$, with 
the index $\bm {[\L]}$ taking on values [1] through [6].  These are cosets involving the 
Vierergruppe and this allows a partitioning of ${\bm {\rm {BC}{}_{4} }} $ (since it contains 
$S_4$) into six distinct subsets or ``corrals,'' that contain four permutations in 
unordered quartets
\begin{figure}[ht]
\begin{center}
\begin{picture}(70,60)
\put(-20,-48){\includegraphics[height = 140\unitlength]{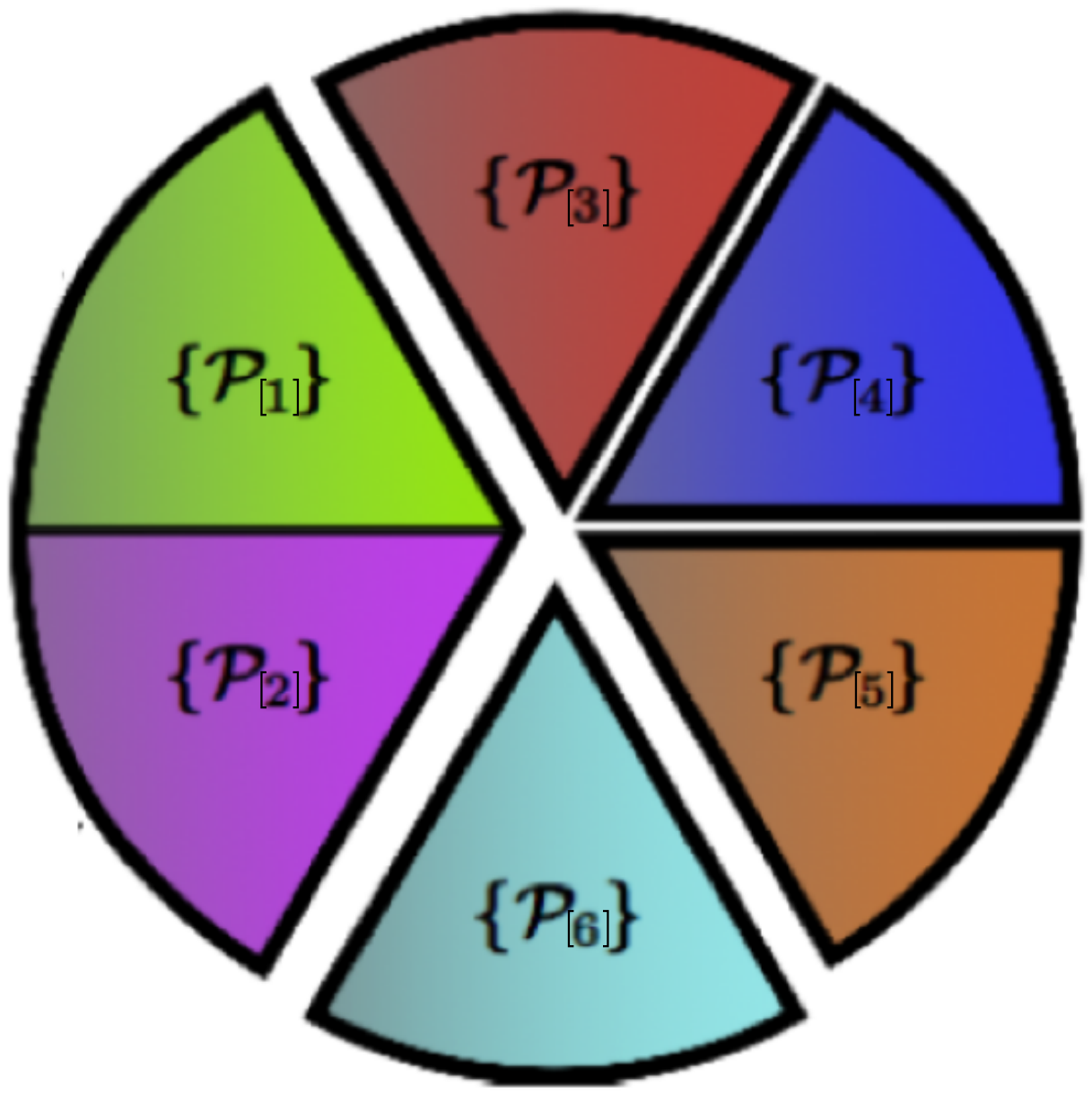}}
\end{picture}
\end{center}
\label{f:pie0}
\end{figure}
\vskip-6pt
$~~~~~~~~~~$ $~~~~~~~~~~$ $~~~~~~~$ $~~~~~$ {\bf {Figure}} {\bf {5:}} 
$ {\bm \{} {\bm {Z}_2}  \times {\bm {Z}_2}  {\bm \} } $ Partitioning of $\bm S_4$
\newline $~$ \newline 
as shown in Fig.\ 5.  All of the 384 elements associated with Fig.\ 1. now reside inside 
96 quartets which are equally distributed among all six partitions (i.e. 16 quartets per 
corral).

We now turn to the assignments of the  ``Boolean Factors'' to the permutation
elements.  In order to do this, we first observe there exits 16 sets of  ``Boolean Factors'' 
that can be assigned to each of the permutation partition factors and  faithfully represent  
${\rm BC}{}_4$.  

Each  ``Boolean Factor'' is equivalent to a real diagonal matrix that squares to the identity.
In the work of  \cite{permutadnk},  a convention was created whereby each of the  ``Boolean 
Factors'' could be unambiguously specified by a single real natural number.  Applying this 
convention to the ``Boolean Factor'' shown (\ref{BooN4dia})
\be \begin{array}{ccccccccc}
 {\bm {\cal S}}_{1} &=&  \, {\bm {\rm I}}{}_{2 \times 2} & \otimes & {\bm {\rm I}}{}_{2 \times 2} 
 & , &  &  & ~~~ \\
  {\bm {\cal S}}_{2} &=&  {\bm {\rm I}}{}_{2 \times 2} & \otimes &{\bm \s}^3  & , &   & &  ~~~ \\
  {\bm {\cal S}}_{3} &=& {\bm \s}^3 & \otimes & {\bm {\rm I}}{}_{2 \times 2}  & , &   & &  ~~~ \\
  {\bm {\cal S}}_{4} &=&  {\bm \s}^3  & \otimes &{\bm \s}^3  & , &   & &    ~~~ \\ 
\end{array} \label{BooN4dia}  \ee
we see these are mapped into the efficient notation as $(0)_b$,  $(10)_b$, $(12)_b$, and 
$(6)_b$, respectively.  

In Appendix B, the Boolean Factors appropriate to the fiducial choice of the quartets 
of permutations set out in Eq.\ (\ref{PermSets}) are listed.  As an example for how to 
use the list, it is instructive to construct an example in an explicit manner.  As noted in
the list, there are 16 appropriate choices of Boolean Factors for each of the fiducial
permutation quartets in (\ref{PermSets}).  

\subsection{Illustrative Discussion}

In order to show how all this formalism works, let us ``derive'' a set of L-matrices by 
making the choice ${\bm {\rm L}}{ {({{\bm S}_{\bm {\cal P}_{1}}}[12] \, \cdot \, {\bm {\cal 
P}_{[1]}} )}} $\footnote{The assignment of signed permutations and their association 
with the so-called ``chiral,'' ``tensor,'' and ``vector'' sets of L-matrices is incorrectly 
stated in the works of \cite{G&G1,G&G2} and are corrected in (\ref{Lp2}), (\ref{Lp4}), 
and (\ref{Lp6}) here.}.  From the list in Appendix B, we find the Boolean Factor quartet 
and from Eq.\ (\ref{PermSets}) we have the permutation element quartet.  So we need 
to calculate the ``dot product'' indicated by,
\be \eqalign{
 {\bm {\rm L}}{ {({{\bm S}_{\bm {\cal P}_{1}}}[12] \, \cdot \, {\bm {\cal P}_{[1]}} )}} ~&=~ 
( {\bm \{} (10)_b  , (12)_b , (6)_b  , (0)_b {\bm  \} }\, {\bm \cdot}  \, {\bm \{}   \,
 (243), ~ (123 ), ~ (134), ~ (142) \, {\bm \} } )   \cr
 ~&=~ {\bm \{}   \, (10)_b (243), ~ (12)_b (123 ), ~ (6)_b (134), ~ (0)_b
 (142) \, {\bm \} }  ~~~,  
} \label{Lp1} \ee
which implies
\be \eqalign{
{\bm {\rm L}}{}_{1}{ {({{\bm S}_{\bm {\cal P}_{1
}}}[12] \, \cdot \, {\bm {\cal P}_{[1]}} )}} ~&=~   \left[\begin{array}{cccc}
1&0&0&0\\ 0&0&0&-1\\ 0&1&0&0\\ 0&0&-1&0
\end{array}\right]   ~~~, ~~~
{\bm {\rm L}}{}_{2}{ {({{\bm S}_{\bm {\cal P}_{1
}}}[12] \, \cdot \, {\bm {\cal P}_{[1]}} )}} ~=~   \left[\begin{array}{cccc}
0&1&0&0\\ 0&0&1&0\\ -1&0&0&0\\ 0&0&0&-1
\end{array}\right]  ~~~, \cr
{\bm {\rm L}}{}_{3}{ {({{\bm S}_{\bm {\cal P}_{1
}}}[12] \, \cdot \, {\bm {\cal P}_{[1]}} )}}  ~&=~    \left[\begin{array}{cccc}
0&0&1&0\\ 0&-1&0&0\\ 0&0&0&-1\\ 1&0&0&0
\end{array}\right]    ~~~,~~~  
{\bm {\rm L}}{}_{4}{ {({{\bm S}_{\bm {\cal P}_{1
}}}[12] \, \cdot \, {\bm {\cal P}_{[1]}} )}} ~=~    \left[\begin{array}{cccc}
~0&~0&~0&~1\\ 1&0&0&0\\ 0&0&1&0\\ 0&1&0&0 
\end{array}\right]     ~~~~.
} \label{Lp2} \ee
These are precisely the L-matrices associated with the adinkra shown in 
Fig.\ 1.  For other choices of the Boolean Factor ${\bm S}_{\bm {\cal P}_{1
}} [\a]$, it is possible to generate other representations of the L-matrices
associated with the  $ {\bm {\cal P}_{[1]}}$ element.

Another set of such matrices can be constructed from ${\bm {\rm L}}{ {({{\bm S}_{\bm {
\cal P}_{2}}}[15] \, \cdot \, {\bm {\cal P}_{[2]}} )}} $ which implies
\be \eqalign{
 {\bm {\rm L}}{ {({{\bm S}_{\bm {\cal P}_{2}}}[15] \, \cdot \, {\bm {\cal P}_{[2]}} )}} ~&=~ 
( {\bm \{} 14)_b  , (4)_b , (8)_b  , (2)_b {\bm  \} }\, {\bm \cdot}  \, {\bm \{}   \,
  (234 ), ~ (124), ~ (132), ~ (143) \, {\bm \} } )   \cr
 ~&=~ {\bm \{}   \, (14)_b (124), ~ (4)_b (124 ), ~ (8)_b (132), ~ (2)_b
 (143) \, {\bm \} }  ~~~,  
} \label{Lp3} \ee
and this yields 
\be \eqalign{
{\bm {\rm L}}{}_{1}{ {({{\bm S}_{\bm {\cal P}_{2
}}}[15] \, \cdot \, {\bm {\cal P}_{[2]}} )}} ~&=~   \left[\begin{array}{cccc}
1&0&0&0\\ 0&0&-1&0\\ 0&0&0&-1\\ 0&-1&0&0
\end{array}\right]   ~~~, ~~~
{\bm {\rm L}}{}_{2}{ {({{\bm S}_{\bm {\cal P}_{2
}}}[15] \, \cdot \, {\bm {\cal P}_{[2]}} )}} ~=~   \left[\begin{array}{cccc}
0&1&0&0\\ 0&0&0&1\\ 0&0&-1&0\\ 1&0&0&0
\end{array}\right]  ~~~, \cr
{\bm {\rm L}}{}_{3}{ {({{\bm S}_{\bm {\cal P}_{2
}}}[15] \, \cdot \, {\bm {\cal P}_{[2]}} )}}  ~&=~    \left[\begin{array}{cccc}
0&0&1&0\\ 1&0&0&0\\ 0&1&0&0\\ 0&0&0&-1
\end{array}\right]    ~~~~~~~~,~~~  
{\bm {\rm L}}{}_{4}{ {({{\bm S}_{\bm {\cal P}_{2
}}}[15] \, \cdot \, {\bm {\cal P}_{[2]}} )}} ~=~    \left[\begin{array}{cccc}
~0&~0&~0&~1\\ 0&-1&0&0\\ 1&0&0&0\\ 0&0&1&0 
\end{array}\right]     ~~~~.
} \label{Lp4} \ee

A third set of such matrices can be constructed from ${\bm {\rm L}}{ {({{\bm S}_{\bm {
\cal P}_{3}}}[12] \, \cdot \, {\bm {\cal P}_{[3]}} )}} $ which implies
\be \eqalign{
 {\bm {\rm L}}{ {({{\bm S}_{\bm {\cal P}_{3}}}[12] \, \cdot \, {\bm {\cal P}_{[3]}} )}} ~&=~ 
( {\bm \{} 10)_b  , (12)_b , (0)_b  , (6)_b {\bm  \} }\, {\bm \cdot}  \, {\bm \{}   \,
 (1243), ~ (23 ), ~ (14), ~ (1342) \, {\bm \} } )   \cr
 ~&=~ {\bm \{}   \, (10)_b (1243), ~ (12)_b (23 ), ~ (0)_b (14), ~ (6)_b
 (1342) \, {\bm \} }  ~~~,  
} \label{Lp5} \ee
and this yields 
\be \eqalign{
{\bm {\rm L}}{}_{1}{ {({{\bm S}_{\bm {\cal P}_{3
}}}[12] \, \cdot \, {\bm {\cal P}_{[3]}} )}} ~&=~   \left[\begin{array}{cccc}
0&1&0&0\\ 0&0&0&-1\\ 1&0&0&0\\ 0&0&-1&0
\end{array}\right]   ~~~, ~~~
{\bm {\rm L}}{}_{2}{ {({{\bm S}_{\bm {\cal P}_{3
}}}[12] \, \cdot \, {\bm {\cal P}_{[3]}} )}} ~=~   \left[\begin{array}{cccc}
1&0&0&0\\ 0&0&1&0\\ 0&-1&0&0\\ 0&0&0&-1
\end{array}\right]  ~~~, \cr
{\bm {\rm L}}{}_{3}{ {({{\bm S}_{\bm {\cal P}_{3
}}}[12] \, \cdot \, {\bm {\cal P}_{[3]}} )}}  ~&=~    \left[\begin{array}{cccc}
0&0&0&1\\ 0&1&0&0\\ 0&0&1&0\\ 1&0&0&0
\end{array}\right]    ~~~~~~~~,~~~  
{\bm {\rm L}}{}_{4}{ {({{\bm S}_{\bm {\cal P}_{3
}}}[12] \, \cdot \, {\bm {\cal P}_{[3]}} )}} ~=~    \left[\begin{array}{cccc}
~0&~0&~1&~0\\ -1&0&0&0\\ 0&0&0&-1\\ 0&1&0&0 
\end{array}\right]     ~~~~.
} \label{Lp6} \ee
The matrices that appear in (\ref{Lp2}), (\ref{Lp4}), and (\ref{Lp6}) are, respectively,
the ones we have traditionally referred to as the ``chiral," ``tensor," 
and ``vector" supermultiplets L-matrices \cite{G-1} as we initially derived these by
application of a reduction process to the corresponding usual 4D, 
$\cal N$ = 1 supermultiplets.  

For each of the cases, we can next calculate the fermionic holoraumy
matrices.  We find,
\be \eqalign{
{\bm {\Tilde V}}{}_{12}{ {({{\bm S}_{\bm {\cal P}_{1}}}[12] \, 
\cdot \, {\bm {\cal P}_{[1]}} )}} ~&=~  +
{\bm {\Tilde V}}{}_{34}{ {({{\bm S}_{\bm {\cal P}_{1}}}[12] \, 
\cdot \, {\bm {\cal P}_{[1]}} )}} ~=~  + \, {\bm {\a^2}}
   ~~~, \cr
{\bm {\Tilde V}}{}_{13}{ {({{\bm S}_{\bm {\cal P}_{1}}}[12] \, 
\cdot \, {\bm {\cal P}_{[1]}} )}} ~&=~   -
{\bm {\Tilde V}}{}_{24}{ {({{\bm S}_{\bm {\cal P}_{1}}}[12] \, 
\cdot \, {\bm {\cal P}_{[1]}} )}} ~=~     + \, {\bm {\a^3}} ~~~, \cr
{\bm {\Tilde V}}{}_{14}{ {({{\bm S}_{\bm {\cal P}_{1}}}[12] \, 
\cdot \, {\bm {\cal P}_{[1]}} )}} ~&=~    +
{\bm {\Tilde V}}{}_{23}{ {({{\bm S}_{\bm {\cal P}_{1}}}[12] \, 
\cdot \, {\bm {\cal P}_{[1]}} )}} ~=~ + \,  {\bm {\a^1}}   ~~~, \cr
} \ee
for the chiral supermultiplet fermionic holoraumy matrices, 
\be \eqalign{
{\bm {\Tilde V}}{}_{12}{ {({{\bm S}_{\bm {\cal P}_{2}}}[15] \, 
\cdot \, {\bm {\cal P}_{[2]}} )}} ~&=~  -
{\bm {\Tilde V}}{}_{34}{ {({{\bm S}_{\bm {\cal P}_{2}}}[15] \, 
\cdot \, {\bm {\cal P}_{[2]}} )}} ~=~  + \,    {\bm {\b^3}} 
 ~~~, \cr
{\bm {\Tilde V}}{}_{13}{ {({{\bm S}_{\bm {\cal P}_{2}}}[15] \, 
\cdot \, {\bm {\cal P}_{[2]}} )}} ~&=~  +
{\bm {\Tilde V}}{}_{24}{ {({{\bm S}_{\bm {\cal P}_{2}}}[15] \, 
\cdot \, {\bm {\cal P}_{[2]}} )}} ~=~  + \,    {\bm {\b^2}}    ~~~, \cr
{\bm {\Tilde V}}{}_{14}{ {({{\bm S}_{\bm {\cal P}_{2}}}[15] \, 
\cdot \, {\bm {\cal P}_{[2]}} )}} ~&=~  - 
{\bm {\Tilde V}}{}_{23}{ {({{\bm S}_{\bm {\cal P}_{2}}}[15] \, 
\cdot \, {\bm {\cal P}_{[2]}} )}} ~=~   + \,    {\bm {\b^1}}    ~~~, \cr
} \ee
for the tensor supermultiplet fermionic holoraumy matrices, and
\be \eqalign{
{\bm {\Tilde V}}{}_{12}{ {({{\bm S}_{\bm {\cal P}_{3}}}[12] \, 
\cdot \, {\bm {\cal P}_{[3]}} )}} ~&=~ -
{\bm {\Tilde V}}{}_{34}{ {({{\bm S}_{\bm {\cal P}_{3}}}[12] \, 
\cdot \, {\bm {\cal P}_{[3]}} )}} ~=~  - {\bm {\b^3}} 
  ~~~, \cr
{\bm {\Tilde V}}{}_{13}{ {({{\bm S}_{\bm {\cal P}_{3}}}[12] \, 
\cdot \, {\bm {\cal P}_{[3]}} )}} ~&=~   +
{\bm {\Tilde V}}{}_{24}{ {({{\bm S}_{\bm {\cal P}_{3}}}[12] \, 
\cdot \, {\bm {\cal P}_{[3]}} )}} ~=~ + {\bm {\b^2}}   ~~~, \cr
{\bm {\Tilde V}}{}_{14}{ {({{\bm S}_{\bm {\cal P}_{3}}}[12] \, 
\cdot \, {\bm {\cal P}_{[3]}} )}} ~&=~  -
{\bm {\Tilde V}}{}_{23}{ {({{\bm S}_{\bm {\cal P}_{3}}}[12] \, 
\cdot \, {\bm {\cal P}_{[3]}} )}} ~=~   - {\bm {\b^1}}   ~~~, \cr
}   \label{Vsss}
\ee
for the vector supermultiplet fermionic holoraumy matrices.  These results
may be used in the formula that appears in (\ref{Gdgt1}) to replicates 
the 3 $\times$ 3 matrix in the upper left-hand corner of (\ref{Gdgt3b})
when evaluated over the L-matrices in (\ref{Lp2}), (\ref{Lp4}), and
(\ref{Lp6}) respectively and taken in this order.

The results in (\ref{Vsss}) can also be used in conjunction with the three
Fiducial Adinkra formulae in 
(\ref{aLbE}), (\ref{Veq}) and (\ref{Gdgt2}) to show that only some of the 
non-vanishing values of $\ell$'s and $\Tilde {\ell}$'s associated with 
(\ref{Lp2}), (\ref{Lp4}), and (\ref{Lp6}) and these are show in the Table 2 (all
values not shown are equal to zero also).
\be
\begin{array}{|c|c|c|c|c|c|c|c|c|c|c|c|c|} \hline
~~{\rm{Representation}}~~ & \ell_{12}{}^{\Hat 2} & \ell _{13}{}^{\Hat 3} & \ell_{14}{}^{\Hat 1}&
\ell_{23}{}^{\Hat 1}& \ell_{24}{}^{\Hat 3}&\ell _{34}{}^{\Hat 2}
&  {\Tilde \ell}{}_{12}{}^{\Hat 3} & {\Tilde \ell}{}_{13}{}^{\Hat 2}  & {\Tilde  \ell}{}_{14}{}^{\Hat 1} 
&  {\Tilde \ell}{}_{23}{}^{\Hat 1} & {\Tilde \ell}{}_{24}{}^{\Hat 2} & {\Tilde \ell}{} _{34}{}^{\Hat 3}
 \\  \hline
{ {({{\bm S}_{\bm {\cal P}_{1}}}[12] \, \cdot \, {\bm {\cal P}_{[1]}} )}}
&1&1&1&1&-1&1&0&0&0&0&0&0   \\  \hline
{ {({{\bm S}_{\bm {\cal P}_{2}}}[15] \, \cdot \, {\bm {\cal P}_{[2]}} )}}
&0&0&0&0&0&0&1&1&1&-1&1&-1  \\  \hline
{ {({{\bm S}_{\bm {\cal P}_{3}}}[12] \, \cdot \, {\bm {\cal P}_{[3]}} )}} 
&0&0&0&0&0&0&-1&1&-1&1&1 &1 \\ \hline
\end{array}
\ee
$~~~~~$ $~~~~~~~~~~~~~$ {\bf {Table}} {\bf {2:}} A Selection of $\ell$
and $\Tilde \ell$ Values For The Three Fiducial Adinkras  
\newline $~$ \newline
The calculation of the Gadget values between these representations
can be easily carried out by regarding the rows containing either
entries of 1/3, 0, or 1 as components of vectors and calculating their 
dot products followed by multiplying these dot products by 1/6.  This
yields
\be  \eqalign{
{~~~~~~~~~}
{}{\cal G}[{ {({{\bm S}_{\bm {\cal P}_{1}}}[12] \, \cdot \, {\bm {\cal P}_{[1]}} )}}
,  { {({{\bm S}_{\bm {\cal P}_{1}}}[12] \, \cdot \, {\bm {\cal P}_{[1]}} )}}]~=~ 
 \fracm 16 \{~& 
1 \,+\,  1 \,+\,  1 \,+\,  1 \,+\,  (-1)^2 \,+\,  1 \,+\,    0 \,+\,  0  \,+\, 
 \, \cr 
 &0 \,+\,  0 \,+\,  0 \,+\,  0  ~ \, \}
~\,~=~ 1 {~\,~\,} ~~~,  \cr
{\cal G}[{ {({{\bm S}_{\bm {\cal P}_{1}}}[12] \, \cdot \, {\bm {\cal P}_{[1]}} )}}
, { {({{\bm S}_{\bm {\cal P}_{2}}}[15] \, \cdot \, {\bm {\cal P}_{[2]}} )}}  ~=~ 
 \fracm 16 \{~& 
 0 \,+\,  0 \,+\,  0 \,+\,  0 \,+\,  0 \,+\,  0 \,+\,  0 \,+\,  0  \,+\,    \cr
& 0  \,+\,  0 \,+\,  0 \,+\,  0  ~ \, \}
~\,~=~ 0 {~\,~\,}  ~~~,  \cr
{\cal G}[{ {({{\bm S}_{\bm {\cal P}_{2}}}[15] \, \cdot \, {\bm {\cal P}_{[2]}} )}},
{ {({{\bm S}_{\bm {\cal P}_{3}}}[12] \, \cdot \, {\bm {\cal P}_{[3]}} )}}]   ~=~ 
 \fracm 16 \{~& 
 0 \,+\,  0 \,+\,  0 \,+\,  0 \,+\,  0 \,+\,  0  \,+\, (-1) \,+\,  (1) \,+\,  \cr
 &  (-1) \,+\,  (-1) \,+\,  (1) \,+\,  (-1) ~ \, \}
  ~=~   - \fracm 13 {~\,~\,}  ~,
}\ee
as three examples.

The connection to the concept of ``useful inequivalence'' \cite{Neqv} comes 
from reduction from 4D considerations to 2D considerations.
The 4D, $\cal N$ = 1 chiral supermultiplet can be reduced to become 
the 2D,  $\cal N$ = 2 chiral supermultiplet which also yields the matrices 
in (\ref{Lp2}).  The 4D, $\cal N$ = 1 vector supermultiplet can be reduced 
to become the 2D, $\cal N$ = 2 twisted chiral supermultiplet \cite{GHR} 
which, in a similar manner, yields the matrices in (\ref{Lp4}).  When one 
examines the Boolean Factors that appear in (\ref{Lp1}) and (\ref{Lp3}), it
can be seen that both use the same quartet of Boolean Factors, though 
the order is different.  

For the skeptical reader, let us dwell on this matter for a bit.  To our
knowledge, the initiation of the topic of usefully inequivalent supermultiplets
in the physics literature can be traced back to the work in \cite{G-HR}, 
where one of the current authors (SJG) gave the first prescription for 
extracting the 2D, $\cal N$ = 2 ``twisted chiral supermultiplet'' via 
dimensional reduction.  In a follow-up work \cite{GHR}, the significance
and even the name ``twisted chiral supermultiplet,''
was introduced.  But of even greater importance it was shown that
in 2D, $\cal N$ = 2 non-linear sigma models with {\em {both}} chiral 
and twisted chiral supermultiplets the geometry of the associated target 
space manifolds are non-Riemannian and contain torsion.  This latter 
result is impossible within the context of sigma-models constructed {\em 
{solely}} from chiral supermultiplets or only twisted chiral supermultiplets 
alone.  In fact prior to the work of \cite{G-HR}, it was thought that all 
non-linear sigma-models with 2D, $\cal N$ = 2 must possess a
target space geometry that is Riemannian.  So the work in \cite{GHR}
established in the physics literature, the principle of ``useful inequivalence''
between SUSY representations precisely by showing how this 
matters at the level of building actions.

The next portion of the cornerstone for our statements can be seen
through the calculations in Appendix B and Appendix C of the work
seen in \cite{HoLoRmY2}.  This work carefully performed an analysis 
of the earlier work in \cite{G-HR} with regard to its implication for 
adinkras.  It was explicitly rederived that the reduction of 4D, $\cal N$ = 
1 vector supermultiplet in \cite{G-HR} leads the 2D, $\cal N$ = 2 
twisted chiral supermultiplet in the Majorana conventions that we
use to derive adinkras.  Subsequently, a reduction of both the 2D, $\cal 
N$ = 2 chiral and twisted chiral supermultiplets were shown to lead to 
adinkras that lie in different corrals. 

Thus, the distinction of the L-matrices related to the 2D, $\cal N$ = 2 
chiral supermultiplet and those related 2D, $\cal N$ = 2 twisted chiral 
supermultiplet is that the permutation quartets utilized are very different 
as each arises from distinct partitions of $S_4$.  To our knowledge, this insight
into the mathematical origins of the distinctions between 2D, $\cal N$ = 2 
chiral and twisted chiral supermultiplets is a unique observation
arising from the use of adinkras.

We thus have an example to prove that the different
corrals used to construct different supermultiplets are directly related to 
the possibility of  ``useful inequivalence.''  Though we are able to make 
this observation, we do not know the breadth to which this relationship 
is realized.  This is a topic for future study.

\newpage
\section{Expanding By Including Complements For The Coxeter Group 
$\bm {{\rm BC}{}_4}$ Quartets}
\label{s6v}
$~~~~$
At this stage, we have distributed all of the elements of ${\rm BC}{}_4$ 
among the partitions.  This, however, does not saturate the number of 
adinkras that were found by the code enabled search.  There are more 
quartets whose existence is due to ``complement flips.''  In order to define 
``complement flips,''  it is first convenient to define ``complement pairs'' of 
``Boolean Factors'.'   Given a  ``Boolean Factor''  $(\#)_b$, its complement 
is given by $(15 - \#)_b$.  In order to illustrate this, a few examples 
suffice.  The contents of this chapter have appeared previously 
\cite{G&G2}, however, for the sake of the convenience of the reader, we 
include these here.
    
As we have already described, it takes a quartet of  ``Boolean Factor'' 
to construct a representation of the Garden Algebra.  We now make 
an observation.

{\it {If a specified  ``Boolean Factor'' quartet (together with a permutation 
partition) satisfies the Garden Algebra, then replacing any of the  
``Boolean Factors'' by their complements leads to another  ``Boolean Factor'' 
quartet that satisfies the Garden Algebra.}}

Let us illustrate the import of this by examining the  ``Boolean 
Factor'' quartet
$\{(0)_b  , (6)_b , (12)_b  , (10)_b  \}$ and all of its  ``Boolean 
Factor'' quartet
complements shown below.
\be\label{e:CM1Anton}\eqalign{
  &\{(0)_b  , (6)_b , (12)_b  , (10)_b  \} :\cr
  &\{(0)_b  , (6)_b , (12)_b  , (5)_b  \} , \{(0)_b  , (6)_b , (3)_b  , (5)_b  \} , 
  \{(0)_b  , (6)_b , (3)_b  , (10)_b  \}~~ , \cr
  &\{(0)_b  , (9)_b , (12)_b  , (5)_b  \} , \{(0)_b  , (9)_b , (3)_b  , (5)_b  \} , 
  \{(0)_b  , (9)_b , (3)_b  , (10)_b  \} ~~, \cr
  & \{(0)_b  , (9)_b , (12)_b  , (10)_b  \} , \{(15)_b  , (9)_b , (12)_b  , (5)_b  \} , 
  \{(15)_b  , (9)_b , (3)_b  , (5)_b  \} ~~, \cr
  & \{(15)_b  , (9)_b , (3)_b  , (10)_b  \} , \{(15)_b  , (9)_b , (12)_b  , (10)_b  \} ,  
  \{(15)_b  , (9)_b , (12)_b  , (5)_b  \} ~~,\cr 
  & \{(15)_b  , (6)_b , (3)_b  , (5)_b  \} , 
  \{(15)_b  , (6)_b , (3)_b  , (10)_b  \} , \{(15)_b  , (6)_b , (12)_b  , (10)_b  \} ~~~. {~}
}\ee    
On the first line of this expression we have the specified  ``Boolean Factor'' 
quartet and under this we list all of its  ``Boolean Factor'' quartet complements.

For the first listed complement, only the fourth  ``Boolean Factor'' entry or the ``fourth 
color'' was replaced by its complement.  This is what is meant by a single ``color flip.'' 
For the second listed complement, the third and fourth  ``Boolean Factor'' entries 
or the ``third color'' and ``fourth color'' were replaced by their complements.  This 
is what is meant by ``flipping' two colors.  For the third listed complement, only the 
third  ``Boolean Factor'' entry or the ``third color'' was replaced by its complement.  This 
is again a ``flipping' of one color.

Concentrating now once more only on the ``Boolean Factor''  quartet $\{(0)_b  , (6)_b , 
(12)_b  , (10)_b  \}$, we can see among the complements one is related to
it in a special manner.  The complement ``Boolean Factor''  quartet $\{(15)_b  , (9)_b , 
(3)_b  , (5)_b  \} $ has all four of its colors flipped with respect to the initial 
``Boolean Factor''  quartet.  Thus we call $\{(0)_b  , (6)_b , (12)_b  , (10)_b  \}$
and $\{(15)_b  , (9)_b , (3)_b  , (5)_b  \} $ antipodal ``Boolean Factor''  quartet pairs.
Among the sixteen  ``Boolean Factor''  quartets shown in (\ref{e:CM1Anton}) eight
such pairs occur.  Thus, for each value of $\a$, one must specify which of the
complements is used to construct and L-matrix.  For this purpose, we use an
index $\b_a$ which takes on eight values.
    
Given two quartets $ (\,{\rm L}^{(\cal R)}_\rI\,)_i{}^\hj$ and $ (\,{\rm L}^{({\cal 
R}^{\prime})}_\rI\,)_i{}^\hj$ where the elements in the second set are related 
to the first by replacing at least one ``Boolean Factor''  by its complement, there 
exist a 4 $\times$ 4 ``Boolean Factor''  matrix denoted by $ \left[ {\cal A}({\cal 
R}, \, {\cal R}^{\prime}) \right] {}_\rI  \,  {}^\rJ $ which acts on the color space
of the links (i. e. the indices of the $I$, $J$, ... type) such that
\be { \eqalign{
 (\,{\rm L}^{(\cal R)}_\rI\,)_i{}^\hj ~=~  \left[ {\cal A}({\cal R}, \, {\cal R}^{\prime}) \right]
 {}_\rI  \,  {}^\rJ (\,{\rm L}^{({\cal R}^{\prime})}_\rJ\, )_i{}^\hj  ~~~,
} }  \label{Ant1}
\ee
and as a consequence of (\ref{Ant1}) we see
 \be { \eqalign{
 (\,{\rm R}^{(\cal R)}_\rI\,)_\hj {}^i ~=~  \left[ {\cal A}({\cal R}, \, {\cal 
 R}^{\prime}) \right]{}_\rI  \,  {}^\rJ (\,{\rm R}^{({\cal R}^{\prime})}_\rJ\, )_\hj {}^i  ~~~.
} }  \label{Ant2}
\ee
It should also be noted that the definition of the complements imply that the 
representations ${\cal R}^{\prime}$ and ${\cal R}$ that appear in (\ref{Ant1}) 
and (\ref{Ant2}) must belong to the same partition sector shown in diagram 
shown in Fig.\ 2.  The equations in (\ref{GarDVs}), (\ref{Ant1}) and
(\ref{Ant2}) imply 
\be   \eqalign{
{\ell}_{{\rm I}{\rm J}}^{\,({\cal R}) \, {\hat a}} &=~  \left[ {\cal A}({\cal R}, \, {\cal 
R}^{\prime}) \right] {}_\rI  \,  {}^\rK  \,  \left[ {\cal A}({\cal R}, \, {\cal R}^{\prime}) 
\right] {}_\rJ  \,  {}^\rL  \,  {\ell}_{{\rm K}{\rm L}}^{\, ({\cal R}^{\prime}) \,{\hat  a}}  
\cr
{\Tilde \ell}_{{\rm I}{\rm J}}^{\,({\cal R}) \, {\hat a}} &=~  \left[ {\cal A}({\cal R}, \, {\cal 
R}^{\prime}) \right] {}_\rI  \,  {}^\rK  \,  \left[ {\cal A}({\cal R}, \, {\cal R}^{\prime}) 
\right] {}_\rJ  \,  {}^\rL  \,  {\Tilde \ell}_{{\rm K}{\rm L}}^{\, ({\cal R}^{\prime}) \,{\hat  a}}  
 ~~~. }   \ee

Let us observe the distinction between the Boolean quartet factors that appear in
(B.1) - (B.6) and all of their complements
is not intrinsic, but is an artifact of the choices made to discuss this aspect
of the construction.  It may be possible to provide a more symmetrical treatment
of the (B.1) - (B.6) and all of their complements.
However, we have not been able to create such a formulation.
    
Now the meaning of the ``representation label,'' first written in (\ref{GarDNAlg2}), can be 
explicitly discussed.  Each value of $\cal R$ corresponds to a specification of the pairs of 
indices ($\L$, $\a | \b_a$).  This implies there are 6 $\times$ 16 $\times$ 8 =  6 $\times$ 
128  =  768 quartets which satisfy the Garden Algebra conditions.  Notice that 1,536/762 
= 2 which shows the algorithmic counting did not remove antipodal ``Boolean Factor''  
quartets.

Let us observe that the $\L$ in the paragraph denotes to which partition the
element resides, the $\a$ label denotes which of the Boolean Factors identified in the
work of \cite{permutadnk}, and finally $\a | \b_a$
denotes the complementary Boolean
Factors listed in Appendix B.  The order in which the Boolean Factor appears
within each quartet matters and the $\a | \b_a$ label indicates the order from
the results listed in this appendix.

Finally, let us note all discussions in this chapter are totally disconnected from 
considerations of four dimensional supersymmetry representations.  We have 
simply enunciated the rich mathematical structure imposed on the Coxeter Group 
${\rm BC}{}_4$ when analyzed through the lens of the ``Garden Algebra''
${\cal GR}$(4,4).

 \newpage
\section{Expanding to Ordered Coxeter Group $\bm {BC_4}$ Quartets}
\label{s5v}
$~~~~$
The discussion in the preceding chapters (and all our previous works) only considered
the quartets without consideration of the order in which the permutations appeared
with the quartets.  In this subsequent discussion we will explore which feature are
modified when consideration of ordered quartets is undertaken.   There are more 
quartets whose existence is due to ``color flips.''  In order to define ``color flips,'' 
let us make note of the previous assumptions used through the analysis so far.

However, what these multiplication tables imply is the existence of thirty permutation
operators ${\bm {\Pi}}{}_{1}$, $\dots$, ${\bm {\Pi}}{}_{30}$ defined to act on the
quartet entries such that
\be  \eqalign{
(12) \, \bm {\{ {\cal V}_{(4)} \}} ~&=~  {\bm {\Pi}}{}_{1}{\bm [} (34) \, \bm {\{  {\cal 
V}_{(4)} \}} {\bm ]} ~~\,~=~  {\bm {\Pi}}{}_{2}{\bm [}(1324) \, \bm {\{  {\cal V}_{(4)} 
\}} {\bm ]}~=~  {\bm {\Pi}}{}_{3}{\bm [}(1423) \, \bm {\{ {\cal V}_{(4)} \}}  {\bm ]} 
~~~~,  \cr
(13) \, \bm {\{ {\cal V}_{(4)} \}} ~&=~  {\bm {\Pi}}{}_{4}{\bm [}(24) \, \bm {\{ {\cal 
V}_{(4)} \}}  ~~\,~~=~  {\bm {\Pi}}{}_{5}{\bm [}(1234) \, \bm {\{  {\cal V}_{(4)} \}} 
{\bm ]}~=~ {\bm {\Pi}}{}_{6}{\bm [}(1432) \, \bm {\{ {\cal V}_{(4)} \}}  {\bm ]} 
~~~~,  \cr
(14) \, \bm {\{ {\cal V}_{(4)} \}} ~&=~  {\bm {\Pi}}{}_{7}{\bm [}(23) \, \bm {\{  {\cal 
V}_{(4)} \}} {\bm ]} ~~\,~=~  {\bm {\Pi}}{}_{8}{\bm [}(1243) \, \bm {\{  {\cal V}_{(4)} 
\}} {\bm ]}~=~ {\bm {\Pi}}{}_{9}{\bm [}(1342) \, \bm {\{ {\cal V}_{(4)} \}}  {\bm ]} 
~~~~,  \cr
(123) \, \bm {\{ {\cal V}_{(4)} \}} ~&=~  {\bm {\Pi}}{}_{10}{\bm [}(134) \, \bm {\{  
{\cal V}_{(4)} \}} {\bm ]}~=~  {\bm {\Pi}}{}_{11}{\bm [}(142) \, \bm {\{ {\cal V}_{(4)} \}}
~=~  {\bm {\Pi}}{}_{12}{\bm [}(243) \, \bm {\{ {\cal V}_{(4)} \}} {\bm ]} 
~~~\,~~,  \cr
(124) \, \bm {\{ {\cal V}_{(4)} \}} ~&=~  {\bm {\Pi}}{}_{13}{\bm [}(132) \, \bm {\{ 
{\cal V}_{(4)} \}} {\bm ]}~=~  {\bm {\Pi}}{}_{14}{\bm [}(143) \, \bm {\{ {\cal V}_{
(4)} \}} ~=~  {\bm {\Pi}}{}_{15}{\bm [}(234) \, \bm {\{ {\cal V}_{(4)} \}}  {\bm ]}
~~~\,~~.
} \label{pVLsets}
\ee

\be  \eqalign{ {~~~~~~~~~~~~~~}
\bm {\{ {\cal V}_{(4)} \}} \, (12) ~&=~  {\bm {\Pi}}{}_{16}{\bm [} \bm {\{ {\cal V}_{
(4)} \}} \, (34)  {\bm ]}~=~  {\bm {\Pi}}{}_{17}{\bm [} \bm {\{ {\cal V}_{(4)} \}} \,
(1324) {\bm ]} \,  ~=~  {\bm {\Pi}}{}_{18}{\bm [} \bm {\{ {\cal V}_{(4)} \}} \, (1423)  
{\bm ]}  ~~~~, \cr
\bm {\{ {\cal V}_{(4)} \}} \, (13) ~&=~  {\bm {\Pi}}{}_{19}{\bm [}\bm {\{ {\cal V}_{
(4)} \}} \,(24)  {\bm ]}~=~  {\bm {\Pi}}{}_{20}{\bm [} \bm {\{ {\cal V}_{(4)} \}} \,
(1234) {\bm ]} ~=~  {\bm {\Pi}}{}_{21}{\bm [}\bm {\{ {\cal V}_{(4)} \}} \, (1432)   
{\bm ]}  ~~~~,  \cr
\bm {\{ {\cal V}_{(4)} \}} \, (14) ~&=~  {\bm {\Pi}}{}_{22}{\bm [} \bm {\{ {\cal V}_{
(4)} \}} \, (23) {\bm ]} ~=~  {\bm {\Pi}}{}_{23}{\bm [} \bm {\{ {\cal V}_{(4)} \}} \, 
(1243)  {\bm ]}~=~  {\bm {\Pi}}{}_{23}{\bm [} \bm {\{ {\cal V}_{(4)} \}} \, (1342)  
{\bm ]}~~~~,  \cr
\bm {\{ {\cal V}_{(4)} \}} \, (123)  {\bm ]}~&=~  {\bm {\Pi}}{}_{25}{\bm [} \bm {\{ {
\cal V}_{(4)} \}} \, (134) {\bm ]} ~=~  {\bm {\Pi}}{}_{26}{\bm [} \bm {\{ {\cal V}_{
(4)} \}} \, (142)  {\bm ]}~=~  {\bm {\Pi}}{}_{27}{\bm [} \bm {\{ {\cal V}_{(4)} \}} \, 
(243)  {\bm ]}~~~\,~~,  \cr
\bm {\{ {\cal V}_{(4)} \}} \, (124) ~&=~  {\bm {\Pi}}{}_{28}{\bm [} \bm {\{ {\cal V}_{
(4)} \}} \, (132)  {\bm ]}~=~  {\bm {\Pi}}{}_{29}{\bm [} \bm {\{ {\cal V}_{(4)} \}} 
\, (143)  {\bm ]} ~=~  {\bm {\Pi}}{}_{30}{\bm [} \bm {\{ {\cal V}_{(4)} \}} \, (234)  
{\bm ]} ~~~\,~~.
} \label{pVRsets}
\ee
It is instructive to illustrate the action of the ${\bm {\Pi}}$-operators.  As they are 
permutations also, cycle notation can be used to denote them.  The ${\bm {\Pi}
}$-operator ${\bm{(34)_q}} $ can be apply to the quartet set $ \bm {\{ {\cal V}_{
(4)} \}} $ where we find
\be \eqalign{
{\bm{(34)_q}} \, \bm {\{ {\cal V}_{(4)} \}} ~&=~ {\bm{(34)_q}} \, {\bm \{} (), ~ (12)(34), 
~ (13)(24), ~ (14)(23) {\bm \} }    \cr
~&=~  {\bm \{} (), ~ (12)(34), ~ (14)(23), ~ (13)(24) {\bm \} }   ~~~.
} \label{Pi} \ee
As there is no need to find the explicit forms of all the ${\bm {\Pi}}$-operators, we
dispense with further discussion on this.  The important point about these
equations is that they show even if ordered quartets of elements are considered,
the notion of the six\footnote{We continue to include the corral defined by
$\bm {\{ {\cal V}_{(4)} \}}$ itself.} ``corrals'' continues to have a mathematically 
well defined meaning.

In light of the results for the Gadget matrix elements reported in chapter two,
we believe working out one explicit example will demonstrate both the use
of complements as well as the use of the ${\bm {\Pi}}$-operators\footnote{We here
acknowledge conversations with K.\ Iga who first gave arguments on the 
importance of this case.}.  Let us arbitrarily pick the ${\bm \{} {\bm {\cal P}}{}_{
[3]}  {\bm \} } $ corral as well as the fifteenth of the appropriate Boolean Factors 
in $(B.3)$ so that we have
\be  \eqalign{
{~~}&  ~~~~ {\bm {\rm L}}{}_{1} ~~~\,\,~~ {\bm {\rm L}}{}_{2}
 ~~~\,\,~~ {\bm {\rm L}}{}_{3}  ~~~\,\,~~ {\bm {\rm L}}{}_{4}  \cr
{\bm \{} {\bm {\cal P}}{}_{[3]}  {\bm \} } ~=~&{\bm \{}  (1243), ~ (23 ),~ (14),  ~  (1342) 
{\bm \} }   ~~~, 
} \label{P-3}
\ee
and as well
\be
\eqalign{
{~~~~~~~~~~~~~~}
{{\bm S}_{\bm {\cal P}}}_3[15] ~=~ 
 \{ (14)_b , (2)_b , (8)_b , (4)_b \}   ~~. ~~~~~~~~~~
 }  \label{S-14}
 \ee
Thus we are led to
\be \eqalign{
 {\bm {\rm L}}{ {({{\bm S}_{\bm {\cal P}_{3}}}[15] \, \cdot \, {\bm {\cal P}_{[3]}} )}} ~&=~ 
( {\bm \{} (14)_b , (2)_b , (8)_b , (4)_b {\bm  \} }\, {\bm \cdot}  \, {\bm \{}   \,
 (1243), ~ (23 ), ~ (14), ~ (1342) \, {\bm \} } )   \cr
 ~&=~ {\bm \{}   \, (14)_b (1243), ~ (2)_b (23 ), ~ (8)_b (14), ~ (4)_b
 (1342) \, {\bm \} }  ~~~,  
} \label{Lp5CP1} \ee
and we can apply the ${\bm {\Pi}}$-operator that interchanges the final two entries of
the quartet.  We can also use cycle-notation to indicate this to obtain
\be \eqalign{
 {\bm {\rm L}}[{{\bm{(34)_q}} {({{\bm S}_{\bm {\cal P}_{3}}}[15] \, \cdot \, {\bm {\cal P}_{
 [3]}} )}}] ~&=~ {\bm \{}   \, (14)_b (1243), ~ (2)_b (23 ), ~ (4)_b (1342), ~ (8)_b (14) 
 \, {\bm \} }  ~~~.
} \label{Lp5CP2} \ee
Finally we apply the complement flip $(2 \leftrightarrow 13)_b$ that switches $(2)_b$ 
with $(13)_b$ to obtain
\be \eqalign{ {~~~~}
{\bm {\rm L}}[  (2 \leftrightarrow 13)_b{{\bm{(34)_q}}  {({{\bm S}_{\bm {\cal P}_{3}}}[15] 
\, \cdot \, {\bm {\cal P}_{[3]}} )}}] ~&=~ {\bm \{}   \, (14)_b (1243), ~ (13)_b (23 ), ~ 
(4)_b (1342), ~ (8)_b (14)  \, {\bm \} }  ~~~,
} \label{Lp5CP3} \ee
and as a way to write a more compact notation we introduce
with $(13)_b$ to obtain
\be \eqalign{
{\bm {\Hat {\rm L}}}
[ { {({{\bm S}_{\bm {\cal P}_{3}}}[15] 
\, \cdot \, {\bm {\cal P}_{[3]}} )}}] ~&\equiv~ 
{\bm { {\rm L}}}
[  (2 \leftrightarrow 13)_b {{{\bm{(34)_q}} } {({{\bm S}_{\bm {\cal P}_{3}}}[15] 
\, \cdot \, {\bm {\cal P}_{[3]}} )}}]  ~~~,
} \label{Lp5CP4} \ee
and this yields 
\be \eqalign{
{\bm {\Hat {\rm L}}}{}_{1}{ {({{\bm S}_{\bm {\cal P}_{3
}}}[15] \, \cdot \, {\bm {\cal P}_{[3]}} )}} ~&=~   \left[\begin{array}{cccc}
0&1&0&0\\ 0&0&0&-1\\ -1&0&0&0\\ 0&0&-1&0
\end{array}\right]   ~~~, ~~~
{\bm {\Hat {\rm L}}}{}_{2}{ {({{\bm S}_{\bm {\cal P}_{3
}}}[15] \, \cdot \, {\bm {\cal P}_{[3]}} )}} ~=~   \left[\begin{array}{cccc}
-1&0&0&0\\ 0&0&1&0\\ 0&-1&0&0\\ 0&0&0&-1
\end{array}\right]  ~~~, \cr
{\bm {\Hat {\rm L}}}{}_{3}{ {({{\bm S}_{\bm {\cal P}_{3
}}}[15] \, \cdot \, {\bm {\cal P}_{[3]}} )}}  ~&=~
\left[\begin{array}{cccc}
~0&~0&~1&~0\\ 1&0&0&0\\ 0&0&0&-1\\ 0&1&0&0 
\end{array}\right]     ~~~~~~~~,~~~  
{\bm {\Hat {\rm L}}}{}_{4}{ {({{\bm S}_{\bm {\cal P}_{3
}}}[15] \, \cdot \, {\bm {\cal P}_{[3]}} )}} ~=~    
\left[\begin{array}{cccc}
0&0&0&1\\ 0&1&0&0\\ 0&0&1&0\\ -1&0&0&0
\end{array}\right]    
~~~~.
} \label{Lp6CP5} \ee
These lead to fermonic holoraumy matrices based on the ${\bm {\Hat {\rm L}}}$
function to be given by
\be \eqalign{
{\bm {\Tilde V}}{}_{12}[ {\bm {\Hat {\rm L}}}{ {({{\bm S}_{\bm {\cal P}_{1}}}[15] \, 
\cdot \, {\bm {\cal P}_{[1]}} )}}] ~&=~  +
{\bm {\Tilde V}}{}_{34}[ {\bm {\Hat {\rm L}}} { {({{\bm S}_{\bm {\cal P}_{1}}}[15] \, 
\cdot \, {\bm {\cal P}_{[1]}} )}}]  ~=~  {\bm {\a^2}}
   ~~~, \cr
{\bm {\Tilde V}}{}_{13}[ {\bm {\Hat {\rm L}}} { {({{\bm S}_{\bm {\cal P}_{1}}}[15] \, 
\cdot \, {\bm {\cal P}_{[1]}} )}}]  ~&=~   -
{\bm {\Tilde V}}{}_{24}[ {\bm {\Hat {\rm L}}} { {({{\bm S}_{\bm {\cal P}_{1}}}[15] \, 
\cdot \, {\bm {\cal P}_{[1]}} )}}] ~=~     {\bm {\a^1}} ~~~, \cr
{\bm {\Tilde V}}{}_{14}[ {\bm {\Hat {\rm L}}} { {({{\bm S}_{\bm {\cal P}_{1}}}[15] \, 
\cdot \, {\bm {\cal P}_{[1]}} )}} ] ~&=~    +
{\bm {\Tilde V}}{}_{23}[ {\bm {\Hat {\rm L}}} { {({{\bm S}_{\bm {\cal P}_{1}}}[15] \, 
\cdot \, {\bm {\cal P}_{[1]}} )}} ] ~=~ - \, {\bm {\a^3}}   ~~~. \cr
}  \label{Lp6CP6} \ee
Using the formula in (\ref{Gdgt1}) and picking the two representations to be $ ({ {\cal R}})$ =
${ {({{\bm S}_{\bm {\cal P}_{1}}}[12] \, \cdot \, {\bm {\cal P}_{[1]}} )}}$ and $( {\cal R}^{
\prime})$ =  $ {\bm {\Hat {\rm L}}} { {({{\bm S}_{\bm {\cal P}_{1}}}[15] \, \cdot \, {\bm 
{\cal P}_{[3]}} )}}$ leads to
\be  \eqalign{
{~~~~~~~~~~}
{{\cal G}} [  ({ {\cal R}}) , ( {\cal R}^{\prime}) ] ~&=~ \fracm 1{24}  \, 
{\rm {Tr}} \, {\bm {\Big[}} \, 
{\bm {\Tilde V}}{}_{12}{ {({{\bm S}_{\bm {\cal P}_{1}}}[12] \, \cdot \, {\bm {\cal P}_{[1]}} )}} ~
{\bm {\Tilde V}}{}_{12}[ {\bm {\Hat {\rm L}}}{ {({{\bm S}_{\bm {\cal P}_{1}}}[15] \, \cdot \, {\bm {\cal P}_{[1]}} )}}]   \cr
~&~ {~~~~~~\,~~~~~+~~}
{\bm {\Tilde V}}{}_{34}{ {({{\bm S}_{\bm {\cal P}_{1}}}[12] \, \cdot \, {\bm {\cal P}_{[1]}} )}} ~
{\bm {\Tilde V}}{}_{34}[ {\bm {\Hat {\rm L}}} { {({{\bm S}_{\bm {\cal P}_{1}}}[15] \, \cdot \, {\bm {\cal P}_{[1]}} )}}]   \cr
~&~ {~~~~~~\,~~~~~+~~}
{\bm {\Tilde V}}{}_{13}{ {({{\bm S}_{\bm {\cal P}_{1}}}[12] \, \cdot \, {\bm {\cal P}_{[1]}} )}} ~
{\bm {\Tilde V}}{}_{13}[ {\bm {\Hat {\rm L}}} { {({{\bm S}_{\bm {\cal P}_{1}}}[15] \, \cdot \, {\bm {\cal P}_{[1]}} )}}]   \cr
~&~ {~~~~~~\,~~~~~+~~}
{\bm {\Tilde V}}{}_{24}{ {({{\bm S}_{\bm {\cal P}_{1}}}[12] \, \cdot \, {\bm {\cal P}_{[1]}} )}} ~
{\bm {\Tilde V}}{}_{24}[ {\bm {\Hat {\rm L}}} { {({{\bm S}_{\bm {\cal P}_{1}}}[15] \, \cdot \, {\bm {\cal P}_{[1]}} )}}]  \cr
~&~ {~~~~~~\,~~~~~+~~}
{\bm {\Tilde V}}{}_{14}{ {({{\bm S}_{\bm {\cal P}_{1}}}[12] \, \cdot \, {\bm {\cal P}_{[1]}} )}}  ~
{\bm {\Tilde V}}{}_{14}[ {\bm {\Hat {\rm L}}} { {({{\bm S}_{\bm {\cal P}_{1}}}[15] \, \cdot \, {\bm {\cal P}_{[1]}} )}} ]  \cr
~&~ {~~~~~~\,~~~~~+~~}
{\bm {\Tilde V}}{}_{23}{ {({{\bm S}_{\bm {\cal P}_{1}}}[12] \, \cdot \, {\bm {\cal P}_{[1]}} )}} ~
{\bm {\Tilde V}}{}_{23}[ {\bm {\Hat {\rm L}}} { {({{\bm S}_{\bm {\cal P}_{1}}}[15] \, \cdot \, {\bm {\cal P}_{[1]}} )}} ] \, {\bm {\Big ]}}  \cr
~&=~ \fracm 1{24}  \, {\rm {Tr}} \, {\bm {\Big[}} \,  {\bm {\a^2}} \,  {\bm {\a^2}} ~+~ {\bm {\a^2}} \,  
{\bm {\a^2}} ~+~ \,  {\bm {\a^3}} \,  {\bm {\a^1}}  ~+~ \,  {\bm {\a^3}} \,  {\bm {\a^1}} 
 ~-~ \,  {\bm {\a^1}} \,  {\bm {\a^3}}  ~-~ \,  {\bm {\a^1}} \,  {\bm {\a^3}} \, 
 {\bm {\Big ]}}  \cr
~&=~ \fracm 1{24}   \, {\bm {\big[}} \,  4~+~ 4 ~+~ 0 ~+~ 0~-~ 0 ~-~ 0
 \, {\bm {\big ]}} ~=~ \fracm 1{3}   ~~~,
}     \label{GdgtVV}
\ee
which is one of the allowed values over the 36,864 $\times$ 36,864 matrix elements.  One can trace
back through the calculation that it is the insertion of the ${\bm {\Pi}}$-operator in
the quartet that is responsible for the appearances of the traces ${\rm {Tr}} \, {\bm 
{\big[}} {\bm {\a^1}} \,  {\bm {\a^3}} {\bm {\big]}}$ on the penultimate line in this calculation.

\newpage
\section{A Counting Intermezzo}
\label{s4vuU}
$~~~~$
It is useful here to step back and do a bit of counting to clearly see the magnitude of the
task of calculation every matrix element that arises (\ref{Gdgt1}) or (\ref{Gdgt2}) from all
possible choices of the adinkra representations $({\cal R})$ and $({\cal R}^{\prime})$.

${~~~~~~~~~~~~~~}$ ${~~~~~~~~~~~~~~}$ {\underline {BC}}${}_4$ {\underline {Contains}} {\underline 
{384}} {\underline {Elements}}

${~~~~~~~~~~~~~~}$ ${~~~~~~~~~~~~~~}$ leads to 96 unordered quartets

${~~~~~~~~~~~~~~}$ ${~~~~~~~~~~~~~~}$ leads to 2,304 ordered quartets

${~~~~~~~~~~~~~~}$ ${~~~~~~~~~~~~~~}$ {\underline {Diadem}} {\underline {Se}p{\underline {aration}
{\underline {Divides}} {\underline {B}}y {\underline {Six}}

${~~~~~~~~~~~~~~}$ ${~~~~~~~~~~~~~~}$ leads to 16 unordered quartets/corral

${~~~~~~~~~~~~~~}$ ${~~~~~~~~~~~~~~}$ leads to 384 ordered quartets/corral

${~~~~~~~~~~~~~~}$ ${~~~~~~~~~~~~~~}$ {\underline {Complements}}
{\underline {Multi}}p{\underline {l}}y 
{\underline {B}}y {\underline {Sixteen}}

${~~~~~~~~~~~~~~}$ ${~~~~~~~~~~~~~~}$ leads to 256 unordered quartets/corral

${~~~~~~~~~~~~~~}$ ${~~~~~~~~~~~~~~}$ leads to 6,144 ordered quartets/corral

${~~~~~~~~~~~~~~}$ ${~~~~~~~~~~~~~~}$ {\underline {Total}} {\underline {Summin}}g 
{\underline {Over}} {\underline {All}} {\underline {Corrals}}

${~~~~~~~~~~~~~~}$ ${~~~~~~~~~~~~~~}$ 1,536 total unordered quartets

${~~~~~~~~~~~~~~}$ ${~~~~~~~~~~~~~~}$ 36,864 total ordered quartets

${~~~~~~~~~~~~~~}$ ${~~~~~~~~~~~~}$ ========================

${~~~~~~~~~~~~~~}$ ${~~~~~~~~~~~~~~}$ 36,864  x 36,864 = 1,358,954,496

${~~~~~~~~~~~~~~}$ ${~~~~~~~~~~~~~~}$ Sym[36,864  x 36,864] = 679,495,680

\noindent
So the task of calculating (\ref{Gdgt1}) or (\ref{Gdgt2}) corresponds to 
calculating 1,358,954,496 matrix elements or given the fact that the
Gadget is symmetrical 679,495,680 matrix elements.  

Before the era
of modern computing technology, to carry out such a multitude of
calculations was simply impossible.  In the next chapter, the codes
that were developed to tackle this problem are described.   The
approach was to develop four different codes, using different languages,
to attack the problem over the ``small ${\rm BC}{}_4$ library.''  As these
were each developed independently, we relied on the consensus of
final results to de-bug any errors that may have occurred.  This was
successfully done to carry out the evaluation of the 9,216 matrix
elements of the 96 $\times$ 96 Gadget values over the  ``small ${\rm BC}{}_4$ library''
of representations.  These results were graphically shown in Fig.\ 3
and are analytically reported in the tables of Appendix D. 

\newpage
\section{The Codes}
\label{codes}

\subsection{Program, libraries.py}

The Program, libraries.py, First uses the numpy library to create 2 5D nested arrays 
\texttt{lib} and \texttt{libtil} to store the values of $\ell _{nm}^{(i)j}$ and $\tilde{\ell} _{
nm}^{(i)j}$ respectively.

When the Latex document containing the libraries is properly formatted(in the style 
of the file provided), the program will iterate though all the libraries ${\cal{P}}_k$ in 
the file and store all  $\ell$ and $\tilde{\ell}$ values into \  \texttt{lib[k][i][n][m][j]} and 
\texttt{libtil[k][i][n][m][j]}. 
	
The main loop then iterates though all $\ell$ values in all dictionaries and calculates 
all the gadgets, $\cal G[R,R']$, using the function \texttt{gadget(r,rprime,lr,lrprime)} it 
takes as input the fallowing parameters. \texttt{r}: the $\cal R$ value. \texttt{rprime}: 
the $\cal R'$ value. \texttt{lr}: the library number for the  $\cal R'$ value(k) and 
\texttt{lrprime} the library number for the  $\cal R'$ value and uses them to compute 
the gadget.
	
After finishing the gadget calculations other contributors completed other pieces 
of software using a different format of for displacing the gadget, using r and rprime 
values ranging all the way up to 96 instead of ranging to 16 and being associated 
a dictionary number. As a sanity check it became necessary to confirm all programs 
produced the same gadget values. This was done in a section of code commented 
out prior to publication in this program. In order to allow other contributors to do 
the same the method of printing the gadget values to a file was adapted to list 
gadgets calculated with \texttt{r} and \texttt{rprime} values ranging up to 96. In 
order to accomplish this with the existing \texttt{gadget} function the following 
loop was utilized in the main function.
	
	\begin{scriptsize}
	\begin{verbatim}
		hori=1
		verti=1
		for vertc in range(1,97):
		    for horc in range(1,97):
			    f.write(str(gadget(vertc-(verti-1)*16,horc-(hori-1)*16,verti,hori))+",")
			    if(horc%16==0):
			        hori+=1
		    hori=1
		    if(vertc%16==0):
		        verti+=1
		    f.write("\n")
	\end{verbatim}
	\end{scriptsize}
The values of the gadget are then written to the file object f which will write to 
a csv spreadsheet file.

Additional Documentation is provided in the comments embedded into libraries.py. 
	
\subsection{C++ Code to Calculate Gadget Values given $\ell$/ $\Tilde {\ell}$ Coefficients} 

This code takes an input file `data.txt' containing the $\ell$ and $\Tilde \ell$ coefficients
which the gadget may be calculated with by summing over the coefficients related to a
specific pair (($\cal R$), (${\cal R}{}^{\prime}$)) of adinkras that are related by the gadget 
value. The code will then output a 96$\times$96 array of gadget values into `results.txt.'
In order to do this calculation, the code makes use of the fact that many of the $\ell$ and 
$\Tilde {\ell}$ values are 0 and has a look up function to either return 0 or the $\ell$/ $\Tilde {\ell}$
value that is provided in the `data.txt'  input file.

	\begin{scriptsize}
	\begin{verbatim}
	         #include <ios t ream>
         #include <f s t ream>
         #include <s t r ing>
	
using namespace std;

int neg(int n) {if(n == 0) return 1; else return 0;} //"negates" integer values as if they were booleans

int lookup(int r, int l, int a, int arr[96][7])
    //Returns the desired l-value to be used in the gadget calculation
{
    if(l == 0) //L12
    {
        if (a != 2) return 0;
        else return neg(arr[r][0])*arr[r][1];
    }
    else if (l == 1) //L13
    {
        if(a != 3) return 0;
        else return neg(arr[r][0])*arr[r][2];
    }
    else if(l == 2) //L14
    {
        if (a != 1) return 0;
        else return neg(arr[r][0])*arr[r][3];
    }
    else if(l == 3) //L23
    {
        if (a != 1) return 0;
        else return neg(arr[r][0])*arr[r][4];
    }
    else if(l == 4) //L24
    {
        if (a != 3) return 0;
        else return neg(arr[r][0])*arr[r][5];
    }
    else if(l == 5) //L34
    {
        if(a != 2) return 0;
        else return neg(arr[r][0])*arr[r][6];
    }
    else if(l == 6) //~L12
    {
        if(a != 3) return 0;
        else return arr[r][0]*arr[r][1];
    }
    else if(l == 7) //~L13
    {
        if(a != 2) return 0;
        else return arr[r][0]*arr[r][2];
    }
    else if(l == 8) //~L14
    {
        if(a != 1) return 0;
        else return arr[r][0]*arr[r][3];
    }
    else if(l == 9) //~L23
    {
        if(a != 1) return 0;
        else return arr[r][0]*arr[r][4];
    }
    else if(l == 10) //~L24
    {
        if(a != 2) return 0;
        else return arr[r][0]*arr[r][5];
    }
    else if(l == 11) //~L34
    {
        if(a != 3) return 0;
        else return arr[r][0]*arr[r][6];
    }

}

double g(int n, int m, int arr[96][7])
{
    // This method will return the correct value for the G function with r = n, and r' = m
    double result = 0;
    for(int a = 1; a <= 3; a++) for(int l = 0; l < 12; l++) result += lookup(n, l, a, arr)*lookup(m, l, a, arr);
    return result/6;
}

void read(std::fstream& data, int arr[96][7])
{
    // This method will read data into the arr in the proper format
    for(int i = 0; i < 96; i++) for(int j = 0; j < 7; j++) data >> arr[i][j];
}

int main()
{
    // This method will calculate the gadget values and put them in "results.txt"
    string line;
    int n, m;
    int arr[96][7];
    std::fstream data("data.txt");
    std::fstream results("results.txt");
    if(data.is_open())
    {
        read(data, arr);
        data.close();
    } else cout << "error: unable to open data file";
    if(results.is_open())
    {
        for(n=0;n<96;n++) {for(m=0;m<96;m++) results << g(n,m,arr) << "\t"; results << "\n";}
        cout << "Calculated gadgets";
    }
    else cout << "error: unable to open results file";
    return 0;
}
         
	\end{verbatim}
	\end{scriptsize}

\subsection{MATLAB Program}

The MATLAB program for calculating values of the Gadget from the ``small" 
BC${}_{4}$ library is composed of two elements: the function ``GFunction.m" 
and the script ``GadgetCalculationCode.m".  The Excel file ``Computational 
Project Data Full.xlsm", which contains the BC${}_{4}$ library arranged 
properly for the code, is also required. 

The function ``GFunction.m" performs the operation described in Equation 4.13. 
It takes as input the two indexes $(\cal R)$ and $({\cal R}{}^{\prime})$, as well 
as two data arrays labeled L1 and L2 which contain the $\ell$ and $\Tilde \ell$ 
library elements respectively.  It outputs the value of the Gadget element indexed 
by $(\cal R)$ and $({\cal R}{}^{\prime})$. In explaining this code's calculation 
process, it is best to begin with the result in (\ref{Gdgt2}).

The calculation is considerably simplified by taking advantage of a helpful 
property of the ``small" BC${}_{4}$ library.  For all values of index $(\cal R)$ 
within the set of a single given value of index IJ, there is only a single value 
of index \^{a} with nonzero library elements.  For example, for $IJ=1$ 
(representing subscript $12$), all nonzero $\ell$ values have index $\hat{a}=2
$, while all nonzero $\Tilde \ell$ values have index $\hat{a}=3$.  Similar 
patterns are observed to hold for all six index $IJ$ values within this library.

Taking advantage of this property, the data sets L1 and L2 contain only the library 
values with the appropriate \^{a} index for each corresponding $IJ$ value.  This 
allowed all library elements necessary for calculating elements of G to be described 
by only two indexes.  In the L1 and L2 arrays, superscript index $(\cal R)$ is contained 
in the row number of each cell, and subscript index $IJ$ is contained in the column 
number.  With these arrays, the Gadget calculation is reduced to only a single 
summation over the ``hatted $a$'' values.

This is the form of the equation utilized by ``GFunction.m" in calculating the Gadget 
element for given $(\cal R)$ and $({\cal R}{}^{\prime})$.  In order to carry out this 
calculation, the function first establishes a 6-element vector named ``subGadget".  
The $IJ^{th}$ entry of subGadget is then determined by performing $l_{IJ}^{(\cal 
R)}*l_{IJ}^{({\cal R}{}^{\prime})}+\widetilde{l}_{IJ}^{(\cal R)}*\widetilde{l}_{IJ}^{({\cal 
R}{}^{\prime})}$, as $IJ$ is looped from 1 through 6.  Lastly, all six elements of 
subGadget are summed up, and the sum is multiplied by 1/6 to produce the 
Gadget element.  This is the standard procedure for summing over an index 
in MATLAB. 

The script ``GadgetCalculationCode.m" runs the function ``GFunction.m" for all 
$96*96$ combinations of $(\cal R)$ and $({\cal R}{}^{\prime})$, to produce the 
Gadget matrix.  It begins by defining L1 and L2 using data imported from the 
Excel file "Computational Project Data Full.xlsm".  It next establishes a $96*96$ 
matrix named ``Result".  Each element $(({\cal R}),({\cal R}{}^{\prime}))$ of matrix 
``Result" is determined by running ``GFunction.m" with $(\cal R)$ and $({\cal 
R}{}^{\prime})$ as input.  To reduce computation time, following the calculation 
of each $(({\cal R}),({\cal R}{}^{\prime}))$ element, the corresponding element 
at coordinate $(({\cal R}{}^{\prime}),({\cal R}))$ is subsequently filled in with the 
same value.  This is permitted, as the Gadget matrix is symmetric over switching 
$(\cal R)$ and $({\cal R}{}^{\prime})$.

At the end of the computation, the matrix ``Result" is the Gadget matrix, fully 
calculated for the ``small" BC${}_{4}$ library.  Additionally, the script counts 
the number of times each that the values 1, 0, and -1/3 are found in the Gadget.  
The Gadget matrix and the three value counts are saved as variables in the 
MATLAB workspace each time the script is run, where they can be easily 
retrieved and exported to Excel or similar data visualization programs.

\subsection{Mathematica-Gadget Code.nb}

The purpose of this Mathematica code is to calculate all values of the 
Gadget, as defined in (\ref{Gdgt2}), given the $\ell$ 
and $\Tilde {\ell}$ values of the, ``smallâ'' BC${}_{4}$ library in Appendix 
B.  Gadget Code.nb is broken up into 9 distinct annotated sections each 
preforming a set of steps toward achieving this goal. 

The code begins by clearing all associations. In order for Mathematica to store 
the ${\ell}$ and $\Tilde {\ell}$ values from the library in Appendix B, a data 
structure consisting of a multi-dimensional array is first constructed. It will hold 
all the index information pertaining to the ${\ell}$ and $\Tilde {\ell}$ coefficients, 
as well as their values. Each ${\ell}$ and $\Tilde {\ell}$ coefficient has 4 
indices associated with it, (${\cal R}$), $\widehat{a}$, $I$ and $J$. Depending on 
the values of these 4 indices an ${\ell}$ and $\Tilde {\ell}$ will equal either 
1, -1, or 0. The ${\ell}$s and $\Tilde {\ell}$s which equal 0 are not shown in 
the Appendix B library. Two arrays of dimension 6$\times$3$\times$96$\times$1 
are constructed to support all ${\ell}$ and $\Tilde {\ell}$ coefficients. The $
{\ell}$s are mapped to variable L1 as a function of $n$, $a$, and $r1$ as part of 
one array, and the $\Tilde {\ell}$s are mapped to variable L2 as a function of $n$, 
$a$, and $r2$ as part of the another array. The $\widehat{a}$ index maps to the 
variable $a$, with values 1 through 3. The (${\cal R}$) index maps to the variables 
$r1$ and $r2$, with values 1 through 96, $r1$ being associated with the L1 matrix 
and $r2$ being associated with the L2 matrix. In the Appendix B library each of 
the ${\ell}$ and $\Tilde {\ell}$ values are grouped into 6 $\mathbf{P}$ permutation 
sets of (${\cal R}$) 1 to 16, in this code (${\cal R}$) is counted from 1 to 96, simplifying 
encoding which $\mathbf{P}$ set the ${\ell}$ or $\Tilde {\ell}$ coefficient is associated 
with. Additionally, to further simplify things, because $I$ and $J$ appear in the 
same 6 pairs 12, 13, 14, 23, 24, 34, the code interprets $IJ$ as one variable, $n$, with 
values 1 through 6. Each ${\ell}$ or $\Tilde {\ell}$ is represented in its corresponding 
L1 or L2 array such that the first 3 dimensions of the multi-dimensional array encode 
its $n$, $a$, $r1$ or $r2$ index values respectively. The goal is to be able to call for 
example, the 1st,2nd,30th entry in the L1, ${\ell}$ coefficient, array and return back 
an integer value, 1, -1, or 0, equaling that particular ${\ell}$ coefficient, given each of 
its specific indices as encoded by the multi-dimensional array structure.   This allows 
Mathematica to have access to all the ${\ell}$ and $\Tilde {\ell}$ values and perform 
the proper Gadget sum.

With the data structure now established, the data from the library must be imported into 
the code. In order to do this an Excel document, Data Record Structure.xlsx, was used. 
In this Excel document are color coded tables for each ${\ell}$ and $\Tilde {\ell}$ value 
from the library. The tables group the values by their $\mathbf{P}$ set and corresponding 
indices, as defined above. Each ${\ell}$ and $\Tilde {\ell}$ value from the library was 
added to the Excel sheet manually, and the tables were double checked to make sure 
the correct values were properly located. After all library values were inserted into the 
tables, Excel was used to automatically add the additional 0 value ${\ell}$ and $\widehat 
{\ell}$ coefficients which where missing from the library but are required in this data structure. 
The ${\ell}$ and $\Tilde {\ell}$ values where then combined into a single (${\cal R}$) 1 
to 96 group, and divided into two 2 dimensional arrays separating the tilded and non-tilded 
coefficients. These arrays where then copied over to 2 separate files, L1s.csv, and L2s.csv. 
In these .csv files the transpose of the arrays where taken and every set of three rows, 
associated with coefficient values that have the same IJ index, or n index in the code, were 
placed one after the other to the right as columns. The data from the library was now in a 
format with an identical structure to the one mentioned in the previous paragraph. Using 
Mathematica's import function these .csv files, for the ${\ell}$ and $\Tilde {\ell}$ coefficients, 
are now imported as 6$\times$3$\times$96$\times$1 arrays. 

In the third section of the code these 6$\times$3$\times$96$\times$1 arrays of data are defined 
to have the functional structure described in section 1 using Mathematica's ``part" structural 
function. Now using the second equation on page 15, the Gadget as a function $G[({\cal R}), 
\, ({\cal R}^{\prime})]$, (${\cal R}$) and $({\cal R}^{\prime})$ from 1 to 96, is defined as 
$1/6$-th the sum of the products of the L1 and L2 array coefficients, with $n$ from 1 to 6, 
and $a$ from 1 to 3. A test value for the Gadget of $({\cal R}^{})$ = 1 and $({\cal R}^{\prime
})$ = 34 is checked to make sure the result is reasonable. Then in order to calculate all 
96$\times$96 Gadget values for the ``small" BC${}_{4}$ case, using the Array function in 
Mathematica, the Gadget function is mapped onto a 96 by 96 matrix where each $i^{th}
\,j^{th}$ element is the corresponding $G(i^{th},j^{th})$ Gadget value. The Gadget Matrix 
has been determined!  The result is quickly checked to make sure it is a symmetric matrix 
and exported as a .mat file.

In order to better present the result, as seen in Appendix A, the 96$\times$96 Gadget Matrix 
is partitioned into a 6$\times$6$\times$16$\times$16 multi-dimensional matrix, where each 
of the 36 nested 16$\times$16 matrices correspond to the P-Matrices of the Gadget Matrix 
Solution. A check is done and it is determined that the P Matrices themselves are symmetric. 
This is no surprise as the Gadget Matrix solution itself was symmetric. Given the symmetry 
of the P-Matrices only 21 of the 36 total, the Upper Triangular result of the Matrices in the 
P-Matrix array, need be presented in order to give the full result as is seen in Appendix A. In 
the last section of code some interesting values of the Gadget Matrix are calculated, revealing 
the 3 unique elements of the Gadget Matrix, how many of each there are, and how many there 
are as a fraction of the total number of values.


\newpage
\section{Conclusion}
\label{conclusions}

 \vskip,2in

Among the main results of this work are those that were shown in
Table 1 and summarized in the formula immediately below Table 1.   
Therefore, we state our main result in the form of a mathematical 
conjecture.

$~~~$ {\it {Summary of the Minimal Four-Color Gadget Conjecture}}:  \newline \indent $~~~$
{\it {Let }}${(\cal R)}$  {\it {and}} $ {({\cal R}^{\prime})}$
{\it {denote any four color adinkra graphs associated 
with the Coxeter}} \newline \indent $~~~$
 {\it {Group}} ${\bm {\rm BC}_4}.$  {\it {To each such graph, there exist six associated matrices called
``fermionic }} \newline \indent $~~~$  {\it {holoraumy matrices''  denoted 
by}} ${\bm {\Tilde V}_{\rI\rJ}^{(\cal R)}}$ {\it {and computed from closed four-cycles 
in the}} \newline \indent $~~~$  {\it {specified graph.  The adinkra Gadget values 
between all pairs of representations}} ${(\cal R)}$  \newline \indent $~~~$
 {\it {and}} ${(\cal R^{\prime})}$, {\it {given}} {\it {by}}
 \vskip0.1pt 
$$
{{\cal G}} [  ({ {\cal R}}) , ( {\cal R}^{\prime}) ] ~=~ \left[ \, \fracm1{48}  \,   
\right] \, \sum_{\rI , \, \rJ} \,{\rm {Tr}} \,  \left[ \, \bm{{\Tilde V}_{\rI\rJ}}{}^{(
\cal R)}  \, \bm{{\Tilde V}_{\rI\rJ}}{}^{({\cal R}^{\prime})}  \right]  ~~~,
$$
$~~~~~~~~$
 {\it {defines a matrix over the space of representations such that the meromorphic
 }} \newline \indent $~~~~$  {\it {``Summary of the Gadget'' function}}
  $$ \eqalign{
 {\cal S}{}_{\cal G} (z) ~&=~ 
{1 \over  {  z^{p_1}  \, (\, z \,+\, \fracm 13 \,)^{p_2}  \, (\, z \,-\, \fracm 13 \,)^{p_3}
\,   \, (\, z \,-\, 1 \,)^{p_4} }}  ~~~,   \cr
& p_1 ~=~ 1,132,462,080 ~~,~~ p_2 ~=~ 127,401,984  ~~,~~  \cr
&p_3 ~=~  84,934,656
 ~~~~\,~~,~~ p_4 ~=~ 14,155,776   ~\,~~,
} $$
\vskip0.5pt  $~~~$
 {\it {has the following properties:}}
 \vskip0.5pt  $~~~~\,~~~$
 {\it {(a.) the sum of the exponents equals to the square of}} 36,864, 
  \newline \indent $~~~~~~\,~~~~~~~$  {\it {i.e.\ the rank of this matrix,}}
  \newline $~~~~~~~~~~~~~$
 {\it {(b.) the poles of this function are the only non-vanishing
}} \newline \indent $~~~~~~\,~~~~~~~$  {\it {entries that appear in this matrix, and}}
   \newline $~~~~~~~~~~~~~$
 {\it {(c.) the exponent associated with each pole is the multiplicity  
 }} \newline \indent $~~~~~~~~~~~~~$  {\it {with which the value of the pole appears in this matrix.}}
\vskip0.5pt \noindent

We state this as a conjecture but recognizing the computations that underlie this 
paper constitute an exhaustive proof by construction, unless there is an error in 
either our reasoning or in our codes.  Currently we know of no analytical way to 
go from the graph theoretic definition of adinkras with four colors, four open nodes, 
and four closed nodes to these results.  We believe it would be an interesting 
mathematical challenge to create a theorem that replicates these results.  Whether 
this ``Summary of the Gadget'' (i.e.\ ${\cal S}{}_{\cal G} (z) $) function has a deeper 
mathematical significance is an open question.  Should it be possible to create 
such a theorem, it potentially could extend the considerations of this work well 
beyond the class of adinkra graphs constructed on the foundation of ${\rm BC}{}_4$.

 \vspace{.05in}
 \begin{center}
\parbox{4in}{{\it ``Your work is going to fill a large part of your life, and the only 
way to be truly satisfied is to do what you believe is great work.  And the only 
way to do great work is to love what you do.  If you haven't found it yet, keep 
looking.  Don't settle.  As with all matters of the heart, you'll know when you find 
it.'' \\ ${~}$ 
 \\ ${~}$ 
\\ ${~}$ }\,\,-\,\, S.\ Jobs $~~~~~~~~~$}
 \parbox{4in}{
 $~~$}  
 \end{center}
 
\noindent
{\bf Added Note In Proof}\\[.1in] \indent
An updated version of the Python code used for calculating all the ordered 
quartets/tetrads in the BC${}_4$ Coxeter Group and for calculating the 
1.3+ billion Gadget values can be obtained from the webpage below.

https://github.com/vkorotkikh/SUSY-BC4CG-36864-Adinkras-and-1.3billion-GadgetVals 
 
 \noindent
{\bf Acknowledgements}\\[.1in] \indent
We are grateful to recognize the contributions of Miles David Miller-Dickson,
and Benedict A. Mondal for their efforts on the calculations in chapter four 
during the 2016 ``Brown University Adinkra Math/Phys Hangout'' (19-23 Dec. 
2016).

This work was partially supported by the National Science Foundation grant 
PHY-1315155.  Additional acknowledgment is given by F.\ Guyton, D.\ S.\ Kessler,  
S.\ Harmalkar, and V.\ A.\ Meszaros  to the the University of Maryland Center 
for String \& Particle Theory (CSPT), for their participation in the 2016 SSTPRS 
(Student Summer Theoretical Physics Research Session) program.    S.J.G. 
acknowledges the generous support of the Provostial Visiting Professorship
Program and the Department of Physics at Brown University for the very 
congenial and generous hospitality during the period of this work.  This 
research was also supported in part by CSPT.  

\newpage
\noindent
$\bm{ \rm{
Appendix ~A: Multiplications ~  Of~ {\{ {\cal V}_{(4)} \}}~
By ~ Permutation ~Group ~Of ~ Order ~Four~ Elements
}} \label{AppA}$

$~~~~$ In this appendix, we include tables that explicitly carry out the multiplication of
the $\bm {\{ {\cal V}_{(4)} \}} $ subset by all of the elements of the permutation group of 
order four.  The tables presented show the results for left multiplications and for right 
multiplications.


\begin{figure}[ht]
\centering{
	\begin{tabular}{| c | c | c | c | c |}
	\hline
	Cycle &   () &   (12)(34) &   (13)(24) &   (14)(23) \\
	\hline
()  &  ()  &  (12)(34) &  (13)(24) &  (14)(23)         \\
(12)(34)  &  (12)(34) &  () &  (14)(23)  &  (13)(24)      \\
(13)(24)  &  (13)(24)&  (14)(23)  &  ()  &  (12)(34)       \\ 
(14)(23)  &  (14)(23) &  (13)(24)  &  (12)(34) &  ()       \\

	\hline
	\end{tabular}
}
\end{figure}
$~~~~~$ $~~~~~~~~~~~~~~~~~~~~~~~~$ {\bf {Table}} {\bf {3:}} Multiplication Of
$\bm {\{ {\cal V}_{(4)} \}}$ By $\bm {\{ {\cal V}_{(4)} \}}$
\newline $~$ \newline

\begin{figure}[ht]
\centering{
	\begin{tabular}{| c | c | c | c | c |}
	\hline
	Cycle &   () &   (12)(34) &   (13)(24) &   (14)(23) \\
	\hline
(12)  &  (12) &  (34) &  (1324)  &  (1423)       \\
(13)  &  (13) &  (1234) &  (24)  &  (1432)      \\
(14)  &  (14)&  (1243)  &  (1342)  &  (23)       \\ 
(23)  &  (23) &  (1342)  &  (1243) &  (14)       \\
(24)  &  (24) &  (1432)  &  (13)  &  (1234)       \\
(34)  &  (34) &  (12) &  (1423)  &  (1324)       \\
	\hline
	\end{tabular}
}
\end{figure}
$~~~~~$ $~~~~~~~~~~~~~~~~~~~~~~$ {\bf {Table}} {\bf {4:}} Left Multiplication Of
$\bm {\{ {\cal V}_{(4)} \}}$ By 2-Cycles
\newline $~$ \newline

\begin{figure}[ht]
\centering{
	\begin{tabular}{| c | c | c | c | c |}
	\hline
	Cycle &   () &   (12)(34) &   (13)(24) &   (14)(23) \\
	\hline
(123)    &   (123) & (134) & (243) & (142)      \\  
(124)    &    (124) & (143) & (132) & (234)    \\  
(132)    &    (132) & (234) & (124) & (143)    \\  
(134)    &   (134) & (123) & (142) & (243)     \\  
(142)    &   (142) & (243) & (134) & (123)     \\  
(143)    &     (143) & (124) & (234) & (132)    \\  
(234)    &   (234) & (132) & (143) & (124)     \\  
(243)    &    (243) & (142) & (123) &   (134)    \\  
	\hline
	\end{tabular}
}
\end{figure}
$~~~~~$ $~~~~~~~~~~~~~~~~~~~~~~$ {\bf {Table}} {\bf {5:}} Left Multiplication Of
$\bm {\{ {\cal V}_{(4)} \}}$ By 3-Cycles
\newline $~$ \newpage

\begin{figure}[ht]
\centering{
	\begin{tabular}{| c | c | c | c | c |}
	\hline
	Cycle &   () &   (12)(34) &   (13)(24) &   (14)(23) \\
	\hline
(1234)   &   (1234) &  (13) &  (1432) &  (24)   \\
(1243)   &   (1243) &  (14) &  (23) &  (1342)   \\
(1324)   &    (1324) &  (1423) &  (12) &  (34)   \\ 
(1342)   &    (1342) &  (23) &  (14) &  (1243)    \\
(1423)   &    (1423) &  (1324) &  (34) &  (12)    \\
(1432)   &    (1432) &  (24) &  (1234) &  (13)  \\
	\hline
	\end{tabular}
}
\end{figure}
$~~~~~$ $~~~~~~~~~~~~~~~~~~~~~~$ {\bf {Table}} {\bf {6:}} Left Multiplication Of
$\bm {\{ {\cal V}_{(4)} \}}$ By 4-Cycles
\newline $~$ \newline

\begin{figure}[ht]
\centering{
	\begin{tabular}{| c | c | c | c | c |}
	\hline
	() &   (12)(34) &   (13)(24) &   (14)(23) & Cycle   \\
	\hline
(12) &  (34)  &  (1423)  &  (1324)  &  (12)  \\ 
(13) &   (1432) &  (24)  &   (1234) &  (13)  \\   
(14) &  (1342) &  (1243)  &  (23)  &  (14)  \\      
(23) &  (1243) &  (1342)   &  (14) &  (23)  \\    
(24) &   (1234) &  (13)  &   (1432) &  (24)  \\    
(34) &  (12) &  (1324) &  (1423)  & (34)  \\
	\hline
	\end{tabular}
}
\end{figure}
$~~~~~$ $~~~~~~~~~~~~~~~~~~~~~~$ {\bf {Table}} {\bf {7:}} Right Multiplication Of
$\bm {\{ {\cal V}_{(4)} \}}$ By 2-Cycles
\newline $~$ \newline

\begin{figure}[ht]
\centering{
	\begin{tabular}{| c | c | c | c | c |}
	\hline
	() &   (12)(34) &   (13)(24) &   (14)(23) & Cycle   \\
	\hline
(123) &  (243) & (142) &  (134) & (123) \\       
(124) &  (234) & (143) &  (132) & (124) \\     
(132) &  (143) & (234) &  (124) & (132) \\      
(134) &  (142) & (243) &  (123) & (134) \\        
(142) &  (134) & (123) &  (243) & (142) \\       
(143) &  (132) & (124) &  (234) & (143) \\      
(234) &  (124) & (132) &  (143) & (234) \\       
(243) &  (123) & (134) &  (142) & (243) \\ 
	\hline
	\end{tabular}
}
\end{figure}
$~~~~~$ $~~~~~~~~~~~~~~~~~~~~~~$ {\bf {Table}} {\bf {8:}} Right Multiplication Of
$\bm {\{ {\cal V}_{(4)} \}}$ By 3-Cycles
\newline $~$ \newpage

\begin{figure}[ht]
\centering{
	\begin{tabular}{| c | c | c | c | c |}
	\hline
	() &   (12)(34) &   (13)(24) &   (14)(23) & Cycle   \\
	\hline
(1243) & (23)     & (14)      &  (1342)  & (1243)  \\     
(1234) & (24)     & (1432)  &  (13)      & (1234)  \\    
(1324) & (1423) &  (34)     &  (12)      & (1324)  \\    
(1342) &  (14)     &  (23)     &  (1243) & (1342)  \\    
(1432) &  (13)     & (1234) &  (24)      & (1432)  \\    
(1423) &  (1324) &  (12)    &  (34)      &  (1423) \\
	\hline
	\end{tabular}
}
\end{figure}
$~~~~~$ $~~~~~~~~~~~~~~~~~~~~~~$ {\bf {Table}} {\bf {9:}} Right Multiplication Of
$\bm {\{ {\cal V}_{(4)} \}}$ By 4-Cycles
\newline $~$ \newline

\newpage
\noindent
$\bm{ \rm{
Appendix ~B: ``Fiducial"  ~ Boolean ~Factors
}} \label{AppB}$

It suffices to specify the  ``Boolean Factors'' in the same order as the 
permutation quartet factors appear in (\ref{PermSets}).  Thus, for each of the six sectors 
we find

$$
\label{e:CMEvenRep1}\eqalign{
{~~~~~~~~~~~~~~}
{{\bm S}_{\bm {\cal P}}}_1[\a]: & 
 \{(0)_b  , (6)_b , (12)_b  , (10)_b  \}, \{
 (0)_b  , (12)_b  , (10)_b  , (6)_b \}, \{
 (2)_b  , (4)_b , (14)_b , (8)_b \}, \cr
 & \{
 (2)_b  , (14)_b , (8)_b , (4)_b \}, \{
 (4)_b , (2)_b  , (8)_b , (14)_b \}, \{
 (4)_b , (8)_b , (14)_b , (2)_b  \}, \cr
 & \{
 (6)_b , (0)_b  , (10)_b  , (12)_b  \}, \{
 (6)_b , (10)_b  , (12)_b  , (0)_b  \}, \{
 (8)_b , (4)_b , (2)_b  , (14)_b \}, \cr
 &\{
 (8)_b , (14)_b , (4)_b , (2)_b  \}, \{
 (10)_b  , (6)_b , (0)_b  , (12)_b  \}, \{
 (10)_b  , (12)_b  , (6)_b , (0)_b  \}, \cr
 & \{
 (12)_b  , (0)_b  , (6)_b , (10)_b  \}, \{
 (12)_b  , (10)_b  , (0)_b  , (6)_b \}, \{
 (14)_b , (2)_b  , (4)_b , (8)_b \}, \cr
 &\{
 (14)_b , (8)_b , (2)_b  , (4)_b\}  ~~,
 }\eqno(B.1)
 $$

$$ 
\label{e:CMEvenRep2}\eqalign{
 {~~~~~~~~~~~~~~}
 {{\bm S}_{\bm {\cal P}}}_2 [\a]: &  
 \{(0)_b , (10)_b , (6)_b , (12)_b \} , 
 \{ (0)_b , (12)_b , (10)_b , (6)_b \} , 
 \{ (2)_b , (8)_b , (4)_b , (14)_b \} , \cr
 & \{
 (2)_b , (14)_b , (8)_b , (4)_b \} , \{
 (4)_b , (8)_b , (14)_b , (2)_b \} , \{
 (4)_b , (14)_b , (2)_b , (8)_b \} , \cr
 &\{
 (6)_b , (10)_b , (12)_b , (0)_b \} , \{
 (6)_b , (12)_b , (0)_b , (10)_b \} , \{
 (8)_b , (2)_b , (14)_b , (4)_b \} , \cr
 & \{
 (8)_b , (4)_b , (2)_b , (14)_b \} , \{
 (10)_b , (0)_b , (12)_b , (6)_b \} , \{
 (10)_b , (6)_b , (0)_b , (12)_b \} , \cr
 &\{
 (12)_b , (0)_b , (6)_b , (10)_b \} , \{
 (12)_b , (6)_b , (10)_b , (0)_b \} , \{
 (14)_b , (2)_b , (4)_b , (8)_b \} , \cr
 &\{
 (14)_b , (4)_b , (8)_b , (2)_b \}  ~~,
 }    \eqno(B.2)
 $$

$$ 
\label{e:CMEvenRep3}\eqalign{
{~~~~~~~~~~~~~~}
{{\bm S}_{\bm {\cal P}}}_3[\a]:  &  
 \{ (0)_b , (6)_b , (10)_b , (12)_b \} , \{
 (0)_b , (12)_b , (6)_b , (10)_b \} , \{
 (2)_b , (4)_b , (8)_b , (14)_b \}, \cr
 & \{
 (2)_b , (14)_b , (4)_b , (8)_b \} , \{
 (4)_b , (2)_b , (14)_b , (8)_b \} , \{
 (4)_b , (8)_b , (2)_b , (14)_b \} ,  \cr
 &\{
 (6)_b , (0)_b , (12)_b , (10)_b \} , \{
 (6)_b , (10)_b , (0)_b , (12)_b \} , \{
 (8)_b , (4)_b , (14)_b , (2)_b \} , \cr 
 &  \{
 (8)_b , (14)_b , (2)_b , (4)_b \} , \{
 (10)_b , (6)_b , (12)_b , (0)_b \} , \{
 (10)_b , (12)_b , (0)_b , (6)_b \} , \cr
 & \{
 (12)_b , (0)_b , (10)_b , (6)_b \} , \{
 (12)_b , (10)_b , (6)_b , (0)_b \} , \{
 (14)_b , (2)_b , (8)_b , (4)_b \} , \cr
 &\{
 (14)_b , (8)_b , (4)_b , (2)_b \}   ~~,
 }  \eqno(B.3)
 $$

$$
\label{e:CMEvenRep4}\eqalign{
{~~~~~~~~~~~~~~}
{{\bm S}_{\bm {\cal P}}}_4[\a]: &   
\{(0)_b , (10)_b , (12)_b , (6)_b \} , \{
 (0)_b , (12)_b , (6)_b , (10)_b \} , \{
 (2)_b , (8)_b , (14)_b , (4)_b \} , \cr
 & \{
 (2)_b , (14)_b , (4)_b , (8)_b \} , \{
 (4)_b , (8)_b , (2)_b , (14)_b \} , \{
 (4)_b , (14)_b , (8)_b , (2)_b \} , \cr 
 &  \{
 (6)_b , (10)_b , (0)_b , (12)_b \} , \{
 (6)_b , (12)_b , (10)_b , (0)_b \} , \{
 (8)_b , (2)_b , (4)_b , (14)_b \} , \cr
 & \{
 (8)_b , (4)_b , (14)_b , (2)_b \} , \{
 (10)_b , (0)_b , (6)_b , (12)_b \} , \{
 (10)_b , (6)_b , (12)_b , (0)_b \} ,  \cr
 & \{
 (12)_b , (0)_b , (10)_b , (6)_b \} , \{
 (12)_b , (6)_b , (0)_b , (10)_b \} , \{
 (14)_b , (2)_b , (8)_b , (4)_b \} , \cr 
 & \{
 (14)_b , (4)_b , (2)_b , (8)_b \}  ~~,
}  \eqno(B.4)
 $$
       
$$
\label{e:CMEvenRep5}\eqalign{
{~~~~~~~~~~~~~~}
{{\bm S}_{\bm {\cal P}}}_5[\a]: 
& \{ (0)_b , (6)_b , (10)_b , (12)_b \} , \{
 (0)_b , (10)_b , (12)_b , (6)_b \} , \{
 (2)_b , (4)_b , (8)_b , (14)_b \} , \cr
 &  \{
 (2)_b , (8)_b , (14)_b , (4)_b \} , \{
 (4)_b , (2)_b , (14)_b , (8)_b \} , \{
 (4)_b , (14)_b , (8)_b , (2)_b \} , \cr
 &\{
 (6)_b , (0)_b , (12)_b , (10)_b \} , \{
 (6)_b , (12)_b , (10)_b , (0)_b \} , \{
 (8)_b , (2)_b , (4)_b , (14)_b \} , \cr
 &\{
 (8)_b , (14)_b , (2)_b , (4)_b \} , \{
 (10)_b , (0)_b , (6)_b , (12)_b \} , \{
 (10)_b , (12)_b , (0)_b , (6)_b \} ,
 \cr
 & \{
 (12)_b , (6)_b , (0)_b , (10)_b \} , \{
 (12)_b , (10)_b , (6)_b , (0)_b \} , \{
 (14)_b , (4)_b , (2)_b , (8)_b \} , \cr
 &\{
 (14)_b , (8)_b , (4)_b , (2)_b \}  ~~,
  } \eqno(B.5)
 $$
 
$$
\label{e:CMEven6}\eqalign{
{~~~~~~~~~~~~~~}
{{\bm S}_{\bm {\cal P}}}_6 [\a]: &  
 \{ (0)_b , (6)_b , (12)_b , (10)_b \} , \{
 (0)_b , (10)_b , (6)_b , (12)_b \} , \{
 (2)_b , (4)_b , (14)_b , (8)_b \} , \cr
 & \{
 (2)_b , (8)_b , (4)_b , (14)_b \} , \{
 (4)_b , (2)_b , (8)_b , (14)_b \} , \{
 (4)_b , (14)_b , (2)_b , (8)_b \} , \cr
 &\{
 (6)_b , (0)_b , (10)_b , (12)_b \} , \{
 (6)_b , (12)_b , (0)_b , (10)_b \} , \{
 (8)_b , (2)_b , (14)_b , (4)_b \} , \cr
 & \{
 (8)_b , (14)_b , (4)_b , (2)_b \} , \{
 (10)_b , (0)_b , (12)_b , (6)_b \} , \{
 (10)_b , (12)_b , (6)_b , (0)_b \} , \cr
 & \{
 (12)_b , (6)_b , (10)_b , (0)_b \} , \{
 (12)_b , (10)_b , (0)_b , (6)_b \} , \{
 (14)_b , (4)_b , (8)_b , (2)_b \} , 
 \cr 
 &\{
 (14)_b , (8)_b , (2)_b , (4)_b \}  ~~.
 }  \eqno(B.6)
 $$
The notation is designed to elicit the fact that for each choice of  ${\bm {\cal P}_{[\L]}}$,
with the subscript $[\L]$ taking on values $[1]$, $\dots$, $[6]$, there are sixteen possible 
choices of ${{\bm S}_{\bm {\cal P}_{[\L]}}}[\a] $ where the index $\a$ enumerates those 
choices taking on values 1, $\dots$ 16.  

\newpage
\noindent
$\bm{ \rm{
Appendix ~C: ~Adinkra ~ \ell ~ \& ~\Tilde \ell ~Values ~Over ~the 
~``Small ~{BC}_4 ~Library"
}} \label{AppC}
$

In this appendix, for all of the elements of ${\rm BC}{}_4$, the explicit values of the  
$\ell$, and ${\Tilde \ell}$ coefficients
which are related to each of the representation $(\cal R)$ in the order
shown in appendix C.
$$
\begin{array}{cccccc|c}
{\bm {\cal P}_{[1]}}: &&&&&& \chi_{\rm o} \\
\hline
\ell _{12}^{(1)2}=1 & \ell _{13}^{(1)3}=1 & \ell _{14}^{(1)1}=1 & \ell _{23}^{(1)1}=1 & \ell _{24}^{(1)3}=-1 
& \ell_{34}^{(1)2}=1 &1 \\
\tilde{\ell }_{12}^{(2)3}=1 & \tilde{\ell }_{13}^{(2)2}=1 & \tilde{\ell }_{14}^{(2)1}=1 & \tilde{\ell }_{23}^{(2)1}=-1 
& \tilde{\ell }_{24}^{(2)2}=1 & \tilde{\ell }_{34}^{(2)3}=- \,1  & -1 \\
\ell _{12}^{(3)2}=1 & \ell _{13}^{(3)3}=1 & \ell _{14}^{(3)1}=1 & \ell _{23}^{(3)1}=1 & \ell _{24}^{(3)3}=-1 
 & \ell_{34}^{(3)2}= \,1  & 1 \\
\tilde{\ell }_{12}^{(4)3}=1 & \tilde{\ell }_{13}^{(4)2}=1 & \tilde{\ell }_{14}^{(4)1}=1 & \tilde{\ell }_{23}^{(4)1}=-1 
& \tilde{\ell }_{24}^{(4)2}=1 & \tilde{\ell }_{34}^{(4)3}=- \,1 & - 1 \\
\ell _{12}^{(5)2}=1 & \ell _{13}^{(5)3}=1 & \ell _{14}^{(5)1}=1 & \ell _{23}^{(5)1}=1 & \ell _{24}^{(5)3}=-1 & 
\ell_{34}^{(5)2}=1 & 1 \\
\tilde{\ell }_{12}^{(6)3}=1 & \tilde{\ell }_{13}^{(6)2}=1 & \tilde{\ell }_{14}^{(6)1}=1 & \tilde{\ell }_{23}^{(6)1}=-1 
& \tilde{\ell }_{24}^{(6)2}=1 & \tilde{\ell }_{34}^{(6)3}=-1 & -1 \\
\ell _{12}^{(7)2}=1 & \ell _{13}^{(7)3}=1 & \ell _{14}^{(7)1}=1 & \ell _{23}^{(7)1}=1 & \ell _{24}^{(7)3}=-1 & 
\ell_{34}^{(7)2}=1 & 1 \\
\end{array}
$$
$~~~~~$ $~~~~~~~~~~~~~~~~$ {\bf {Table}} {\bf {10:}} Non-vanishing $\ell$, $\Tilde \ell$, and
$\chi_{\rm o} $ Values For Representation ${\bm {\cal P}_{[1]}}$
\newline $~$ \newline
$$
\begin{array}{cccccc|c}
{\bm {\cal P}_{[2]}}: &&&&&& \chi_{\rm o} \\
\hline
 \tilde{\ell }_{12}^{(1)3}=1 & \tilde{\ell }_{13}^{(1)2}=1 & \tilde{\ell }_{14}^{(1)1}=1 & \tilde{\ell }_{23}^{(1)1}=-1 &   \tilde{\ell }_{24}^{(1)2}=1 & \tilde{\ell }_{34}^{(1)3}=-1 & -1 \\
 \ell _{12}^{(2)2}=1 & \ell _{13}^{(2)3}=1 & \ell _{14}^{(2)1}=1 & \ell _{23}^{(2)1}=1 & \ell _{24}^{(2)3}=-1 & \ell_{34}^{(2)2}=1  & 1 \\
 \tilde{\ell }_{12}^{(3)3}=1 & \tilde{\ell }_{13}^{(3)2}=1 & \tilde{\ell }_{14}^{(3)1}=1 & \tilde{\ell }_{23}^{(3)1}=-1 &  \tilde{\ell }_{24}^{(3)2}=1 & \tilde{\ell }_{34}^{(3)3}=-1  & - 1 \\
 \ell _{12}^{(4)2}=1 & \ell _{13}^{(4)3}=1 & \ell _{14}^{(4)1}=1 & \ell _{23}^{(4)1}=1 & \ell _{24}^{(4)3}=-1 & \ell_{34}^{(4)2}=1 &  1 \\
 \ell _{12}^{(5)2}=1 & \ell _{13}^{(5)3}=1 & \ell _{14}^{(5)1}=1 & \ell _{23}^{(5)1}=1 & \ell _{24}^{(5)3}=-1 & \ell_{34}^{(5)2}=1 & 1 \\
 \tilde{\ell }_{12}^{(6)3}=1 & \tilde{\ell }_{13}^{(6)2}=1 & \tilde{\ell }_{14}^{(6)1}=1 & \tilde{\ell }_{23}^{(6)1}=-1 &  \tilde{\ell }_{24}^{(6)2}=1 & \tilde{\ell }_{34}^{(6)3}=-1 & -1 \\
 \ell _{12}^{(7)2}=1 & \ell _{13}^{(7)3}=1 & \ell _{14}^{(7)1}=1 & \ell _{23}^{(7)1}=1 & \ell _{24}^{(7)3}=-1 & \ell_{34}^{(7)2}=1 & 1 \\
\end{array}
$$
$~~~~~$ $~~~~~~~~~~~~~~~~$ {\bf {Table}} {\bf {11:}} Non-vanishing $\ell$, $\Tilde \ell$, and
$\chi_{\rm o} $ Values For Representation ${\bm {\cal P}_{[2]}}$
\newline $~$ \newpage

$$
\begin{array}{cccccc|c}
{\bm {\cal P}_{[3]}}: &&&&&& \chi_{\rm o} \\
\hline
\tilde{\ell }_{12}^{(1)3}=-1 & \tilde{\ell }_{13}^{(1)2}=1 & \tilde{\ell }_{14}^{(1)1}=-1 & \tilde{\ell }_{23}^{(1)1}=1 &
\tilde{\ell }_{24}^{(1)2}=1 & \tilde{\ell }_{34}^{(1)3}=1 & -1 \\
\ell _{12}^{(2)2}=-1 & \ell _{13}^{(2)3}=-1 & \ell _{14}^{(2)1}=1 & \ell _{23}^{(2)1}=1 & \ell _{24}^{(2)3}=1 & 
\ell_{34}^{(2)2}=-1 & 1 \\
\tilde{\ell }_{12}^{(3)3}=-1 & \tilde{\ell }_{13}^{(3)2}=1 & \tilde{\ell }_{14}^{(3)1}=-1 & \tilde{\ell }_{23}^{(3)1}=1 &
\tilde{\ell }_{24}^{(3)2}=1 & \tilde{\ell }_{34}^{(3)3}=1  & - 1 \\
\ell _{12}^{(4)2}=-1 & \ell _{13}^{(4)3}=-1 & \ell _{14}^{(4)1}=1 & \ell _{23}^{(4)1}=1 & \ell _{24}^{(4)3}=1 & 
\ell_{34}^{(4)2}=-1  &  1 \\
\tilde{\ell }_{12}^{(5)3}=-1 & \tilde{\ell }_{13}^{(5)2}=1 & \tilde{\ell }_{14}^{(5)1}=-1 & \tilde{\ell }_{23}^{(5)1}=1 &
\tilde{\ell }_{24}^{(5)2}=1 & \tilde{\ell }_{34}^{(5)3}=1 & -1 \\
\ell _{12}^{(6)2}=-1 & \ell _{13}^{(6)3}=-1 & \ell _{14}^{(6)1}=1 & \ell _{23}^{(6)1}=1 & \ell _{24}^{(6)3}=1 & 
\ell_{34}^{(6)2}=-1 & 1 \\
\tilde{\ell }_{12}^{(7)3}=-1 & \tilde{\ell }_{13}^{(7)2}=1 & \tilde{\ell }_{14}^{(7)1}=-1 & \tilde{\ell }_{23}^{(7)1}=1 &
\tilde{\ell }_{24}^{(7)2}=1 & \tilde{\ell }_{34}^{(7)3}=1 & - 1 \\
\end{array}
$$
$~~~~~$ $~~~~~~~~~~~~~~~~$ {\bf {Table}} {\bf {12:}} Non-vanishing $\ell$, $\Tilde \ell$, and
$\chi_{\rm o} $ Values For Representation ${\bm {\cal P}_{[3]}}$
\newline $~$ \newline

$$
\begin{array}{cccccc|c}
{\bm {\cal P}_{[4]}}: &&&&&& \chi_{\rm o} \\
\hline
\ell _{12}^{(1)2}=1 & \ell _{13}^{(1)3}=1 & \ell _{14}^{(1)1}=1 & \ell _{23}^{(1)1}=1 & \ell _{24}^{(1)3}=-1 & 
\ell_{34}^{(1)2}=1 & 1 \\
\tilde{\ell }_{12}^{(2)3}=1 & \tilde{\ell }_{13}^{(2)2}=1 & \tilde{\ell }_{14}^{(2)1}=1 & \tilde{\ell }_{23}^{(2)1}=-1 &
\tilde{\ell }_{24}^{(2)2}=1 & \tilde{\ell }_{34}^{(2)3}=-1 & -1 \\
\ell _{12}^{(3)2}=1 & \ell _{13}^{(3)3}=1 & \ell _{14}^{(3)1}=1 & \ell _{23}^{(3)1}=1 & \ell _{24}^{(3)3}=-1 & 
\ell_{34}^{(3)2}=1  &  1 \\
\tilde{\ell }_{12}^{(4)3}=1 & \tilde{\ell }_{13}^{(4)2}=1 & \tilde{\ell }_{14}^{(4)1}=1 & \tilde{\ell }_{23}^{(4)1}=-1 &
\tilde{\ell }_{24}^{(4)2}=1 & \tilde{\ell }_{34}^{(4)3}=-1  &  - 1 \\
\tilde{\ell }_{12}^{(5)3}=1 & \tilde{\ell }_{13}^{(5)2}=1 & \tilde{\ell }_{14}^{(5)1}=1 & \tilde{\ell }_{23}^{(5)1}=-1 &
\tilde{\ell }_{24}^{(5)2}=1 & \tilde{\ell }_{34}^{(5)3}=-1   & - 1 \\
\ell_{12}^{(6)2}=1 & \ell _{13}^{(6)3}=1 & \ell _{14}^{(6)1}=1 & \ell _{23}^{(6)1}=1 & \ell _{24}^{(6)3}=-1 & 
\ell_{34}^{(6)2}=1 &  1 \\
\tilde{\ell }_{12}^{(7)3}=1 & \tilde{\ell }_{13}^{(7)2}=1 & \tilde{\ell }_{14}^{(7)1}=1 & \tilde{\ell }_{23}^{(7)1}=-1 &
\tilde{\ell }_{24}^{(7)2}=1 & \tilde{\ell }_{34}^{(7)3}=-1 & - 1 \\
\end{array}
$$
$~~~~~$ $~~~~~~~~~~~~~~~~$ {\bf {Table}} {\bf {13:}} Non-vanishing $\ell$, $\Tilde \ell$, and
$\chi_{\rm o} $ Values For Representation ${\bm {\cal P}_{[4]}}$
\newline $~$ \newline

$$
\begin{array}{cccccc|c}
{\bm {\cal P}_{[5]}}: &&&&&& \chi_{\rm o} \\
\hline
 \ell _{12}^{(1)2}=1 & \ell _{13}^{(1)3}=1 & \ell _{14}^{(1)1}=1 & \ell _{23}^{(1)1}=1 & \ell _{24}^{(1)3}=-1 & 
 \ell_{34}^{(1)2}=1 & 1 \\
 \tilde{\ell }_{12}^{(2)3}=1 & \tilde{\ell }_{13}^{(2)2}=1 & \tilde{\ell }_{14}^{(2)1}=1 & \tilde{\ell }_{23}^{(2)1}=-1 &
\tilde{\ell }_{24}^{(2)2}=1 & \tilde{\ell }_{34}^{(2)3}=-1 & -1 \\
\ell _{12}^{(3)2}=1 & \ell _{13}^{(3)3}=1 & \ell _{14}^{(3)1}=1 & \ell _{23}^{(3)1}=1 & \ell _{24}^{(3)3}=-1 & 
\ell_{34}^{(3)2}=1  &  1 \\
\tilde{\ell }_{12}^{(4)3}=1 & \tilde{\ell }_{13}^{(4)2}=1 & \tilde{\ell }_{14}^{(4)1}=1 & \tilde{\ell }_{23}^{(4)1}=-1 &
\tilde{\ell }_{24}^{(4)2}=1 & \tilde{\ell }_{34}^{(4)3}=-1  &  - 1 \\
\ell_{12}^{(5)2}=1 & \ell _{13}^{(5)3}=1 & \ell _{14}^{(5)1}=1 & \ell _{23}^{(5)1}=1 & \ell _{24}^{(5)3}=-1 & 
\ell_{34}^{(5)2}=1   &  1 \\
\tilde{\ell }_{12}^{(6)3}=1 & \tilde{\ell }_{13}^{(6)2}=1 & \tilde{\ell }_{14}^{(6)1}=1 & \tilde{\ell }_{23}^{(6)1}=-1 &
\tilde{\ell }_{24}^{(6)2}=1 & \tilde{\ell }_{34}^{(6)3}=-1 &  - 1 \\
\ell_{12}^{(7)2}=1 & \ell _{13}^{(7)3}=1 & \ell _{14}^{(7)1}=1 & \ell _{23}^{(7)1}=1 & \ell _{24}^{(7)3}=-1 & 
\ell_{34}^{(7)2}=1 &  1 \\
\end{array}
$$
$~~~~~$ $~~~~~~~~~~~~~~~~$ {\bf {Table}} {\bf {14:}} Non-vanishing $\ell$, $\Tilde \ell$, and
$\chi_{\rm o} $ Values For Representation ${\bm {\cal P}_{[5]}}$
\newline $~$ \newline

$$
\begin{array}{cccccc|c}
{\bm {\cal P}_{[6]}}: &&&&&& \chi_{\rm o} \\
\hline
\tilde{\ell }_{12}^{(1)3}=1 & \tilde{\ell }_{13}^{(1)2}=1 & \tilde{\ell }_{14}^{(1)1}=1 & \tilde{\ell }_{23}^{(1)1}=-1 &
\tilde{\ell }_{24}^{(1)2}=1 & \tilde{\ell }_{34}^{(1)3}=-1 & - 1 \\
\ell _{12}^{(2)2}=1 & \ell _{13}^{(2)3}=1 & \ell _{14}^{(2)1}=1 & \ell _{23}^{(2)1}=1 & \ell _{24}^{(2)3}=-1 & 
\ell_{34}^{(2)2}=1 & 1 \\
\tilde{\ell }_{12}^{(3)3}=1 & \tilde{\ell }_{13}^{(3)2}=1 & \tilde{\ell }_{14}^{(3)1}=1 & \tilde{\ell }_{23}^{(3)1}=-1 &
\tilde{\ell }_{24}^{(3)2}=1 & \tilde{\ell }_{34}^{(3)3}=-1  &  - 1 \\
\ell _{12}^{(4)2}=1 & \ell _{13}^{(4)3}=1 & \ell _{14}^{(4)1}=1 & \ell _{23}^{(4)1}=1 & \ell _{24}^{(4)3}=-1 & 
\ell_{34}^{(4)2}=1  &   1 \\
\tilde{\ell }_{12}^{(5)3}=1 & \tilde{\ell }_{13}^{(5)2}=1 & \tilde{\ell }_{14}^{(5)1}=1 & \tilde{\ell }_{23}^{(5)1}=-1 &
\tilde{\ell }_{24}^{(5)2}=1 & \tilde{\ell }_{34}^{(5)3}=-1   &  - 1 \\
\ell _{12}^{(6)2}=1 & \ell _{13}^{(6)3}=1 & \ell _{14}^{(6)1}=1 & \ell _{23}^{(6)1}=1 & \ell _{24}^{(6)3}=-1 & 
\ell_{34}^{(6)2}=1 &   1 \\
\tilde{\ell }_{12}^{(7)3}=1 & \tilde{\ell }_{13}^{(7)2}=1 & \tilde{\ell }_{14}^{(7)1}=1 & \tilde{\ell }_{23}^{(7)1}=-1 &
\tilde{\ell }_{24}^{(7)2}=1 & \tilde{\ell }_{34}^{(7)3}=-1 &  - 1 \\
\end{array}
$$
$~~~~~$ $~~~~~~~~~~~~~~~~$ {\bf {Table}} {\bf {15:}} Non-vanishing $\ell$, $\Tilde \ell$, and
$\chi_{\rm o} $ Values For Representation ${\bm {\cal P}_{[6]}}$
\newline $~$ \newline
\noindent
The value of the``Kye-Oh'' function when expressed in terms of the $\ell$ and $\Tilde \ell$
parameters of (\ref{Veq}) takes the form
$$ \eqalign{
 { {\chi_{\rm o} ({{\bm S}_{\bm {\cal P}_{[\L]}}}[\a] \, \cdot \, {\bm {\cal P}_{[\L]}} )}} ~&= ~\frc{1}{3}
\sum_{{\hat a} } \, {\Big [} ~ 
{\ell}_{1 \, 2}^{\,({\cal R}) \hat{a}} \,  {\ell}_{3 \, 4}^{\, ({\cal R}) \hat{a}}  ~-~ 
{\ell}_{1 \, 3}^{\,({\cal R}) \hat{a}} \,  {\ell}_{ 2 \, 4}^{\, ({\cal R}) \hat{a}} ~+~
{\ell}_{1\, 4}^{\,({\cal R}) \hat{a}} \,  {\ell}_{2\, 3}^{\, ({\cal R}) \hat{a}}   
~ {\Big ]} ~+~    \cr
~&~~~~ ~\frc{1}{3}
\sum_{{\hat a} } \, {\Big [} ~ 
{{\Tilde {\ell}}}_{1 \, 2}^{\,({\cal R}) \hat{a}} \,  {{\Tilde {\ell}}}_{3 \, 4}^{\, ({\cal R}) \hat{a}}  ~-~ 
{{\Tilde {\ell}}}_{1 \, 3}^{\,({\cal R}) \hat{a}} \,  {{\Tilde {\ell}}}_{ 2 \, 4}^{\, ({\cal R}) \hat{a}} ~+~
{{\Tilde {\ell}}}_{1\, 4}^{\,({\cal R}) \hat{a}} \,  {{\Tilde {\ell}}}_{2\, 3}^{\, ({\cal R}) \hat{a}}   
~ {\Big ]}
~~~.
} \eqno(C.1) $$
Here the representation label $({\cal R})$ corresponds to a specification of $[\L]$ and $\a$.

\newpage
\noindent
$\bm{ \rm{
Appendix ~D: ~Adinkra ~Gadget ~Values ~Over ~the ~``Small ~{BC}_4 
~Library"
}} \label{AppD}
$

\indent
In this appendix, we give the values of the gadget between matrix elements
over all 96 elements of the ``small BC${}_4$ library."

$$
\begin{array}{cc|cccccccccccccccc}
{~} & {\cal S}^{\prime} & 1 & 2  & 3 & 4 & 5 & 6 & 7 & 8  & 9 & 10 & 11 & 12 & 
13 & 14 & 15 & 16  \\
{\cal S}  & {} & {} & {} & {} & {} & {} & {} & {} & {} & {} & {} & {} & {} & {} & {} &  {} 
 \\ \hline
1  & {} &1 & 0 & 1 & 0 & 1 & 0 & 1 & 0 & 0 & 1  & 0  & 1  & 0  & 1  & 0  & 1  \\
2  & {} &0 & 1 & 0 & 1 & 0 & 1 & 0 & 1 & 1 & 0  & 1  & 0  & 1  & 0  & 1  & 0  \\
3  & {} &1 & 0 & 1 & 0 & 1 & 0 & 1 & 0 & 0 & 1  & 0  & 1  & 0  & 1  & 0  & 1  \\
4  & {} &0 & 1 & 0 & 1 & 0 & 1 & 0 & 1 & 1 & 0  & 1  & 0  & 1  & 0  & 1  & 0  \\
5  & {} &1 & 0 & 1 & 0 & 1 & 0 & 1 & 0 & 0 & 1  & 0  & 1  & 0  & 1  & 0  & 1  \\
6  & {} &0 & 1 & 0 & 1 & 0 & 1 & 0 & 1 & 1 & 0  & 1  & 0  & 1  & 0  & 1  & 0  \\
7  & {} &1 & 0 & 1 & 0 & 1 & 0 & 1 & 0 & 0 & 1  & 0  & 1  & 0  & 1  & 0  & 1  \\
8  & {} &0 & 1 & 0 & 1 & 0 & 1 & 0 & 1 & 1 & 0  & 1  & 0  & 1  & 0  & 1  & 0  \\
9  & {} &0 & 1 & 0 & 1 & 0 & 1 & 0 & 1 & 1 & 0  & 1  & 0  & 1  & 0  & 1  & 0  \\
10 & {} &1 & 0 & 1 & 0 & 1 & 0 & 1 & 0 & 0 & 1  & 0  & 1  & 0  & 1  & 0  & 1  \\
11 & {} &0 & 1 & 0 & 1 & 0 & 1 & 0 & 1 & 1 & 0  & 1  & 0  & 1  & 0  & 1  & 0  \\
12 & {} &1 & 0 & 1 & 0 & 1 & 0 & 1 & 0 & 0 & 1  & 0  & 1  & 0  & 1  & 0  & 1  \\
13 & {} &0 & 1 & 0 & 1 & 0 & 1 & 0 & 1 & 1 & 0  & 1  & 0  & 1  & 0  & 1  & 0  \\
14 & {} &1 & 0 & 1 & 0 & 1 & 0 & 1 & 0 & 0 & 1  & 0  & 1  & 0  & 1  & 0  & 1  \\
15 & {} &0 & 1 & 0 & 1 & 0 & 1 & 0 & 1 & 1 & 0  & 1  & 0  & 1  & 0  & 1  & 0  \\
16 & {} &1 & 0 & 1 & 0 & 1 & 0 & 1 & 0 & 0 & 1  & 0  & 1  & 0  & 1  & 0  & 1 
 \\ \hline
\end{array}  {~~~~~~~~~~~~~~~} 
$$
$~~~~~$ $~~~~~~$ {\bf {Table}} {\bf {16:}} Gadget
Values For ${\bm {\cal P}_{[1]}}$ $\times$ ${\bm {\cal P}_{[1]}}$ With Different Boolean
Factors

$$
\begin{array}{cc|cccccccccccccccc}
{~} & {\cal S}^{\prime} & 1 & 2  & 3 & 4 & 5 & 6 & 7 & 8  & 9 & 10 & 11 & 12 & 
13 & 14 & 15 & 16  \\
{\cal S}  & {} & {} & {} & {} & {} & {} & {} & {} & {} & {} & {} & {} & {} & {} & {} &  ${~}$ 
 \\ \hline
1  &  {} &0 & 1 & 0 & 1 & 1 & 0 & 1 & 0 & 0 & 1  & 0  & 1  & 1  & 0  & 1  & 0  \\
2  &  {} &1 & 0 & 1 & 0 & 0 & 1 & 0 & 1 & 1 & 0  & 1  & 0  & 0  & 1  & 0  & 1  \\
3  &  {} &0 & 1 & 0 & 1 & 1 & 0 & 1 & 0 & 0 & 1  & 0  & 1  & 1  & 0  & 1  & 0  \\
4  &  {} &1 & 0 & 1 & 0 & 0 & 1 & 0 & 1 & 1 & 0  & 1  & 0  & 0  & 1  & 0  & 1  \\
5  &  {} &0 & 1 & 0 & 1 & 1 & 0 & 1 & 0 & 0 & 1  & 0  & 1  & 1  & 0  & 1  & 0  \\
6  &  {} &1 & 0 & 1 & 0 & 0 & 1 & 0 & 1 & 1 & 0  & 1  & 0  & 0  & 1  & 0  & 1  \\
7  &  {} &0 & 1 & 0 & 1 & 1 & 0 & 1 & 0 & 0 & 1  & 0  & 1  & 1  & 0  & 1  & 0  \\
8  &  {} &1 & 0 & 1 & 0 & 0 & 1 & 0 & 1 & 1 & 0  & 1  & 0  & 0  & 1  & 0  & 1  \\
9  &  {} &1 & 0 & 1 & 0 & 0 & 1 & 0 & 1 & 1 & 0  & 1  & 0  & 0  & 1  & 0  & 1  \\
10 &  {} &0 & 1 & 0 & 1 & 1 & 0 & 1 & 0 & 0 & 1  & 0  & 1  & 1  & 0  & 1  & 0  \\
11 &  {} &1 & 0 & 1 & 0 & 0 & 1 & 0 & 1 & 1 & 0  & 1  & 0  & 0  & 1  & 0  & 1  \\
12 &  {} &0 & 1 & 0 & 1 & 1 & 0 & 1 & 0 & 0 & 1  & 0  & 1  & 1  & 0  & 1  & 0  \\
13 &  {} &1 & 0 & 1 & 0 & 0 & 1 & 0 & 1 & 1 & 0  & 1  & 0  & 0  & 1  & 0  & 1  \\
14 &  {} &0 & 1 & 0 & 1 & 1 & 0 & 1 & 0 & 0 & 1  & 0  & 1  & 1  & 0  & 1  & 0  \\
15 &  {} &1 & 0 & 1 & 0 & 0 & 1 & 0 & 1 & 1 & 0  & 1  & 0  & 0  & 1  & 0  & 1  \\
16 &  {} &0 & 1 & 0 & 1 & 1 & 0 & 1 & 0 & 0 & 1  & 0  & 1  & 1  & 0  & 1  & 0 
 \\ \hline
\end{array}  {~~~~~~~~~~~~~~~} 
$$
$~~~~~$ $~~~~~~$ {\bf {Table}} {\bf {17:}} Gadget
Values For ${\bm {\cal P}_{[1]}}$ $\times$ ${\bm {\cal P}_{[2]}}$ With Different Boolean
Factors

$$
\begin{array}{cc|cccccccccccccccc}
{~} & {\cal S}^{\prime} & 1 & 2  & 3 & 4 & 5 & 6 & 7 & 8  & 9 & 10 & 11 & 12 & 
13 & 14 & 15 & 16  \\
{\cal S}  & {} & {} & {} & {} & {} & {} & {} & {} & {} & {} & {} & {} & {} & {} & {} &  {}  \\ \hline
1  & {} & 0    & -{\fracm 13} & 0    & -{\fracm 13} & 0    & -{\fracm 13} & 0    & -{\fracm 13} & -{\fracm 13} & 0    & -{\fracm 13} & 0    & -{\fracm 13} & 0    & -{\fracm 13} & 0    \\
2  & {} & -{\fracm 13} & 0    & -{\fracm 13} & 0    & -{\fracm 13} & 0    & -{\fracm 13} & 0    & 0    & -{\fracm 13} & 0    & -{\fracm 13} & 0    & -{\fracm 13} & 0    & -{\fracm 13} \\
3  & {} & 0    & -{\fracm 13} & 0    & -{\fracm 13} & 0    & -{\fracm 13} & 0    & -{\fracm 13} & -{\fracm 13} & 0    & -{\fracm 13} & 0    & -{\fracm 13} & 0    & -{\fracm 13} & 0    \\
4  & {} & -{\fracm 13} & 0    & -{\fracm 13} & 0    & -{\fracm 13} & 0    & -{\fracm 13} & 0    & 0    & -{\fracm 13} & 0    & -{\fracm 13} & 0    & -{\fracm 13} & 0    & -{\fracm 13} \\
5  & {} & 0    & -{\fracm 13} & 0    & -{\fracm 13} & 0    & -{\fracm 13} & 0    & -{\fracm 13} & -{\fracm 13} & 0    & -{\fracm 13} & 0    & -{\fracm 13} & 0    & -{\fracm 13} & 0    \\
6  & {} & -{\fracm 13} & 0    & -{\fracm 13} & 0    & -{\fracm 13} & 0    & -{\fracm 13} & 0    & 0    & -{\fracm 13} & 0    & -{\fracm 13} & 0    & -{\fracm 13} & 0    & -{\fracm 13} \\
\hline
\end{array}  {~~~~~~~~~~~~~~~} 
$$
$~~~~~$ $~~~~~~$ {\bf {Table}} {\bf {18:}} Gadget
Values For ${\bm {\cal P}_{[1]}}$ $\times$ ${\bm {\cal P}_{[3]}}$ With Different Boolean
Factors

$$
\begin{array}{cc|cccccccccccccccc}
{~} & {\cal S}^{\prime} & 1 & 2  & 3 & 4 & 5 & 6 & 7 & 8  & 9 & 10 & 11 & 12 & 
13 & 14 & 15 & 16  \\
{\cal S}  & {} & {} & {} & {} & {} & {} & {} & {} & {} & {} & {} & {} & {} & {} & {} &  {} 
 \\ \hline
1  & {} & 1 & 0 & 1 & 0 & 0 & 1 & 0 & 1 & 1 & 0  & 1  & 0  & 0  & 1  & 0  & 1  \\
2  & {} & 0 & 1 & 0 & 1 & 1 & 0 & 1 & 0 & 0 & 1  & 0  & 1  & 1  & 0  & 1  & 0  \\
3  & {} & 1 & 0 & 1 & 0 & 0 & 1 & 0 & 1 & 1 & 0  & 1  & 0  & 0  & 1  & 0  & 1  \\
4  & {} & 0 & 1 & 0 & 1 & 1 & 0 & 1 & 0 & 0 & 1  & 0  & 1  & 1  & 0  & 1  & 0  \\
5  & {} & 1 & 0 & 1 & 0 & 0 & 1 & 0 & 1 & 1 & 0  & 1  & 0  & 0  & 1  & 0  & 1  \\
6  & {} & 0 & 1 & 0 & 1 & 1 & 0 & 1 & 0 & 0 & 1  & 0  & 1  & 1  & 0  & 1  & 0  \\
7  & {} & 1 & 0 & 1 & 0 & 0 & 1 & 0 & 1 & 1 & 0  & 1  & 0  & 0  & 1  & 0  & 1  \\
8  & {} & 0 & 1 & 0 & 1 & 1 & 0 & 1 & 0 & 0 & 1  & 0  & 1  & 1  & 0  & 1  & 0  \\
9  & {} & 0 & 1 & 0 & 1 & 1 & 0 & 1 & 0 & 0 & 1  & 0  & 1  & 1  & 0  & 1  & 0  \\
10 & {} & 1 & 0 & 1 & 0 & 0 & 1 & 0 & 1 & 1 & 0  & 1  & 0  & 0  & 1  & 0  & 1  \\
11 & {} & 0 & 1 & 0 & 1 & 1 & 0 & 1 & 0 & 0 & 1  & 0  & 1  & 1  & 0  & 1  & 0  \\
12 & {} & 1 & 0 & 1 & 0 & 0 & 1 & 0 & 1 & 1 & 0  & 1  & 0  & 0  & 1  & 0  & 1  \\
13 & {} & 0 & 1 & 0 & 1 & 1 & 0 & 1 & 0 & 0 & 1  & 0  & 1  & 1  & 0  & 1  & 0  \\
14 & {} & 1 & 0 & 1 & 0 & 0 & 1 & 0 & 1 & 1 & 0  & 1  & 0  & 0  & 1  & 0  & 1  \\
15 & {} & 0 & 1 & 0 & 1 & 1 & 0 & 1 & 0 & 0 & 1  & 0  & 1  & 1  & 0  & 1  & 0  \\
16 & {} & 1 & 0 & 1 & 0 & 0 & 1 & 0 & 1 & 1 & 0  & 1  & 0  & 0  & 1  & 0  & 1
 \\ \hline
\end{array}  {~~~~~~~~~~~~~~~} 
$$
$~~~~~$ $~~~~~~$ {\bf {Table}} {\bf {19:}} Gadget
Values For ${\bm {\cal P}_{[1]}}$ $\times$ ${\bm {\cal P}_{[4]}}$ With Different Boolean
Factors

$$
\begin{array}{cc|cccccccccccccccc}
{~} & {\cal S}^{\prime} & 1 & 2  & 3 & 4 & 5 & 6 & 7 & 8  & 9 & 10 & 11 & 12 & 
13 & 14 & 15 & 16  \\
{\cal S}  & {} & {} & {} & {} & {} & {} & {} & {} & {} & {} & {} & {} & {} & {} & {} &  {}  \\ 
\hline
1  & {} & 1 & 0 & 1 & 0 & 1 & 0 & 1 & 0 & 0 & 1  & 0  & 1  & 0  & 1  & 0  & 1  \\
2  & {} & 0 & 1 & 0 & 1 & 0 & 1 & 0 & 1 & 1 & 0  & 1  & 0  & 1  & 0  & 1  & 0  \\
3  & {} & 1 & 0 & 1 & 0 & 1 & 0 & 1 & 0 & 0 & 1  & 0  & 1  & 0  & 1  & 0  & 1  \\
4  & {} & 0 & 1 & 0 & 1 & 0 & 1 & 0 & 1 & 1 & 0  & 1  & 0  & 1  & 0  & 1  & 0  \\
5  & {} & 1 & 0 & 1 & 0 & 1 & 0 & 1 & 0 & 0 & 1  & 0  & 1  & 0  & 1  & 0  & 1  \\
6  & {} & 0 & 1 & 0 & 1 & 0 & 1 & 0 & 1 & 1 & 0  & 1  & 0  & 1  & 0  & 1  & 0  \\
7  & {} & 1 & 0 & 1 & 0 & 1 & 0 & 1 & 0 & 0 & 1  & 0  & 1  & 0  & 1  & 0  & 1  \\
8  & {} & 0 & 1 & 0 & 1 & 0 & 1 & 0 & 1 & 1 & 0  & 1  & 0  & 1  & 0  & 1  & 0  \\
9  & {} & 0 & 1 & 0 & 1 & 0 & 1 & 0 & 1 & 1 & 0  & 1  & 0  & 1  & 0  & 1  & 0  \\
10 & {} & 1 & 0 & 1 & 0 & 1 & 0 & 1 & 0 & 0 & 1  & 0  & 1  & 0  & 1  & 0  & 1  \\
11 & {} & 0 & 1 & 0 & 1 & 0 & 1 & 0 & 1 & 1 & 0  & 1  & 0  & 1  & 0  & 1  & 0  \\
12 & {} & 1 & 0 & 1 & 0 & 1 & 0 & 1 & 0 & 0 & 1  & 0  & 1  & 0  & 1  & 0  & 1  \\
13 & {} & 0 & 1 & 0 & 1 & 0 & 1 & 0 & 1 & 1 & 0  & 1  & 0  & 1  & 0  & 1  & 0  \\
14 & {} & 1 & 0 & 1 & 0 & 1 & 0 & 1 & 0 & 0 & 1  & 0  & 1  & 0  & 1  & 0  & 1  \\
15 & {} & 0 & 1 & 0 & 1 & 0 & 1 & 0 & 1 & 1 & 0  & 1  & 0  & 1  & 0  & 1  & 0  \\
16 & {} & 1 & 0 & 1 & 0 & 1 & 0 & 1 & 0 & 0 & 1  & 0  & 1  & 0  & 1  & 0  & 1 
 \\ \hline
\end{array}  {~~~~~~~~~~~~~~~} 
$$
$~~~~~$ $~~~~~~$ {\bf {Table}} {\bf {20:}} Gadget
Values For ${\bm {\cal P}_{[1]}}$ $\times$ ${\bm {\cal P}_{[5]}}$ With Different Boolean
Factors

$$
\begin{array}{cc|cccccccccccccccc}
{~} & {\cal S}^{\prime} & 1 & 2  & 3 & 4 & 5 & 6 & 7 & 8  & 9 & 10 & 11 & 12 & 
13 & 14 & 15 & 16  \\
{\cal S}  & {} & {} & {} & {} & {} & {} & {} & {} & {} & {} & {} & {} & {} & {} & {} &  {} 
 \\ \hline
1  & {} & 0 & 1 & 0 & 1 & 0 & 1 & 0 & 1 & 1 & 0  & 1  & 0  & 1  & 0  & 1  & 0  \\
2  & {} & 1 & 0 & 1 & 0 & 1 & 0 & 1 & 0 & 0 & 1  & 0  & 1  & 0  & 1  & 0  & 1  \\
3  & {} & 0 & 1 & 0 & 1 & 0 & 1 & 0 & 1 & 1 & 0  & 1  & 0  & 1  & 0  & 1  & 0  \\
4  & {} & 1 & 0 & 1 & 0 & 1 & 0 & 1 & 0 & 0 & 1  & 0  & 1  & 0  & 1  & 0  & 1  \\
5  & {} & 0 & 1 & 0 & 1 & 0 & 1 & 0 & 1 & 1 & 0  & 1  & 0  & 1  & 0  & 1  & 0  \\
6  & {} & 1 & 0 & 1 & 0 & 1 & 0 & 1 & 0 & 0 & 1  & 0  & 1  & 0  & 1  & 0  & 1  \\
7  & {} & 0 & 1 & 0 & 1 & 0 & 1 & 0 & 1 & 1 & 0  & 1  & 0  & 1  & 0  & 1  & 0  \\
8  & {} & 1 & 0 & 1 & 0 & 1 & 0 & 1 & 0 & 0 & 1  & 0  & 1  & 0  & 1  & 0  & 1  \\
9  & {} & 1 & 0 & 1 & 0 & 1 & 0 & 1 & 0 & 0 & 1  & 0  & 1  & 0  & 1  & 0  & 1  \\
10 & {} & 0 & 1 & 0 & 1 & 0 & 1 & 0 & 1 & 1 & 0  & 1  & 0  & 1  & 0  & 1  & 0  \\
11 & {} & 1 & 0 & 1 & 0 & 1 & 0 & 1 & 0 & 0 & 1  & 0  & 1  & 0  & 1  & 0  & 1  \\
12 & {} & 0 & 1 & 0 & 1 & 0 & 1 & 0 & 1 & 1 & 0  & 1  & 0  & 1  & 0  & 1  & 0  \\
13 & {} & 1 & 0 & 1 & 0 & 1 & 0 & 1 & 0 & 0 & 1  & 0  & 1  & 0  & 1  & 0  & 1  \\
14 & {} & 0 & 1 & 0 & 1 & 0 & 1 & 0 & 1 & 1 & 0  & 1  & 0  & 1  & 0  & 1  & 0  \\
15 & {} & 1 & 0 & 1 & 0 & 1 & 0 & 1 & 0 & 0 & 1  & 0  & 1  & 0  & 1  & 0  & 1  \\
16 & {} & 0 & 1 & 0 & 1 & 0 & 1 & 0 & 1 & 1 & 0  & 1  & 0  & 1  & 0  & 1  & 0 
 \\ \hline
\end{array}  {~~~~~~~~~~~~~~~} 
$$
$~~~~~$ $~~~~~~$ {\bf {Table}} {\bf {21:}} Gadget
Values For ${\bm {\cal P}_{[1]}}$ $\times$ ${\bm {\cal P}_{[6]}}$ With Different Boolean
Factors

$$
\begin{array}{cc|cccccccccccccccc}
{~} & {\cal S}^{\prime} & 1 & 2  & 3 & 4 & 5 & 6 & 7 & 8  & 9 & 10 & 11 & 12 & 
13 & 14 & 15 & 16  \\
{\cal S}  & {} & {} & {} & {} & {} & {} & {} & {} & {} & {} & {} & {} & {} & {} & {} &  {}  \\
 \hline
1  & {} & 1 & 0 & 1 & 0 & 0 & 1 & 0 & 1 & 1 & 0  & 1  & 0  & 0  & 1  & 0  & 1  \\
2  & {} & 0 & 1 & 0 & 1 & 1 & 0 & 1 & 0 & 0 & 1  & 0  & 1  & 1  & 0  & 1  & 0  \\
3  & {} & 1 & 0 & 1 & 0 & 0 & 1 & 0 & 1 & 1 & 0  & 1  & 0  & 0  & 1  & 0  & 1  \\
4  & {} & 0 & 1 & 0 & 1 & 1 & 0 & 1 & 0 & 0 & 1  & 0  & 1  & 1  & 0  & 1  & 0  \\
5  & {} & 0 & 1 & 0 & 1 & 1 & 0 & 1 & 0 & 0 & 1  & 0  & 1  & 1  & 0  & 1  & 0  \\
6  & {} & 1 & 0 & 1 & 0 & 0 & 1 & 0 & 1 & 1 & 0  & 1  & 0  & 0  & 1  & 0  & 1  \\
7  & {} & 0 & 1 & 0 & 1 & 1 & 0 & 1 & 0 & 0 & 1  & 0  & 1  & 1  & 0  & 1  & 0  \\
8  & {} & 1 & 0 & 1 & 0 & 0 & 1 & 0 & 1 & 1 & 0  & 1  & 0  & 0  & 1  & 0  & 1  \\
9  & {} & 1 & 0 & 1 & 0 & 0 & 1 & 0 & 1 & 1 & 0  & 1  & 0  & 0  & 1  & 0  & 1  \\
10 & {} & 0 & 1 & 0 & 1 & 1 & 0 & 1 & 0 & 0 & 1  & 0  & 1  & 1  & 0  & 1  & 0  \\
11 & {} & 1 & 0 & 1 & 0 & 0 & 1 & 0 & 1 & 1 & 0  & 1  & 0  & 0  & 1  & 0  & 1  \\
12 & {} & 0 & 1 & 0 & 1 & 1 & 0 & 1 & 0 & 0 & 1  & 0  & 1  & 1  & 0  & 1  & 0  \\
13 & {} & 0 & 1 & 0 & 1 & 1 & 0 & 1 & 0 & 0 & 1  & 0  & 1  & 1  & 0  & 1  & 0  \\
14 & {} & 1 & 0 & 1 & 0 & 0 & 1 & 0 & 1 & 1 & 0  & 1  & 0  & 0  & 1  & 0  & 1  \\
15 & {} & 0 & 1 & 0 & 1 & 1 & 0 & 1 & 0 & 0 & 1  & 0  & 1  & 1  & 0  & 1  & 0  \\
16 & {} & 1 & 0 & 1 & 0 & 0 & 1 & 0 & 1 & 1 & 0  & 1  & 0  & 0  & 1  & 0  & 1 
 \\ \hline
\end{array}  {~~~~~~~~~~~~~~~} 
$$
$~~~~~$ $~~~~~~$ {\bf {Table}} {\bf {21:}} Gadget
Values For ${\bm {\cal P}_{[2]}}$ $\times$ ${\bm {\cal P}_{[2]}}$ With Different Boolean
Factors

$$
\begin{array}{cc|cccccccccccccccc}
{~} & {\cal S}^{\prime} & 1 & 2  & 3 & 4 & 5 & 6 & 7 & 8  & 9 & 10 & 11 & 12 & 
13 & 14 & 15 & 16  \\
{\cal S}  & {} & {} & {} & {} & {} & {} & {} & {} & {} & {} & {} & {} & {} & {} & {} &  {}  \\ \hline
1  & {} & -{\fracm 13} & 0    & -{\fracm 13} & 0    & -{\fracm 13} & 0    & -{\fracm 13} & 0    & 0    & -{\fracm 13} & 0    & -{\fracm 13} & 0    & -{\fracm 13} & 0    & -{\fracm 13} \\
2  & {} & 0    & -{\fracm 13} & 0    & -{\fracm 13} & 0    & -{\fracm 13} & 0    & -{\fracm 13} & -{\fracm 13} & 0    & -{\fracm 13} & 0    & -{\fracm 13} & 0    & -{\fracm 13} & 0    \\
3  & {} & -{\fracm 13} & 0    & -{\fracm 13} & 0    & -{\fracm 13} & 0    & -{\fracm 13} & 0    & 0    & -{\fracm 13} & 0    & -{\fracm 13} & 0    & -{\fracm 13} & 0    & -{\fracm 13} \\
4  & {} & 0    & -{\fracm 13} & 0    & -{\fracm 13} & 0    & -{\fracm 13} & 0    & -{\fracm 13} & -{\fracm 13} & 0    & -{\fracm 13} & 0    & -{\fracm 13} & 0    & -{\fracm 13} & 0    \\
5  & {} & 0    & -{\fracm 13} & 0    & -{\fracm 13} & 0    & -{\fracm 13} & 0    & -{\fracm 13} & -{\fracm 13} & 0    & -{\fracm 13} & 0    & -{\fracm 13} & 0    & -{\fracm 13} & 0    \\
6  & {} & -{\fracm 13} & 0    & -{\fracm 13} & 0    & -{\fracm 13} & 0    & -{\fracm 13} & 0    & 0    & -{\fracm 13} & 0    & -{\fracm 13} & 0    & -{\fracm 13} & 0    & -{\fracm 13} \\
\hline
\end{array}  {~~~~~~~~~~~~~~~} 
$$
$~~~~~$ $~~~~~~$ {\bf {Table}} {\bf {21:}} Gadget
Values For ${\bm {\cal P}_{[2]}}$ $\times$ ${\bm {\cal P}_{[3]}}$ With Different Boolean
Factors

$$
\begin{array}{cc|cccccccccccccccc}
{~} & {\cal S}^{\prime} & 1 & 2  & 3 & 4 & 5 & 6 & 7 & 8  & 9 & 10 & 11 & 12 & 
13 & 14 & 15 & 16  \\
{\cal S}  & {} & {} & {} & {} & {} & {} & {} & {} & {} & {} & {} & {} & {} & {} & {} &  {}  \\ 
\hline
1  & {} & 0 & 1 & 0 & 1 & 1 & 0 & 1 & 0 & 0 & 1  & 0  & 1  & 1  & 0  & 1  & 0  \\
2  & {} & 1 & 0 & 1 & 0 & 0 & 1 & 0 & 1 & 1 & 0  & 1  & 0  & 0  & 1  & 0  & 1  \\
3  & {} & 0 & 1 & 0 & 1 & 1 & 0 & 1 & 0 & 0 & 1  & 0  & 1  & 1  & 0  & 1  & 0  \\
4  & {} & 1 & 0 & 1 & 0 & 0 & 1 & 0 & 1 & 1 & 0  & 1  & 0  & 0  & 1  & 0  & 1  \\
5  & {} & 1 & 0 & 1 & 0 & 0 & 1 & 0 & 1 & 1 & 0  & 1  & 0  & 0  & 1  & 0  & 1  \\
6  & {} & 0 & 1 & 0 & 1 & 1 & 0 & 1 & 0 & 0 & 1  & 0  & 1  & 1  & 0  & 1  & 0  \\
7  & {} & 1 & 0 & 1 & 0 & 0 & 1 & 0 & 1 & 1 & 0  & 1  & 0  & 0  & 1  & 0  & 1  \\
8  & {} & 0 & 1 & 0 & 1 & 1 & 0 & 1 & 0 & 0 & 1  & 0  & 1  & 1  & 0  & 1  & 0  \\
9  & {} & 0 & 1 & 0 & 1 & 1 & 0 & 1 & 0 & 0 & 1  & 0  & 1  & 1  & 0  & 1  & 0  \\
10 & {} & 1 & 0 & 1 & 0 & 0 & 1 & 0 & 1 & 1 & 0  & 1  & 0  & 0  & 1  & 0  & 1  \\
11 & {} & 0 & 1 & 0 & 1 & 1 & 0 & 1 & 0 & 0 & 1  & 0  & 1  & 1  & 0  & 1  & 0  \\
12 & {} & 1 & 0 & 1 & 0 & 0 & 1 & 0 & 1 & 1 & 0  & 1  & 0  & 0  & 1  & 0  & 1  \\
13 & {} & 1 & 0 & 1 & 0 & 0 & 1 & 0 & 1 & 1 & 0  & 1  & 0  & 0  & 1  & 0  & 1  \\
14 & {} & 0 & 1 & 0 & 1 & 1 & 0 & 1 & 0 & 0 & 1  & 0  & 1  & 1  & 0  & 1  & 0  \\
15 & {} & 1 & 0 & 1 & 0 & 0 & 1 & 0 & 1 & 1 & 0  & 1  & 0  & 0  & 1  & 0  & 1  \\
16 & {} & 0 & 1 & 0 & 1 & 1 & 0 & 1 & 0 & 0 & 1  & 0  & 1  & 1  & 0  & 1  & 0 
 \\ \hline
\end{array}  {~~~~~~~~~~~~~~~}  
$$
$~~~~~$ $~~~~~~$ {\bf {Table}} {\bf {21:}} Gadget
Values For ${\bm {\cal P}_{[2]}}$ $\times$ ${\bm {\cal P}_{[4]}}$ With Different Boolean
Factors

$$
\begin{array}{cc|cccccccccccccccc}
{~} & {\cal S}^{\prime} & 1 & 2  & 3 & 4 & 5 & 6 & 7 & 8  & 9 & 10 & 11 & 12 & 
13 & 14 & 15 & 16  \\
{\cal S}  & {} & {} & {} & {} & {} & {} & {} & {} & {} & {} & {} & {} & {} & {} & {} &  {}  \\ 
\hline
1  & {} & 0 & 1 & 0 & 1 & 0 & 1 & 0 & 1 & 1 & 0  & 1  & 0  & 1  & 0  & 1  & 0  \\
2  & {} & 1 & 0 & 1 & 0 & 1 & 0 & 1 & 0 & 0 & 1  & 0  & 1  & 0  & 1  & 0  & 1  \\
3  & {} & 0 & 1 & 0 & 1 & 0 & 1 & 0 & 1 & 1 & 0  & 1  & 0  & 1  & 0  & 1  & 0  \\
4  & {} & 1 & 0 & 1 & 0 & 1 & 0 & 1 & 0 & 0 & 1  & 0  & 1  & 0  & 1  & 0  & 1  \\
5  & {} & 1 & 0 & 1 & 0 & 1 & 0 & 1 & 0 & 0 & 1  & 0  & 1  & 0  & 1  & 0  & 1  \\
6  & {} & 0 & 1 & 0 & 1 & 0 & 1 & 0 & 1 & 1 & 0  & 1  & 0  & 1  & 0  & 1  & 0  \\
7  & {} & 1 & 0 & 1 & 0 & 1 & 0 & 1 & 0 & 0 & 1  & 0  & 1  & 0  & 1  & 0  & 1  \\
8  & {} & 0 & 1 & 0 & 1 & 0 & 1 & 0 & 1 & 1 & 0  & 1  & 0  & 1  & 0  & 1  & 0  \\
9  & {} & 0 & 1 & 0 & 1 & 0 & 1 & 0 & 1 & 1 & 0  & 1  & 0  & 1  & 0  & 1  & 0  \\
10 & {} & 1 & 0 & 1 & 0 & 1 & 0 & 1 & 0 & 0 & 1  & 0  & 1  & 0  & 1  & 0  & 1  \\
11 & {} & 0 & 1 & 0 & 1 & 0 & 1 & 0 & 1 & 1 & 0  & 1  & 0  & 1  & 0  & 1  & 0  \\
12 & {} & 1 & 0 & 1 & 0 & 1 & 0 & 1 & 0 & 0 & 1  & 0  & 1  & 0  & 1  & 0  & 1  \\
13 & {} & 1 & 0 & 1 & 0 & 1 & 0 & 1 & 0 & 0 & 1  & 0  & 1  & 0  & 1  & 0  & 1  \\
14 & {} & 0 & 1 & 0 & 1 & 0 & 1 & 0 & 1 & 1 & 0  & 1  & 0  & 1  & 0  & 1  & 0  \\
15 & {} & 1 & 0 & 1 & 0 & 1 & 0 & 1 & 0 & 0 & 1  & 0  & 1  & 0  & 1  & 0  & 1  \\
16 & {} & 0 & 1 & 0 & 1 & 0 & 1 & 0 & 1 & 1 & 0  & 1  & 0  & 1  & 0  & 1  & 0 
 \\ \hline
\end{array}  {~~~~~~~~~~~~~~~} 
$$
$~~~~~$ $~~~~~~$ {\bf {Table}} {\bf {22:}} Gadget
Values For ${\bm {\cal P}_{[2]}}$ $\times$ ${\bm {\cal P}_{[5]}}$ With Different Boolean
Factors

$$
\begin{array}{cc|cccccccccccccccc}
{~} & {\cal S}^{\prime} & 1 & 2  & 3 & 4 & 5 & 6 & 7 & 8  & 9 & 10 & 11 & 12 & 
13 & 14 & 15 & 16  \\
{\cal S}  & {} & {} & {} & {} & {} & {} & {} & {} & {} & {} & {} & {} & {} & {} & {} &  {}  \\ 
\hline
1  & {} & 1 & 0 & 1 & 0 & 1 & 0 & 1 & 0 & 0 & 1  & 0  & 1  & 0  & 1  & 0  & 1  \\
2  & {} & 0 & 1 & 0 & 1 & 0 & 1 & 0 & 1 & 1 & 0  & 1  & 0  & 1  & 0  & 1  & 0  \\
3  & {} & 1 & 0 & 1 & 0 & 1 & 0 & 1 & 0 & 0 & 1  & 0  & 1  & 0  & 1  & 0  & 1  \\
4  & {} & 0 & 1 & 0 & 1 & 0 & 1 & 0 & 1 & 1 & 0  & 1  & 0  & 1  & 0  & 1  & 0  \\
5  & {} & 0 & 1 & 0 & 1 & 0 & 1 & 0 & 1 & 1 & 0  & 1  & 0  & 1  & 0  & 1  & 0  \\
6  & {} & 1 & 0 & 1 & 0 & 1 & 0 & 1 & 0 & 0 & 1  & 0  & 1  & 0  & 1  & 0  & 1  \\
7  & {} & 0 & 1 & 0 & 1 & 0 & 1 & 0 & 1 & 1 & 0  & 1  & 0  & 1  & 0  & 1  & 0  \\
8  & {} & 1 & 0 & 1 & 0 & 1 & 0 & 1 & 0 & 0 & 1  & 0  & 1  & 0  & 1  & 0  & 1  \\
9  & {} & 1 & 0 & 1 & 0 & 1 & 0 & 1 & 0 & 0 & 1  & 0  & 1  & 0  & 1  & 0  & 1  \\
10 & {} & 0 & 1 & 0 & 1 & 0 & 1 & 0 & 1 & 1 & 0  & 1  & 0  & 1  & 0  & 1  & 0  \\
11 & {} & 1 & 0 & 1 & 0 & 1 & 0 & 1 & 0 & 0 & 1  & 0  & 1  & 0  & 1  & 0  & 1  \\
12 & {} & 0 & 1 & 0 & 1 & 0 & 1 & 0 & 1 & 1 & 0  & 1  & 0  & 1  & 0  & 1  & 0  \\
13 & {} & 0 & 1 & 0 & 1 & 0 & 1 & 0 & 1 & 1 & 0  & 1  & 0  & 1  & 0  & 1  & 0  \\
14 & {} & 1 & 0 & 1 & 0 & 1 & 0 & 1 & 0 & 0 & 1  & 0  & 1  & 0  & 1  & 0  & 1  \\
15 & {} & 0 & 1 & 0 & 1 & 0 & 1 & 0 & 1 & 1 & 0  & 1  & 0  & 1  & 0  & 1  & 0  \\
16 & {} & 1 & 0 & 1 & 0 & 1 & 0 & 1 & 0 & 0 & 1  & 0  & 1  & 0  & 1  & 0  & 1 
 \\ \hline
\end{array}  {~~~~~~~~~~~~~~~} 
$$
$~~~~~$ $~~~~~~$ {\bf {Table}} {\bf {23:}} Gadget
Values For ${\bm {\cal P}_{[2]}}$ $\times$ ${\bm {\cal P}_{[6]}}$ With Different Boolean
Factors

$$
\begin{array}{cc|cccccccccccccccc}
{~} & {\cal S}^{\prime} & 1 & 2  & 3 & 4 & 5 & 6 & 7 & 8  & 9 & 10 & 11 & 12 & 
13 & 14 & 15 & 16  \\
{\cal S}  & {} & {} & {} & {} & {} & {} & {} & {} & {} & {} & {} & {} & {} & {} & {} &  {}  \\ 
\hline
1  & {} & 1 & 0 & 1 & 0 & 1 & 0 & 1 & 0 & 0 & 1  & 0  & 1  & 0  & 1  & 0  & 1  \\
2  & {} & 0 & 1 & 0 & 1 & 0 & 1 & 0 & 1 & 1 & 0  & 1  & 0  & 1  & 0  & 1  & 0  \\
3  & {} & 1 & 0 & 1 & 0 & 1 & 0 & 1 & 0 & 0 & 1  & 0  & 1  & 0  & 1  & 0  & 1  \\
4  & {} & 0 & 1 & 0 & 1 & 0 & 1 & 0 & 1 & 1 & 0  & 1  & 0  & 1  & 0  & 1  & 0  \\
5  & {} & 1 & 0 & 1 & 0 & 1 & 0 & 1 & 0 & 0 & 1  & 0  & 1  & 0  & 1  & 0  & 1  \\
6  & {} & 0 & 1 & 0 & 1 & 0 & 1 & 0 & 1 & 1 & 0  & 1  & 0  & 1  & 0  & 1  & 0  \\
7  & {} & 1 & 0 & 1 & 0 & 1 & 0 & 1 & 0 & 0 & 1  & 0  & 1  & 0  & 1  & 0  & 1  \\
8  & {} & 0 & 1 & 0 & 1 & 0 & 1 & 0 & 1 & 1 & 0  & 1  & 0  & 1  & 0  & 1  & 0  \\
9  & {} & 0 & 1 & 0 & 1 & 0 & 1 & 0 & 1 & 1 & 0  & 1  & 0  & 1  & 0  & 1  & 0  \\
10 & {} & 1 & 0 & 1 & 0 & 1 & 0 & 1 & 0 & 0 & 1  & 0  & 1  & 0  & 1  & 0  & 1  \\
11 & {} & 0 & 1 & 0 & 1 & 0 & 1 & 0 & 1 & 1 & 0  & 1  & 0  & 1  & 0  & 1  & 0  \\
12 & {} & 1 & 0 & 1 & 0 & 1 & 0 & 1 & 0 & 0 & 1  & 0  & 1  & 0  & 1  & 0  & 1  \\
13 & {} & 0 & 1 & 0 & 1 & 0 & 1 & 0 & 1 & 1 & 0  & 1  & 0  & 1  & 0  & 1  & 0  \\
14 & {} & 1 & 0 & 1 & 0 & 1 & 0 & 1 & 0 & 0 & 1  & 0  & 1  & 0  & 1  & 0  & 1  \\
15 & {} & 0 & 1 & 0 & 1 & 0 & 1 & 0 & 1 & 1 & 0  & 1  & 0  & 1  & 0  & 1  & 0  \\
16 & {} & 1 & 0 & 1 & 0 & 1 & 0 & 1 & 0 & 0 & 1  & 0  & 1  & 0  & 1  & 0  & 1 
 \\ \hline
\end{array}  {~~~~~~~~~~~~~~~} 
$$
$~~~~~$ $~~~~~~$ {\bf {Table}} {\bf {24:}} Gadget
Values For ${\bm {\cal P}_{[3]}}$ $\times$ ${\bm {\cal P}_{[3]}}$ With Different Boolean
Factors

$$
\begin{array}{cc|cccccccccccccccc}
{~} & {\cal S}^{\prime} & 1 & 2  & 3 & 4 & 5 & 6 & 7 & 8  & 9 & 10 & 11 & 12 & 
13 & 14 & 15 & 16  \\
{\cal S}  & {} & {} & {} & {} & {} & {} & {} & {} & {} & {} & {} & {} & {} & {} & {} &  {}  \\ \hline
1  & {} & 0    & -{\fracm 13} & 0    & -{\fracm 13} & -{\fracm 13} & 0    & -{\fracm 13} & 0    & 0    & -{\fracm 13} & 0    & -{\fracm 13} & -{\fracm 13} & 0    & -{\fracm 13} & 0    \\
2  & {} & -{\fracm 13} & 0    & -{\fracm 13} & 0    & 0    & -{\fracm 13} & 0    & -{\fracm 13} & -{\fracm 13} & 0    & -{\fracm 13} & 0    & 0    & -{\fracm 13} & 0    & -{\fracm 13} \\
3  & {} & 0    & -{\fracm 13} & 0    & -{\fracm 13} & -{\fracm 13} & 0    & -{\fracm 13} & 0    & 0    & -{\fracm 13} & 0    & -{\fracm 13} & -{\fracm 13} & 0    & -{\fracm 13} & 0    \\
4  & {} & -{\fracm 13} & 0    & -{\fracm 13} & 0    & 0    & -{\fracm 13} & 0    & -{\fracm 13} & -{\fracm 13} & 0    & -{\fracm 13} & 0    & 0    & -{\fracm 13} & 0    & -{\fracm 13} \\
5  & {} & 0    & -{\fracm 13} & 0    & -{\fracm 13} & -{\fracm 13} & 0    & -{\fracm 13} & 0    & 0    & -{\fracm 13} & 0    & -{\fracm 13} & -{\fracm 13} & 0    & -{\fracm 13} & 0    \\
6  & {} & -{\fracm 13} & 0    & -{\fracm 13} & 0    & 0    & -{\fracm 13} & 0    & -{\fracm 13} & -{\fracm 13} & 0    & -{\fracm 13} & 0    & 0    & -{\fracm 13} & 0    & -{\fracm 13} \\
\hline
\end{array}  {~~~~~~~~~~~~~~~} 
$$
$~~~~~$ $~~~~~~$ {\bf {Table}} {\bf {24:}} Gadget
Values For ${\bm {\cal P}_{[3]}}$ $\times$ ${\bm {\cal P}_{[4]}}$ With Different Boolean
Factors
$$
\begin{array}{cc|cccccccccccccccc}
{~} & {\cal S}^{\prime} & 1 & 2  & 3 & 4 & 5 & 6 & 7 & 8  & 9 & 10 & 11 & 12 & 
13 & 14 & 15 & 16  \\
{\cal S}  & {} & {} & {} & {} & {} & {} & {} & {} & {} & {} & {} & {} & {} & {} & {} &  {}  \\ \hline
1  & {} & 0    & -{\fracm 13} & 0    & -{\fracm 13} & 0    & -{\fracm 13} & 0    & -{\fracm 13} & -{\fracm 13} & 0    & -{\fracm 13} & 0    & -{\fracm 13} & 0    & -{\fracm 13} & 0    \\
2  & {} & -{\fracm 13} & 0    & -{\fracm 13} & 0    & -{\fracm 13} & 0    & -{\fracm 13} & 0    & 0    & -{\fracm 13} & 0    & -{\fracm 13} & 0    & -{\fracm 13} & 0    & -{\fracm 13} \\
3  & {} & 0    & -{\fracm 13} & 0    & -{\fracm 13} & 0    & -{\fracm 13} & 0    & -{\fracm 13} & -{\fracm 13} & 0    & -{\fracm 13} & 0    & -{\fracm 13} & 0    & -{\fracm 13} & 0    \\
4  & {} & -{\fracm 13} & 0    & -{\fracm 13} & 0    & -{\fracm 13} & 0    & -{\fracm 13} & 0    & 0    & -{\fracm 13} & 0    & -{\fracm 13} & 0    & -{\fracm 13} & 0    & -{\fracm 13} \\
5  & {} & 0    & -{\fracm 13} & 0    & -{\fracm 13} & 0    & -{\fracm 13} & 0    & -{\fracm 13} & -{\fracm 13} & 0    & -{\fracm 13} & 0    & -{\fracm 13} & 0    & -{\fracm 13} & 0    \\
6  & {} & -{\fracm 13} & 0    & -{\fracm 13} & 0    & -{\fracm 13} & 0    & -{\fracm 13} & 0    & 0    & -{\fracm 13} & 0    & -{\fracm 13} & 0    & -{\fracm 13} & 0    & -{\fracm 13} \\
\hline
\end{array}  {~~~~~~~~~~~~~~~} 
$$
$~~~~~$ $~~~~~~$ {\bf {Table}} {\bf {25:}} Gadget
Values For ${\bm {\cal P}_{[3]}}$ $\times$ ${\bm {\cal P}_{[5]}}$ With Different Boolean
Factors

$$
\begin{array}{cc|cccccccccccccccc}
{~} & {\cal S}^{\prime} & 1 & 2  & 3 & 4 & 5 & 6 & 7 & 8  & 9 & 10 & 11 & 12 & 
13 & 14 & 15 & 16  \\
{\cal S}  & {} & {} & {} & {} & {} & {} & {} & {} & {} & {} & {} & {} & {} & {} & {} &  {}  \\ 
\hline
1  & {} & -{\fracm 13} & 0    & -{\fracm 13} & 0    & -{\fracm 13} & 0    & -{\fracm 13} & 0    & 0    & -{\fracm 13} & 0    & -{\fracm 13} & 0    & -{\fracm 13} & 0    & -{\fracm 13} \\
2  & {} & 0    & -{\fracm 13} & 0    & -{\fracm 13} & 0    & -{\fracm 13} & 0    & -{\fracm 13} & -{\fracm 13} & 0    & -{\fracm 13} & 0    & -{\fracm 13} & 0    & -{\fracm 13} & 0    \\
3  & {} & -{\fracm 13} & 0    & -{\fracm 13} & 0    & -{\fracm 13} & 0    & -{\fracm 13} & 0    & 0    & -{\fracm 13} & 0    & -{\fracm 13} & 0    & -{\fracm 13} & 0    & -{\fracm 13} \\
4  & {} & 0    & -{\fracm 13} & 0    & -{\fracm 13} & 0    & -{\fracm 13} & 0    & -{\fracm 13} & -{\fracm 13} & 0    & -{\fracm 13} & 0    & -{\fracm 13} & 0    & -{\fracm 13} & 0    \\
5  & {} & -{\fracm 13} & 0    & -{\fracm 13} & 0    & -{\fracm 13} & 0    & -{\fracm 13} & 0    & 0    & -{\fracm 13} & 0    & -{\fracm 13} & 0    & -{\fracm 13} & 0    & -{\fracm 13} \\
6  & {} & 0    & -{\fracm 13} & 0    & -{\fracm 13} & 0    & -{\fracm 13} & 0    & -{\fracm 13} & -{\fracm 13} & 0    & -{\fracm 13} & 0    & -{\fracm 13} & 0    & -{\fracm 13} & 0    \\
\hline
\end{array}  {~~~~~~~~~~~~~~~} 
$$
$~~~~~$ $~~~~~~$ {\bf {Table}} {\bf {26:}} Gadget
Values For ${\bm {\cal P}_{[3]}}$ $\times$ ${\bm {\cal P}_{[6]}}$ With Different Boolean
Factors

$$
\begin{array}{cc|cccccccccccccccc}
{~} & {\cal S}^{\prime} & 1 & 2  & 3 & 4 & 5 & 6 & 7 & 8  & 9 & 10 & 11 & 12 & 
13 & 14 & 15 & 16  \\
{\cal S}  & {} & {} & {} & {} & {} & {} & {} & {} & {} & {} & {} & {} & {} & {} & {} &  {} \\
\hline
1  & {} & 1 & 0 & 1 & 0 & 0 & 1 & 0 & 1 & 1 & 0  & 1  & 0  & 0  & 1  & 0  & 1  \\
2  & {} & 0 & 1 & 0 & 1 & 1 & 0 & 1 & 0 & 0 & 1  & 0  & 1  & 1  & 0  & 1  & 0  \\
3  & {} & 1 & 0 & 1 & 0 & 0 & 1 & 0 & 1 & 1 & 0  & 1  & 0  & 0  & 1  & 0  & 1  \\
4  & {} & 0 & 1 & 0 & 1 & 1 & 0 & 1 & 0 & 0 & 1  & 0  & 1  & 1  & 0  & 1  & 0  \\
5  & {} & 0 & 1 & 0 & 1 & 1 & 0 & 1 & 0 & 0 & 1  & 0  & 1  & 1  & 0  & 1  & 0  \\
6  & {} & 1 & 0 & 1 & 0 & 0 & 1 & 0 & 1 & 1 & 0  & 1  & 0  & 0  & 1  & 0  & 1  \\
7  & {} & 0 & 1 & 0 & 1 & 1 & 0 & 1 & 0 & 0 & 1  & 0  & 1  & 1  & 0  & 1  & 0  \\
8  & {} & 1 & 0 & 1 & 0 & 0 & 1 & 0 & 1 & 1 & 0  & 1  & 0  & 0  & 1  & 0  & 1  \\
9  & {} & 1 & 0 & 1 & 0 & 0 & 1 & 0 & 1 & 1 & 0  & 1  & 0  & 0  & 1  & 0  & 1  \\
10 & {} & 0 & 1 & 0 & 1 & 1 & 0 & 1 & 0 & 0 & 1  & 0  & 1  & 1  & 0  & 1  & 0  \\
11 & {} & 1 & 0 & 1 & 0 & 0 & 1 & 0 & 1 & 1 & 0  & 1  & 0  & 0  & 1  & 0  & 1  \\
12 & {} & 0 & 1 & 0 & 1 & 1 & 0 & 1 & 0 & 0 & 1  & 0  & 1  & 1  & 0  & 1  & 0  \\
13 & {} & 0 & 1 & 0 & 1 & 1 & 0 & 1 & 0 & 0 & 1  & 0  & 1  & 1  & 0  & 1  & 0  \\
14 & {} & 1 & 0 & 1 & 0 & 0 & 1 & 0 & 1 & 1 & 0  & 1  & 0  & 0  & 1  & 0  & 1  \\
15 & {} & 0 & 1 & 0 & 1 & 1 & 0 & 1 & 0 & 0 & 1  & 0  & 1  & 1  & 0  & 1  & 0  \\
16 & {} & 1 & 0 & 1 & 0 & 0 & 1 & 0 & 1 & 1 & 0  & 1  & 0  & 0  & 1  & 0  & 1 
 \\ \hline
\end{array}  {~~~~~~~~~~~~~~~} 
$$
$~~~~~$ $~~~~~~$ {\bf {Table}} {\bf {27:}} Gadget
Values For ${\bm {\cal P}_{[4]}}$ $\times$ ${\bm {\cal P}_{[4]}}$ With Different Boolean
Factors

$$
\begin{array}{cc|cccccccccccccccc}
{~} & {\cal S}^{\prime} & 1 & 2  & 3 & 4 & 5 & 6 & 7 & 8  & 9 & 10 & 11 & 12 & 
13 & 14 & 15 & 16  \\
{\cal S}  & {} & {} & {} & {} & {} & {} & {} & {} & {} & {} & {} & {} & {} & {} & {} &  {} \\
\hline
1  & {}  & 1 & 0 & 1 & 0 & 1 & 0 & 1 & 0 & 0 & 1  & 0  & 1  & 0  & 1  & 0  & 1  \\
2  & {}  & 0 & 1 & 0 & 1 & 0 & 1 & 0 & 1 & 1 & 0  & 1  & 0  & 1  & 0  & 1  & 0  \\
3  & {}  & 1 & 0 & 1 & 0 & 1 & 0 & 1 & 0 & 0 & 1  & 0  & 1  & 0  & 1  & 0  & 1  \\
4  & {}  & 0 & 1 & 0 & 1 & 0 & 1 & 0 & 1 & 1 & 0  & 1  & 0  & 1  & 0  & 1  & 0  \\
5  & {}  & 0 & 1 & 0 & 1 & 0 & 1 & 0 & 1 & 1 & 0  & 1  & 0  & 1  & 0  & 1  & 0  \\
6  & {}  & 1 & 0 & 1 & 0 & 1 & 0 & 1 & 0 & 0 & 1  & 0  & 1  & 0  & 1  & 0  & 1  \\
7  & {}  & 0 & 1 & 0 & 1 & 0 & 1 & 0 & 1 & 1 & 0  & 1  & 0  & 1  & 0  & 1  & 0  \\
8  & {}  & 1 & 0 & 1 & 0 & 1 & 0 & 1 & 0 & 0 & 1  & 0  & 1  & 0  & 1  & 0  & 1  \\
9  & {}  & 1 & 0 & 1 & 0 & 1 & 0 & 1 & 0 & 0 & 1  & 0  & 1  & 0  & 1  & 0  & 1  \\
10 & {}  & 0 & 1 & 0 & 1 & 0 & 1 & 0 & 1 & 1 & 0  & 1  & 0  & 1  & 0  & 1  & 0  \\
11 & {}  & 1 & 0 & 1 & 0 & 1 & 0 & 1 & 0 & 0 & 1  & 0  & 1  & 0  & 1  & 0  & 1  \\
12 & {}  & 0 & 1 & 0 & 1 & 0 & 1 & 0 & 1 & 1 & 0  & 1  & 0  & 1  & 0  & 1  & 0  \\
13 & {}  & 0 & 1 & 0 & 1 & 0 & 1 & 0 & 1 & 1 & 0  & 1  & 0  & 1  & 0  & 1  & 0  \\
14 & {}  & 1 & 0 & 1 & 0 & 1 & 0 & 1 & 0 & 0 & 1  & 0  & 1  & 0  & 1  & 0  & 1  \\
15 & {}  & 0 & 1 & 0 & 1 & 0 & 1 & 0 & 1 & 1 & 0  & 1  & 0  & 1  & 0  & 1  & 0  \\
16 & {}  & 1 & 0 & 1 & 0 & 1 & 0 & 1 & 0 & 0 & 1  & 0  & 1  & 0  & 1  & 0  & 1 
 \\ \hline
\end{array}  {~~~~~~~~~~~~~~~} 
$$
$~~~~~$ $~~~~~~$ {\bf {Table}} {\bf {28:}} Gadget
Values For ${\bm {\cal P}_{[4]}}$ $\times$ ${\bm {\cal P}_{[5]}}$ With Different Boolean
Factors

$$
\begin{array}{cc|cccccccccccccccc}
{~} & {\cal S}^{\prime} & 1 & 2  & 3 & 4 & 5 & 6 & 7 & 8  & 9 & 10 & 11 & 12 & 
13 & 14 & 15 & 16  \\ 
{\cal S}  & {} & {} & {} & {} & {} & {} & {} & {} & {} & {} & {} & {} & {} & {} & {} &  {} \\
 \hline
1  & {}  & 0 & 1 & 0 & 1 & 0 & 1 & 0 & 1 & 1 & 0  & 1  & 0  & 1  & 0  & 1  & 0  \\
2  & {}  & 1 & 0 & 1 & 0 & 1 & 0 & 1 & 0 & 0 & 1  & 0  & 1  & 0  & 1  & 0  & 1  \\
3  & {}  & 0 & 1 & 0 & 1 & 0 & 1 & 0 & 1 & 1 & 0  & 1  & 0  & 1  & 0  & 1  & 0  \\
4  & {}  & 1 & 0 & 1 & 0 & 1 & 0 & 1 & 0 & 0 & 1  & 0  & 1  & 0  & 1  & 0  & 1  \\
5  & {}  & 1 & 0 & 1 & 0 & 1 & 0 & 1 & 0 & 0 & 1  & 0  & 1  & 0  & 1  & 0  & 1  \\
6  & {}  & 0 & 1 & 0 & 1 & 0 & 1 & 0 & 1 & 1 & 0  & 1  & 0  & 1  & 0  & 1  & 0  \\
7  & {}  & 1 & 0 & 1 & 0 & 1 & 0 & 1 & 0 & 0 & 1  & 0  & 1  & 0  & 1  & 0  & 1  \\
8  & {}  & 0 & 1 & 0 & 1 & 0 & 1 & 0 & 1 & 1 & 0  & 1  & 0  & 1  & 0  & 1  & 0  \\
9  & {}  & 0 & 1 & 0 & 1 & 0 & 1 & 0 & 1 & 1 & 0  & 1  & 0  & 1  & 0  & 1  & 0  \\
10 & {}  & 1 & 0 & 1 & 0 & 1 & 0 & 1 & 0 & 0 & 1  & 0  & 1  & 0  & 1  & 0  & 1  \\
11 & {}  & 0 & 1 & 0 & 1 & 0 & 1 & 0 & 1 & 1 & 0  & 1  & 0  & 1  & 0  & 1  & 0  \\
12 & {}  & 1 & 0 & 1 & 0 & 1 & 0 & 1 & 0 & 0 & 1  & 0  & 1  & 0  & 1  & 0  & 1  \\
13 & {}  & 1 & 0 & 1 & 0 & 1 & 0 & 1 & 0 & 0 & 1  & 0  & 1  & 0  & 1  & 0  & 1  \\
14 & {}  & 0 & 1 & 0 & 1 & 0 & 1 & 0 & 1 & 1 & 0  & 1  & 0  & 1  & 0  & 1  & 0  \\
15 & {}  & 1 & 0 & 1 & 0 & 1 & 0 & 1 & 0 & 0 & 1  & 0  & 1  & 0  & 1  & 0  & 1  \\
16 & {}  & 0 & 1 & 0 & 1 & 0 & 1 & 0 & 1 & 1 & 0  & 1  & 0  & 1  & 0  & 1  & 0 
 \\ \hline
\end{array}  {~~~~~~~~~~~~~~~} 
$$
$~~~~~$ $~~~~~~$ {\bf {Table}} {\bf {28:}} Gadget
Values For ${\bm {\cal P}_{[4]}}$ $\times$ ${\bm {\cal P}_{[6]}}$ With Different Boolean
Factors

$$
\begin{array}{cc|cccccccccccccccc}
{~} & {\cal S}^{\prime} & 1 & 2  & 3 & 4 & 5 & 6 & 7 & 8  & 9 & 10 & 11 & 12 & 
13 & 14 & 15 & 16  \\ \hline
{\cal S}  & {} & {} & {} & {} & {} & {} & {} & {} & {} & {} & {} & {} & {} & {} & {} &  {} \\
1  & {}  & 1 & 0 & 1 & 0 & 1 & 0 & 1 & 0 & 0 & 1  & 0  & 1  & 0  & 1  & 0  & 1  \\
2  & {}  & 0 & 1 & 0 & 1 & 0 & 1 & 0 & 1 & 1 & 0  & 1  & 0  & 1  & 0  & 1  & 0  \\
3  & {}  & 1 & 0 & 1 & 0 & 1 & 0 & 1 & 0 & 0 & 1  & 0  & 1  & 0  & 1  & 0  & 1  \\
4  & {}  & 0 & 1 & 0 & 1 & 0 & 1 & 0 & 1 & 1 & 0  & 1  & 0  & 1  & 0  & 1  & 0  \\
5  & {}  & 1 & 0 & 1 & 0 & 1 & 0 & 1 & 0 & 0 & 1  & 0  & 1  & 0  & 1  & 0  & 1  \\
6  & {}  & 0 & 1 & 0 & 1 & 0 & 1 & 0 & 1 & 1 & 0  & 1  & 0  & 1  & 0  & 1  & 0  \\
7  & {}  & 1 & 0 & 1 & 0 & 1 & 0 & 1 & 0 & 0 & 1  & 0  & 1  & 0  & 1  & 0  & 1  \\
8  & {}  & 0 & 1 & 0 & 1 & 0 & 1 & 0 & 1 & 1 & 0  & 1  & 0  & 1  & 0  & 1  & 0  \\
9  & {}  & 0 & 1 & 0 & 1 & 0 & 1 & 0 & 1 & 1 & 0  & 1  & 0  & 1  & 0  & 1  & 0  \\
10 & {}  & 1 & 0 & 1 & 0 & 1 & 0 & 1 & 0 & 0 & 1  & 0  & 1  & 0  & 1  & 0  & 1  \\
11 & {}  & 0 & 1 & 0 & 1 & 0 & 1 & 0 & 1 & 1 & 0  & 1  & 0  & 1  & 0  & 1  & 0  \\
12 & {}  & 1 & 0 & 1 & 0 & 1 & 0 & 1 & 0 & 0 & 1  & 0  & 1  & 0  & 1  & 0  & 1  \\
13 & {}  & 0 & 1 & 0 & 1 & 0 & 1 & 0 & 1 & 1 & 0  & 1  & 0  & 1  & 0  & 1  & 0  \\
14 & {}  & 1 & 0 & 1 & 0 & 1 & 0 & 1 & 0 & 0 & 1  & 0  & 1  & 0  & 1  & 0  & 1  \\
15 & {}  & 0 & 1 & 0 & 1 & 0 & 1 & 0 & 1 & 1 & 0  & 1  & 0  & 1  & 0  & 1  & 0  \\
16 & {}  & 1 & 0 & 1 & 0 & 1 & 0 & 1 & 0 & 0 & 1  & 0  & 1  & 0  & 1  & 0  & 1 
 \\ \hline
\end{array}  {~~~~~~~~~~~~~~~} 
$$
$~~~~~$ $~~~~~~$ {\bf {Table}} {\bf {28:}} Gadget
Values For ${\bm {\cal P}_{[5]}}$ $\times$ ${\bm {\cal P}_{[5]}}$ With Different Boolean
Factors

$$
\begin{array}{cc|cccccccccccccccc}
{~} & {\cal S}^{\prime} & 1 & 2  & 3 & 4 & 5 & 6 & 7 & 8  & 9 & 10 & 11 & 12 & 
13 & 14 & 15 & 16  \\
{\cal S}  & {} & {} & {} & {} & {} & {} & {} & {} & {} & {} & {} & {} & {} & {} & {} &  {} \\
\hline
1  & {}  & 0 & 1 & 0 & 1 & 0 & 1 & 0 & 1 & 1 & 0  & 1  & 0  & 1  & 0  & 1  & 0  \\
2  & {}  & 1 & 0 & 1 & 0 & 1 & 0 & 1 & 0 & 0 & 1  & 0  & 1  & 0  & 1  & 0  & 1  \\
3  & {}  & 0 & 1 & 0 & 1 & 0 & 1 & 0 & 1 & 1 & 0  & 1  & 0  & 1  & 0  & 1  & 0  \\
4  & {}  & 1 & 0 & 1 & 0 & 1 & 0 & 1 & 0 & 0 & 1  & 0  & 1  & 0  & 1  & 0  & 1  \\
5  & {}  & 0 & 1 & 0 & 1 & 0 & 1 & 0 & 1 & 1 & 0  & 1  & 0  & 1  & 0  & 1  & 0  \\
6  & {}  & 1 & 0 & 1 & 0 & 1 & 0 & 1 & 0 & 0 & 1  & 0  & 1  & 0  & 1  & 0  & 1  \\
7  & {}  & 0 & 1 & 0 & 1 & 0 & 1 & 0 & 1 & 1 & 0  & 1  & 0  & 1  & 0  & 1  & 0  \\
8  & {}  & 1 & 0 & 1 & 0 & 1 & 0 & 1 & 0 & 0 & 1  & 0  & 1  & 0  & 1  & 0  & 1  \\
9  & {}  & 1 & 0 & 1 & 0 & 1 & 0 & 1 & 0 & 0 & 1  & 0  & 1  & 0  & 1  & 0  & 1  \\
10 & {}  & 0 & 1 & 0 & 1 & 0 & 1 & 0 & 1 & 1 & 0  & 1  & 0  & 1  & 0  & 1  & 0  \\
11 & {}  & 1 & 0 & 1 & 0 & 1 & 0 & 1 & 0 & 0 & 1  & 0  & 1  & 0  & 1  & 0  & 1  \\
12 & {}  & 0 & 1 & 0 & 1 & 0 & 1 & 0 & 1 & 1 & 0  & 1  & 0  & 1  & 0  & 1  & 0  \\
13 & {}  & 1 & 0 & 1 & 0 & 1 & 0 & 1 & 0 & 0 & 1  & 0  & 1  & 0  & 1  & 0  & 1  \\
14 & {}  & 0 & 1 & 0 & 1 & 0 & 1 & 0 & 1 & 1 & 0  & 1  & 0  & 1  & 0  & 1  & 0  \\
15 & {}  & 1 & 0 & 1 & 0 & 1 & 0 & 1 & 0 & 0 & 1  & 0  & 1  & 0  & 1  & 0  & 1  \\
16 & {}  & 0 & 1 & 0 & 1 & 0 & 1 & 0 & 1 & 1 & 0  & 1  & 0  & 1  & 0  & 1  & 0 \\
\hline
\end{array}  {~~~~~~~~~~~~~~~} 
$$
$~~~~~$ $~~~~~~$ {\bf {Table}} {\bf {29:}} Gadget
Values For ${\bm {\cal P}_{[5]}}$ $\times$ ${\bm {\cal P}_{[6]}}$ With Different Boolean
Factors

$$
\begin{array}{cc|cccccccccccccccc}
{~} & {\cal S}^{\prime} & 1 & 2  & 3 & 4 & 5 & 6 & 7 & 8  & 9 & 10 & 11 & 12 & 
13 & 14 & 15 & 16  \\
{\cal S}  & {} & {} & {} & {} & {} & {} & {} & {} & {} & {} & {} & {} & {} & {} & {} &  {} \\
\hline
1  & {}  & 1 & 0 & 1 & 0 & 1 & 0 & 1 & 0 & 0 & 1  & 0  & 1  & 0  & 1  & 0  & 1  \\
2  & {}  & 0 & 1 & 0 & 1 & 0 & 1 & 0 & 1 & 1 & 0  & 1  & 0  & 1  & 0  & 1  & 0  \\
3  & {}  & 1 & 0 & 1 & 0 & 1 & 0 & 1 & 0 & 0 & 1  & 0  & 1  & 0  & 1  & 0  & 1  \\
4  & {}  & 0 & 1 & 0 & 1 & 0 & 1 & 0 & 1 & 1 & 0  & 1  & 0  & 1  & 0  & 1  & 0  \\
5  & {}  & 1 & 0 & 1 & 0 & 1 & 0 & 1 & 0 & 0 & 1  & 0  & 1  & 0  & 1  & 0  & 1  \\
6  & {}  & 0 & 1 & 0 & 1 & 0 & 1 & 0 & 1 & 1 & 0  & 1  & 0  & 1  & 0  & 1  & 0  \\
7  & {}  & 1 & 0 & 1 & 0 & 1 & 0 & 1 & 0 & 0 & 1  & 0  & 1  & 0  & 1  & 0  & 1  \\
8  & {}  & 0 & 1 & 0 & 1 & 0 & 1 & 0 & 1 & 1 & 0  & 1  & 0  & 1  & 0  & 1  & 0  \\
9  & {}  & 0 & 1 & 0 & 1 & 0 & 1 & 0 & 1 & 1 & 0  & 1  & 0  & 1  & 0  & 1  & 0  \\
10 & {}  & 1 & 0 & 1 & 0 & 1 & 0 & 1 & 0 & 0 & 1  & 0  & 1  & 0  & 1  & 0  & 1  \\
11 & {}  & 0 & 1 & 0 & 1 & 0 & 1 & 0 & 1 & 1 & 0  & 1  & 0  & 1  & 0  & 1  & 0  \\
12 & {}  & 1 & 0 & 1 & 0 & 1 & 0 & 1 & 0 & 0 & 1  & 0  & 1  & 0  & 1  & 0  & 1  \\
13 & {}  & 0 & 1 & 0 & 1 & 0 & 1 & 0 & 1 & 1 & 0  & 1  & 0  & 1  & 0  & 1  & 0  \\
14 & {}  & 1 & 0 & 1 & 0 & 1 & 0 & 1 & 0 & 0 & 1  & 0  & 1  & 0  & 1  & 0  & 1  \\
15 & {}  & 0 & 1 & 0 & 1 & 0 & 1 & 0 & 1 & 1 & 0  & 1  & 0  & 1  & 0  & 1  & 0  \\
16 & {}  & 1 & 0 & 1 & 0 & 1 & 0 & 1 & 0 & 0 & 1  & 0  & 1  & 0  & 1  & 0  & 1 \\
\hline
\end{array}  {~~~~~~~~~~~~~~~} 
$$
$~~~~~$ $~~~~~~$ {\bf {Table}} {\bf {30:}} Gadget
Values For ${\bm {\cal P}_{[6]}}$ $\times$ ${\bm {\cal P}_{[6]}}$ With Different Boolean
Factors

\newpage
\noindent
$\bm{ \rm{
Appendix ~E: ~Cycle ~  Labelling~ Conventions
}} \label{AppE}
$

$~~~~$ In this appendix we wish to discuss a point about using cycle notation
to describe permutation matrices.  There is an ambiguity in notation
that we need to address in view of some our past works.  

Let us begin by writing a permutation in the form of a matrix.  For the
purposes of our discussion we will concentrate on $\bm M$ and
$\bm N$ 
where
$$
{\bm M}  ~=~
\left[\begin{array}{cccc}
~1 & ~0 &  ~0  &  ~0 \\
~0 & ~0 &  ~0  &  ~ 1 \\
~0 & ~1 &  ~0  &  ~0 \\
~0 & ~0 &  ~ 1  &  ~0 \\
\end{array}\right]   ~~~,~~~
{\bm N}  ~=~
\left[\begin{array}{cccc}
~0 & ~1 &  ~0  &  ~0 \\
~0 & ~0 &  ~0  &  ~ 1 \\
~1 & ~0 &  ~0  &  ~0 \\
~0 & ~0 &  ~ 1  &  ~0 \\
\end{array}\right]    ~~~,~~~
{\bm O}  ~=~
\left[\begin{array}{cccc}
~1 & ~0 &  ~0  &  ~0 \\
~0 & ~0 &  ~1  &  ~ 0 \\
~0 & ~1 &  ~0  &  ~0 \\
~0 & ~0 &  ~ 0  &  ~1 \\
\end{array}\right]   ~~.
\eqno(E.1)
$$

If we use the convention the numbers to appear in the cycle should denote 
the presence of the non-vanishing entries as read from an upward to 
downward direction along each column from the top, but which does not 
appear as a diagonal entry.  We can refer to this as ``the read down 
convention.''

Applying this rule to $\bm M$ we see that the second column has a non-vanishing
entry in the third row, the third column has a non-vanishing entry in the fourth 
row and the fourth column has a non-vanishing entry in the second row.  This 
suggest a notation for the matrices $\bm M$ in the form of (234) as we take 
each column from left to right.  We can apply the same logic to $\bm N$ to
suggest a notation name (1342) and for $\bm O$ notation name (23). 

However, there is another possible convention.

We could use the convention the numbers to appear in the cycle should 
denote the presence of the non-vanishing entries as read from left side to 
right side along each row from the left, but which does not appear as a 
diagonal entry.  We can refer to this as ``the read across convention.''

Applying this rule to $\bm M$ we see that the second row has a non-vanishing
entry in the fourth column, the fourth row has a non-vanishing entry in the 
third column, and third row has a non-vanishing entry in the second column.  
This suggest a notation for the matrix $\bm M$ in the form of (243) as we take 
each column from top to bottom.  We can apply the same logic to $\bm N$ to
suggest a notation name (1243) and for $\bm O$ notation name (23).

This discussion illustrates that for 2-cycles like $\bm O$, either convention
leads to the same name.  However, for 3-cycles and 4-cycles, the notational
names are different for the same matrix depending on the convention used.
However, there is a simply ``translation'' between the two conventions.  If the 
notation of a permutation is given by one expression in the read down
convention, the notation for the same permutation can be found by reading
in a ``backward ordering'' for the notation in the other convention.  When 
(234) is read backward it becomes (432) or (using cyclicity) (243).  Similarly, 
when (1342) is read backward, it becomes (2431) or (using cyclicity) (1243).  
For 2-cycles both conventions lead to the same expression.
 
The L-matrices for each of the respective representations $(CM)$, $(TM)$,
and $(VM)$ are given by
$$  {
{\bm {\rm L}}{}^{(CM)}_{1}   ~=~
\left[\begin{array}{cccc}
~1 & ~~0 &  ~~0  &  ~~0 \\
~0 & ~~0 &  ~~0  &  ~-\, 1 \\
~0 & ~~1 &  ~~0  &  ~~0 \\
~0 & ~~0 &  ~-\, 1  &  ~~0 \\
\end{array}\right] ~~~,~~~
{\bm {\rm L}}{}^{(CM)}_{2}   ~=~
\left[\begin{array}{cccc}
~0 & ~~1 &  ~~0  &  ~ \, \, 0 \\
~0 & ~~ 0 &  ~~1  &  ~~0 \\
-\, 1 & ~~ 0 &  ~~0  &  ~~0 \\
~ 0 & ~~~0 &  ~~0  &   -\, 1 \\
\end{array}\right]  ~~~, }
$$
$$  {
  {\bm {\rm L}}{}^{(CM)}_{3}   ~=~
\left[\begin{array}{cccc}
~0 & ~~0 &  ~~1  &  ~~0 \\
~0 & ~- \, 1 &  ~~0  &  ~~0 \\
~0 & ~~0 &  ~~0  &  -\, 1 \\
~1 & ~~0 &  ~~0  &  ~~0 \\
\end{array}\right] ~~~,~~~
  {\bm {\rm L}}{}^{(CM)}_{4}   ~=~
\left[\begin{array}{cccc}
~0 & ~~0 &  ~~0  &  ~ \, \, 1 \\
~1 & ~~ 0 &  ~~0  &  ~~0 \\
~0 & ~~ 0 &  ~~1  &  ~~0 \\
~ 0 & ~~~1 &  ~~0  &   ~~0  \\
\end{array}\right]  ~~.  }
 \label{chiD0F2}
$$

$$  { 
  {\bm {\rm L}}{}^{(TM)}_{1}   ~=~
\left[\begin{array}{cccc}
~1 & ~0 &  ~0  &  ~0 \\
~0 & ~0 &  -\, 1  &  ~ 0 \\
~0 & ~0 &  ~0  &  -\,1 \\
~0 & -\,1 &  ~ 0  &  ~0 \\
\end{array}\right] ~~~,~~~
  {\bm {\rm L}}{}^{(TM)}_{2}   ~=~
\left[\begin{array}{cccc}
~0 & ~1 &  ~0  &  ~  0 \\
~0 & ~ 0 &  ~0  &  ~ 1 \\
~0 & ~ 0 &  -\,1  &  ~ 0 \\
 ~ 1 & ~0 &  ~0  &   ~ 0 \\
\end{array}\right]  ~~~,
}    $$
$$ {~~~~} { 
  {\bm {\rm L}}{}^{(TM)}_{3}   ~=~
\left[\begin{array}{cccc}
~0 & ~0 &  ~1  &  ~0 \\
~1 & ~0 &  ~ 0  &  ~ 0 \\
~0 & ~1 &  ~0  &   ~0 \\
~0 & ~0 &  ~ 0  &  -\, 1 \\
\end{array}\right] ~~~~~~,~~~
  {\bm {\rm L}}{}^{(TM)}_{4}   ~=~
\left[\begin{array}{cccc}
~0 & ~0 &  ~0  &  ~  1 \\
~0 & -\, 1 &  ~0  &  ~ 0 \\
~1 & ~ 0 &  ~0  &  ~ 0 \\
 ~0 & ~0 &  ~1  &   ~ 0 \\
\end{array}\right]  ~~~,  }
\label{tenD0F}
$$

$$ { 
  {\bm {\rm L}}{}^{(VM)}_{1}   ~=~
\left[\begin{array}{cccc}
~0 & ~1 &  ~ 0  &  ~ 0 \\
~0 & ~0 &  ~0  &  -\,1 \\
~1 & ~0 &  ~ 0  &  ~0 \\
~0 & ~0 &  -\, 1  &  ~0 \\
\end{array}\right] ~~~,~~~
  {\bm {\rm L}}{}^{(VM)}_{2}   ~=~
\left[\begin{array}{cccc}
~1 & ~ 0 &  ~0  &  ~ 0 \\
~0 & ~ 0 &  ~1  &  ~ 0 \\
 ~0 & - \, 1 &  ~0  &   ~ 0 \\
~0 & ~0 &  ~0  &  -\, 1 \\
\end{array}\right]  ~~~, }
$$
$$ {~~~~} { 
  {\bm {\rm L}}{}^{(VM)}_{3}   ~=~
\left[\begin{array}{cccc}
~0 & ~0 &  ~ 0  &  ~ 1 \\
~0 & ~1 &  ~0  &   ~0 \\
~0 & ~0 &  ~ 1  &  ~0 \\
~1 & ~0 &  ~0  &  ~0 \\
\end{array}\right] ~~~~~~,~~~
  {\bm {\rm L}}{}^{(VM)}_{4}   ~=~
\left[\begin{array}{cccc}
~0 & ~0 &  ~1  &  ~ 0 \\
-\,1 & ~ 0 &  ~0  &  ~ 0 \\
 ~0 & ~0 &  ~0  &   - \, 1 \\
~0 & ~1 &  ~0  &  ~  0 \\
\end{array}\right]  ~~~. }
\eqno(E.2)
$$
Using the ``read down convention'' these matrices in (E.2) become the expressions given in
(E.3).
$$  \eqalign{
{\bm {\rm L}}{}^{(CM)}_{1}   ~&=~ (10)_b (234) ~~,~~ {\bm {\rm L}}{}^{(CM)}_{2}
~=~ (12)_b (132) ~~,~~ {\bm {\rm L}}{}^{(CM)}_{3} ~=~ (6)_b (143)
~~,~~ {\bm {\rm L}}{}^{(CM)}_{4} ~=~ (0)_b (124)
~~~,    \cr
{\bm {\rm L}}{}^{(TM)}_{1}   ~&=~  (14)_b (243)  ~~,~~ {\bm {\rm L}}{}^{(TM)}_{2}
~=~  (4)_b (142)  ~~,~~ {\bm {\rm L}}{}^{(TM)}_{3} ~=~  (8)_b (123) 
~~,~~ {\bm {\rm L}}{}^{(TM)}_{4} ~=~  (2)_b (134) 
~~~,   \cr
{\bm {\rm L}}{}^{(VM)}_{1}   ~&=~  (10)_b (1342)  ~~,~~ {\bm {\rm L}}{}^{(VM)}_{2}
~=~  (4)_b (23)  ~~,~~ {\bm {\rm L}}{}^{(VM)}_{3} ~=~  (0)_b (14) 
~~,~~ {\bm {\rm L}}{}^{(VM)}_{4} ~=~  (6)_b (1243) 
~~~. }
\eqno(E.3)
$$
Using the ``read across convention'' these matrices in (E.2) become the expressions given in
(E.4).
$$  \eqalign{
{\bm {\rm L}}{}^{(CM)}_{1}   ~&=~ (10)_b (243) ~~,~~ {\bm {\rm L}}{}^{(CM)}_{2}
~=~ (12)_b (123) ~~,~~ {\bm {\rm L}}{}^{(CM)}_{3} ~=~ (6)_b (134)
~~,~~ {\bm {\rm L}}{}^{(CM)}_{4} ~=~ (0)_b (142)
~~~,    \cr
{\bm {\rm L}}{}^{(TM)}_{1}   ~&=~  (14)_b (234)  ~~,~~ {\bm {\rm L}}{}^{(TM)}_{2}
~=~  (4)_b (124)  ~~,~~ {\bm {\rm L}}{}^{(TM)}_{3} ~=~  (8)_b (132) 
~~,~~ {\bm {\rm L}}{}^{(TM)}_{4} ~=~  (2)_b (143) 
~~~,   \cr
{\bm {\rm L}}{}^{(VM)}_{1}   ~&=~  (10)_b (1243)  ~~,~~ {\bm {\rm L}}{}^{(VM)}_{2}
~=~  (4)_b (23)  ~~,~~ {\bm {\rm L}}{}^{(VM)}_{3} ~=~  (0)_b (14) 
~~,~~ {\bm {\rm L}}{}^{(VM)}_{4} ~=~  (6)_b (1342) 
~~~. }
\eqno(E.4)
$$
All the expressions, tables, etc. in this work are written in the read across convention.

\newpage
$$~~$$

\end{document}